\documentclass[letterpaper,titlepage,11pt]{article}

\usepackage{jheppub}

\usepackage[T1]{fontenc}
\usepackage[utf8]{inputenc} % pdfLaTeX için
\usepackage[turkish,english]{babel}

\usepackage{amsmath,amssymb,amsfonts}
\usepackage{esint}
\usepackage{empheq}
\usepackage{multirow}
\usepackage{enumerate}
\usepackage{needspace}

\usepackage{xcolor}
\usepackage{tikz}

\usepackage{textcomp}
\usepackage{bbm}

\usepackage{lineno}
\usepackage[active]{srcltx}

\usepackage{mathrsfs}
\usepackage{booktabs}

\usepackage{caption}

\usepackage{tikz}
\usetikzlibrary{arrows.meta,positioning}
\DeclareMathOperator{\tr}{tr}

\newcommand{\Lt}{{\tt L}}

\newcommand{\St}{{\tt S}}
\newcommand{\Ut}{{\tt U}}
\newcommand{\ut}{{\tt u}}
\title{\boldmath{Holonomies and Boundary Symmetries in the Discrete Warped Chern--Simons Gravity}}
\subheader{ITU Preprint,\,\\ \today}
\newcommand{\itu}{\dagger}

\author[\itu]{H.~T.~\"Ozer}
\emailAdd{ozert@itu.edu.tr}

\author[\itu]{Ayt\"ul~Filiz}
\emailAdd{aytulfiliz@itu.edu.tr}

\affiliation[\itu]{Istanbul Technical University,\,Faculty of Science and Letters,
\,Physics Department,\\34469 Maslak,\,Istanbul,Turkey.}
\abstract{
\begin{abstract}
\noindent We investigate a discrete warped Chern--Simons description of three-dimensional warped gravity based on boundary holonomies and monodromy sectors. Starting from the lower-spin $\St \Lt(2,\mathbb R)\oplus \Ut(1)$ gauge structure associated with warped $\mathrm{AdS}_3$ holography and warped conformal field theories (WCFTs), we construct a discrete boundary framework in which ordered products of link holonomies replace continuous gauge connections along noncontractible cycles. In this setting, boundary monodromies become the primary gauge-invariant observables characterizing the physical sectors of the theory. 
We show that the hyperbolic, elliptic, and parabolic sectors naturally arise from the conjugacy classes of the discrete $\St \Lt(2,\mathbb R)$ monodromy, while the additional $\Ut(1)$ holonomy supplies the warped contribution to the boundary charges. Using these monodromy invariants, we derive a discrete entropy relation entirely from boundary holonomy data without relying on a smooth geometric thermal background. The resulting entropy reproduces the characteristic warped black-hole and WCFT structure in the continuum limit.
We further demonstrate that the continuum warped holonomy conditions are recovered from the large-lattice limit of the ordered boundary products, establishing a direct correspondence between discrete monodromies and continuous Wilson loops. 
Our analysis suggests that warped gravitational thermodynamics may be understood from a fundamentally holonomy-based perspective in which boundary monodromy sectors provide an alternative
organizational description of the physical states within the discrete warped framework.
\\

\end{abstract}
}

\begin{document}
\maketitle
\flushbottom
%%%%%%%%%%%%%%%%%%%%%%%%%%%%%%%%%%%%%%%%%%%%%%%%%%%%%%%%%%%%%%%%%%%%%%%%%%%%%%%%%%%%%%%%%%%%
%%%%%%%%%%%%%%%%%%%%%%%%%%%%%%%%%%%%%%%%%%%%%%%%%%%%%%%%%%%%%%%%%%%%%%%%%%%%%%%%%%%%%%%%%%%%
%%%%%%%%%%%%        INTRODUCTION        %%%%%%%%%%%%%%%%%%%%%%%%%%%%%%%%%%%%%%%%%%%%%%%%%%%%
%%%%%%%%%%%%%%%%%%%%%%%%%%%%%%%%%%%%%%%%%%%%%%%%%%%%%%%%%%%%%%%%%%%%%%%%%%%%%%%%%%%%%%%%%%%%
%%%%%%%%%%%%%%%%%%%%%%%%%%%%%%%%%%%%%%%%%%%%%%%%%%%%%%%%%%%%%%%%%%%%%%%%%%%%%%%%%%%%%%%%%%%%
\section{Introduction}
\label{sec:intro}
%1
Three-dimensional gravity provides a remarkable framework in which
gravitational dynamics may be described entirely in gauge-theoretic terms
through the Chern--Simons (CS) formalism~\cite{Witten1991}. In
asymptotically AdS$_3$ backgrounds, this perspective establishes a deep
relation between bulk gauge fields, boundary degrees of freedom, and
asymptotic symmetry structures governed by infinite-dimensional symmetry
algebras~\cite{BrownHenneaux1986}. Within this framework, global and
thermodynamic properties of gravitational configurations are naturally
encoded in holonomies of the gauge connection around noncontractible
cycles, providing a direct gauge-theoretic characterization of black hole
sectors and boundary observables.
%2

Warped AdS$_3$ holography extends this structure beyond the relativistic
conformal regime through boundary conditions leading to a
Virasoro$\oplus\,\ut(1)$ asymptotic symmetry algebra rather than the
conventional Virasoro$\times$Virasoro
structure~\cite{CompereSongStrominger,CompereDetournay}. This holographic
setting is closely related to warped conformal field theories (WCFTs),
which constitute non-Lorentz-invariant quantum field theories
characterized by Virasoro--Kac--Moody symmetries and chiral scaling
behavior~\cite{HofmanStrominger,DetournayHartmanHofman}. Warped black
holes and their thermal sectors provide further examples of this structure
in three-dimensional gravity~\cite{AnninosWarped}. A minimal
gauge-theoretic realization of warped holography is furnished by
$\St\Lt(2,\mathbb{R})\oplus\Ut(1)$ Chern--Simons theory, commonly referred
to as lower-spin gravity, where boundary currents, warped thermal sectors,
and holographic charges are encoded directly in the gauge
connection~\cite{HofmanRollier,AfsharLowerSpin,AzeyanagiDetournayRiegler2019}.
%3

An important feature of warped Chern--Simons gravity is that thermodynamic
quantities are governed by gauge-invariant holonomy data associated with
nontrivial cycles of the boundary connection. In warped holography and
higher-spin gravity, thermal entropy is encoded through holonomy
conditions, Wilson lines, and conjugacy structures of the underlying gauge
connection~\cite{GutperleKraus,HenneauxRey,AmmonGutperleKrausPerlmutter,AzeyanagiDetournayRiegler2019}.
Thermodynamic consistency and smoothness conditions are therefore
controlled directly by generalized holonomies rather than local geometric
quantities alone. These observations strongly suggest that boundary
monodromies and holonomy sectors constitute the fundamental carriers of
thermal information in warped Chern--Simons gravity.
%4

Motivated by this viewpoint, it is natural to ask whether warped
thermodynamic structures may already be formulated directly at the discrete
level in terms of boundary holonomies themselves. Lattice gauge theory and
discrete gauge constructions provide a natural framework for such an
approach, where group-valued holonomies are treated as natural
gauge-invariant variables from the
outset~\cite{Wilson1974,Kogut1979,Rothe2012,Dittrich2008}. Related
holonomy-based studies in lower-dimensional gravity have further
demonstrated that boundary symmetry structures and monodromy sectors may
be described directly in terms of discrete gauge data~\cite{Ozer:2026qlp}.
%5

Unlike our previous discrete JT/BF construction, where the emphasis was
placed primarily on two-dimensional BF gravity, affine boundary symmetries,
and the emergence of asymptotic structures from discrete boundary
holonomies, the present work investigates a qualitatively different
holographic setting based on the lower-spin
$\St\Lt(2,\mathbb{R})\oplus\Ut(1)$ Chern--Simons structure associated with
warped holography and WCFTs. The present framework incorporates an
additional Abelian holonomy sector whose interplay with the non-Abelian
monodromy modifies the effective thermodynamic structure through warped
charge contributions absent in the JT/BF setting. It furthermore introduces
warped thermal holonomies, Wilson-loop observables, and
chamber-dependent thermodynamic sectors that do not arise in the JT/BF
framework. A further conceptual difference is that the present work focuses
directly on the organization of warped thermodynamic sectors in terms of
boundary monodromy data, rather than being governed solely by the emergence
of boundary symmetry structures.
%6

The present work develops a discrete boundary description of warped
Chern--Simons gravity in terms of group-valued holonomies and boundary
monodromies, where these gauge-invariant data characterize the global
boundary sectors of the theory rather than appearing only as derived
quantities associated with an underlying smooth geometry. This constitutes
a shift in emphasis relative to the conventional continuum formulation, in
which smooth gauge connections and regularity conditions are typically
taken as the starting point, while holonomies arise as derived observables.
In the discrete setting developed here, greater emphasis is instead placed
on boundary monodromies and their conjugacy classes as the structures
organizing the global sectors of the theory. Smooth warped geometries then
emerge in the large-lattice limit as continuum realizations of the
corresponding monodromy sectors.
%7

The central contribution of this work lies not in the entropy formula
itself --- which reproduces the warped Cardy
structure~\cite{DetournayHartmanHofman} and the holonomy-based thermal
entropy~\cite{AzeyanagiDetournayRiegler2019} already established in the
continuum lower-spin theory --- but in the \emph{derivation framework}. We
demonstrate that the same entropy formula is already fully encoded in
discrete boundary monodromy data, without postulating a smooth thermal
geometry, a Euclidean regularity condition, or a continuous gauge
connection as a logical prerequisite. The organizing principle underlying
this result is the \emph{monodromy-first} perspective, in which the warped
thermodynamic sector is reconstructed entirely from the conjugacy class of
the boundary $\mathrm{SL}(2,\mathbb{R})$ holonomy and the Abelian warped
charge alone.
%8

The purpose of the present work is not to replace the continuum warped
description, but to reorganize its thermodynamic content directly at the
level of boundary holonomy data. An important consequence of this viewpoint
is that the hyperbolic, elliptic, and parabolic sectors arise naturally as
distinct boundary monodromy chambers of the same underlying gauge system,
admitting a unified treatment within a single discrete framework; the
continuum warped geometries are then recovered only as smooth large-lattice
realizations of these underlying sectors.
%9

Within this framework, warped asymptotic symmetries, thermal sectors, and
entropy emerge directly from the structure of discrete boundary monodromies
and their associated holonomy chambers. The continuous warped
Virasoro--Kac--Moody symmetry structure then appears naturally as the
continuum realization of the underlying discrete boundary holonomy sector.
%10

This paper is organized as follows. In
Section~\ref{ContinuousWarpedCSGravity}, we review the continuous
lower-spin warped Chern--Simons theory, including warped boundary
conditions, Virasoro--Kac--Moody asymptotic symmetries, holonomy
conditions, and the associated WCFT interpretation.
Section~\ref{DiscreteWarpedCSGeometry} develops the discrete warped
Chern--Simons geometry and introduces the corresponding boundary lattice,
group-valued link variables, and discrete flatness constraints. In
Section~\ref{BoundaryMonodromiesAndWarpedSectors}, we analyze boundary
monodromies, conjugacy classes, stabilizer structures, and the resulting
warped holonomy chambers.
Section~\ref{HolonomyEntropyInTheDiscreteTheory} is devoted to the
derivation of warped entropy directly from boundary monodromy invariants
together with its continuum matching to the continuous thermal warped entropy. 
In Section~\ref{WilsonLoopsAndDiscreteThermalCycles}, we
investigate Wilson loops and discrete thermal cycles, emphasizing the role
of noncontractible boundary holonomies in warped thermodynamics.
Section~\ref{QuantumWarpedHolonomySectors} discusses the quantum
interpretation of the resulting monodromy chambers, including the chamber
decomposition of the Hilbert space and the associated density of states.
Finally, Section~\ref{DiscussionAndOutlook} contains our conclusions and
future directions. Additional technical derivations and supplementary
discussions are collected in the appendices.

%%%%%%%%%%%%%%%%%%%%%%%%%%%%%%%%%%%%%%%%%%%%%%%%%%%%%%%%%%%%%%%%%%%%%%%%%%%%%%%%%%%%%%%%%%
%%%%%%%%%%%%%%%%%%%%%%%%%%%%%%%%%%%%%%%%%%%%%%%%%%%%%%%%%%%%%%%%%%%%%%%%%%%%%%%%%%%%%%%%%%
%%%%%%%%%%           SECTION-2        %%%%%%%%%%%%%%%%%%%%%%%%%%%%%%%%%%%%%%%%%%%%%%%%%%%%
%%%%%%%%%%%%%%%%%%%%%%%%%%%%%%%%%%%%%%%%%%%%%%%%%%%%%%%%%%%%%%%%%%%%%%%%%%%%%%%%%%%%%%%%%%
%%%%%%%%%%%%%%%%%%%%%%%%%%%%%%%%%%%%%%%%%%%%%%%%%%%%%%%%%%%%%%%%%%%%%%%%%%%%%%%%%%%%%%%%%%

\section{Continuous Warped Chern--Simons Gravity}
\label{ContinuousWarpedCSGravity}

This section reviews the continuous lower-spin description of warped gravity that will serve as the reference point for the discrete construction developed in the following sections. The relevant gauge structure is based on $\St \Lt(2,\mathbb R)\oplus \Ut(1)$ \cite{CompereSongStrominger, CompereDetournay, HofmanRollier}, where the $\St \Lt(2,\mathbb{R})$ sector carries the gravitational part of the warped background and the additional $\Ut(1)$ sector accounts for the warped current contribution. The purpose of this review is not to rederive all details of warped black-hole physics, but to isolate the ingredients that will later
admit a direct discrete counterpart: boundary conditions, canonical charges, Virasoro--Kac--Moody asymptotic symmetry, holonomy constraints, and entropy \cite{DetournayHartmanHofman, AzeyanagiDetournayRiegler2019}.

The central point is that the thermodynamic information of warped black holes can be expressed in terms of gauge data along noncontractible cycles. In the continuous theory, this information is obtained from the
holonomy of the connection around thermal and angular cycles \cite{GutperleKraus, AmmonGutperleKrausPerlmutter}. In the discrete theory, the corresponding role will be played by ordered products of boundary link variables. Thus, the continuous discussion below should be read as the smooth reference limit of a more fundamental monodromy-based description. We first introduce the lower-spin $\St \Lt(2,\mathbb R)\oplus \Ut(1)$
Chern--Simons theory and fix the algebraic conventions. We then recall the warped boundary conditions and the resulting asymptotic symmetry algebra \cite{CompereSongStrominger, BrownHenneaux1986, Carlip:2005zn}. After that, we summarize the holonomy conditions leading to the thermal entropy. Finally, we explain the WCFT interpretation of the entropy formula and identify the role of the vacuum charges and tilt parameter
\cite{HofmanStrominger,DetournayHartmanHofman}.
%%%%%%%%%%%%%%%%%%%%%%%%%%%%%%%%%%%%%%%%%%%%%%%%%%%%%%%%%%%%%%%%%%%%%%%%%%%%%%%%%%%%%%%%%%
%%%%%%%%%%%%%%%%%%%%%%%%%%%%%%%%%%%%%%%%%%%%%%%%%%%%%%%%%%%%%%%%%%%%%%%%%%%%%%%%%%%%%%%%%%
%%%%%%%%%%           SECTION-2.1      %%%%%%%%%%%%%%%%%%%%%%%%%%%%%%%%%%%%%%%%%%%%%%%%%%%%
%%%%%%%%%%%%%%%%%%%%%%%%%%%%%%%%%%%%%%%%%%%%%%%%%%%%%%%%%%%%%%%%%%%%%%%%%%%%%%%%%%%%%%%%%%
%%%%%%%%%%%%%%%%%%%%%%%%%%%%%%%%%%%%%%%%%%%%%%%%%%%%%%%%%%%%%%%%%%%%%%%%%%%%%%%%%%%%%%%%%%

\subsection{Lower-Spin \texorpdfstring{$\St \Lt(2,\mathbb R)\oplus \Ut(1)$}{SL(2,R)+U(1)} Chern--Simons Theory}
\label{LowerSpinCSTheory}

We begin with the continuous lower-spin gauge theory describing warped $\mathrm{AdS}_3$ geometries
\cite{CompereSongStrominger,CompereDetournay,HofmanRollier,AzeyanagiDetournayRiegler2019,Merbis:2014vja}. The theory is governed by a Chern--Simons action based on the direct sum algebra  $\mathfrak{sl}(2,\mathbb R)\oplus \mathfrak{u}(1)$, where the non-Abelian sector captures the gravitational part of the warped background while the additional Abelian sector generates the warped current contribution. Unlike ordinary $\mathrm{AdS}_3$ gravity, whose asymptotic structure is controlled entirely by two Virasoro sectors, warped geometries require the simultaneous presence of a Virasoro algebra and an affine $\ut(1)$ current algebra. 

The corresponding bulk action is given by
\cite{AzeyanagiDetournayRiegler2019}
\begin{equation}
I_{\text{CS}}
=
\frac{k}{4\pi}
\int_{\mathcal M}
\tr\!\left(
A\wedge dA
+
\frac{2}{3}A\wedge A\wedge A
\right)
+
\frac{\kappa}{8\pi}
\int_{\mathcal M}
C\wedge dC \, ,
\label{LowerSpinCSAction}
\end{equation}
where $A$ denotes the $\St \Lt(2,\mathbb R)$ gauge connection, $C$ is the $\Ut(1)$ connection, $k$ is the Chern--Simons level associated with the non-Abelian sector, and $\kappa$ determines the normalization of the Abelian current sector. The manifold $\mathcal M$ is taken to possess cylindrical topology with coordinates $(t,\rho,\phi)$, where $\phi$ parametrizes the angular boundary cycle.

The $\mathfrak{sl}(2,\mathbb R)$ generators are denoted by $L_n$ with $n=\{-1,0,1\}$ and satisfy
\begin{equation}
[L_n,L_m]
=
(n-m)L_{n+m} \, .
\label{SL2Algebra}
\end{equation}
The additional Abelian generator $S$ obeys
\begin{equation}
[L_n,S]=0,
\qquad
[S,S]=0 \, .
\label{U1Algebra}
\end{equation}
The invariant bilinear form is chosen such that
\begin{equation}
\tr(L_1L_{-1})=-1,
\qquad
\tr(L_0L_0)=\frac12,
\qquad
\tr(SS)=1 \, .
\label{InvariantBilinearForm}
\end{equation}
These conventions fix the normalization of the canonical boundary charges and determine the corresponding Virasoro and affine current levels appearing in the asymptotic symmetry algebra.

The equations of motion obtained from Eq.~\eqref{LowerSpinCSAction} are simply the flatness conditions
\begin{equation}
F=dA+A\wedge A=0,
\qquad
dC=0 \, .
\label{LowerSpinFlatnessConditions}
\end{equation}
Thus, the theory possesses no local propagating bulk degrees of freedom, and all physically relevant information is encoded in boundary data and global holonomy structures. This topological character will play a central role in the discrete construction developed later in Section~\ref{DiscreteWarpedCSGeometry}, where continuous gauge connections will be replaced by boundary link holonomies.

For later convenience, we partially fix the radial dependence of the gauge fields by writing
\begin{equation}
A=b^{-1}(a+d)b,
\qquad
C=c \, ,
\label{RadialGaugeChoice}
\end{equation}
where $b=b(\rho)=e^{\rho L_0}$ carries the radial dependence while $a$ and $c$ depend only on the boundary coordinates $(t,\phi)$. This decomposition isolates the boundary dynamical sector and simplifies the asymptotic symmetry analysis.

The lower-spin structure summarized above provides the continuous warped gauge system that will serve as the reference point for the discrete theory constructed in the following sections. In particular, the flatness conditions, boundary holonomies, and asymptotic current structure will admit direct discrete counterparts formulated in terms of ordered products of boundary link variables.

%%%%%%%%%%%%%%%%%%%%%%%%%%%%%%%%%%%%%%%%%%%%%%%%%%%%%%%%%%%%%%%%%%%%%%%%%%%%%%%%%%%%%%%%%%
%%%%%%%%%%%%%%%%%%%%%%%%%%%%%%%%%%%%%%%%%%%%%%%%%%%%%%%%%%%%%%%%%%%%%%%%%%%%%%%%%%%%%%%%%%
%%%%%%%%%%           SECTION-2.2      %%%%%%%%%%%%%%%%%%%%%%%%%%%%%%%%%%%%%%%%%%%%%%%%%%%%
%%%%%%%%%%%%%%%%%%%%%%%%%%%%%%%%%%%%%%%%%%%%%%%%%%%%%%%%%%%%%%%%%%%%%%%%%%%%%%%%%%%%%%%%%%
%%%%%%%%%%%%%%%%%%%%%%%%%%%%%%%%%%%%%%%%%%%%%%%%%%%%%%%%%%%%%%%%%%%%%%%%%%%%%%%%%%%%%%%%%%
\subsection{Boundary Conditions and Asymptotic Symmetries}
\label{BoundaryConditionsAndAsymptoticSymmetries}

We now specify the boundary conditions describing the warped sector of the lower-spin theory introduced in Section~\ref{LowerSpinCSTheory}. The purpose of these boundary conditions is to isolate the phase space associated with warped $\mathrm{AdS}_3$ geometries while simultaneously preserving a nontrivial asymptotic symmetry structure generated by a Virasoro algebra together with an affine $\ut(1)$ current algebra
\cite{CompereSongStrominger,CompereDetournay,DetournayHartmanHofman}.

Using the radial decomposition introduced in Eq.~\eqref{RadialGaugeChoice}, we write the reduced boundary connections as \cite{AzeyanagiDetournayRiegler2019}
\begin{equation}
a=a_t\,dt+a_\phi\,d\phi,
\qquad
c=c_t\,dt+c_\phi\,d\phi \, .
\label{BoundaryConnections}
\end{equation}
The warped boundary conditions are chosen as
\begin{align}
a_\phi
&=
L_1
-
\mathcal L L_{-1} \, ,
\label{WarpedBoundaryConditionAphi}
\\
a_t
&=
\mu L_1
-
\mu\mathcal L L_{-1} \, ,
\label{WarpedBoundaryConditionAt}
\\
c_\phi
&=
\frac{4\pi}{\kappa}\mathcal K\,S \, ,
\label{WarpedBoundaryConditionCphi}
\\
c_t
&=
\left(
\nu
+
\frac{4\pi}{\kappa}\mu\mathcal K
\right)S \, .
\label{WarpedBoundaryConditionCt}
\end{align}
Here $\mathcal L(t,\phi)$ and $\mathcal K(t,\phi)$ denote the state-dependent functions characterizing the physical sector, while $\mu$ and $\nu$ play the role of fixed chemical potentials.

The equations of motion Eq.~\eqref{LowerSpinFlatnessConditions} determine the time evolution of the boundary data. For constant chemical potentials, one obtains
\begin{align}
\partial_t\mathcal L
&=
\mu\,\partial_\phi\mathcal L
+
2\mathcal L\,\partial_\phi\mu
-
\frac{k}{4\pi}\partial_\phi^3\mu
+
\mathcal K\,\partial_\phi\nu \, ,
\label{TimeEvolutionL}
\\
\partial_t\mathcal K
&=
\mu\,\partial_\phi\mathcal K
+
\mathcal K\,\partial_\phi\mu
+
\frac{\kappa}{4\pi}\partial_\phi\nu \, .
\label{TimeEvolutionK}
\end{align}
These relations already exhibit the characteristic coupling between the Virasoro and affine current sectors expected in warped holography.

The asymptotic symmetries are generated by gauge transformations preserving the boundary conditions above. Their action on the state-dependent functions is given by
\begin{align}
\delta\mathcal L
&=
\epsilon\,\partial_\phi\mathcal L
+
2\mathcal L\,\partial_\phi\epsilon
+
\mathcal K\,\partial_\phi\sigma
-
\frac{k}{4\pi}\partial_\phi^3\epsilon \, ,
\label{VariationL}
\\
\delta\mathcal K
&=
\epsilon\,\partial_\phi\mathcal K
+
\mathcal K\,\partial_\phi\epsilon
+
\frac{\kappa}{4\pi}\partial_\phi\sigma \, ,
\label{VariationK}
\end{align}
where $\epsilon(\phi)$ and $\sigma(\phi)$ denote the infinitesimal gauge parameters associated with the $\St\Lt(2,\mathbb R)$ and $\Ut(1)$ sectors respectively.

The corresponding canonical boundary charge takes the form
\begin{equation}
Q[\epsilon,\sigma]
=
\int d\phi
\left(
\mathcal L\,\epsilon
+
\mathcal K\,\sigma
\right) .
\label{CanonicalBoundaryCharge}
\end{equation}
Using Eq.~\eqref{CanonicalBoundaryCharge} together with the variations Eq.~\eqref{VariationL} and Eq.~\eqref{VariationK}, one obtains the asymptotic symmetry algebra
\begin{align}
[L_n,L_m]
&=
(n-m)L_{n+m}
+
\frac{c}{12}n(n^2-1)\delta_{n+m,0} \, ,
\label{VirasoroAlgebra}
\\
[L_n,K_m]
&=
-mK_{n+m} \, ,
\label{MixedWarpedAlgebra}
\\
[K_n,K_m]
&=
\frac{\kappa}{2}n\delta_{n+m,0} \, ,
\label{AffineU1Algebra}
\end{align}
with central charge
\begin{equation}
c=6k \, .
\label{WarpedCentralCharge}
\end{equation}

The algebra Eq.~\eqref{VirasoroAlgebra}--Eq.~\eqref{AffineU1Algebra} defines the characteristic Virasoro--Kac--Moody structure of warped conformal symmetry. In contrast to ordinary $\mathrm{AdS}_3$ gravity, where the asymptotic symmetry algebra consists of two independent Virasoro sectors, the warped theory contains a semidirect product between Virasoro transformations and an affine $\ut(1)$ current sector. This structure will later admit a direct discrete realization in terms of boundary monodromies and their associated conjugacy classes.
%%%%%%%%%%%%%%%%%%%%%%%%%%%%%%%%%%%%%%%%%%%%%%%%%%%%%%%%%%%%%%%%%%%%%%%%%%%%%%%%%%%%%%%%%%
%%%%%%%%%%%%%%%%%%%%%%%%%%%%%%%%%%%%%%%%%%%%%%%%%%%%%%%%%%%%%%%%%%%%%%%%%%%%%%%%%%%%%%%%%%
%%%%%%%%%%           SECTION-2.3      %%%%%%%%%%%%%%%%%%%%%%%%%%%%%%%%%%%%%%%%%%%%%%%%%%%%
%%%%%%%%%%%%%%%%%%%%%%%%%%%%%%%%%%%%%%%%%%%%%%%%%%%%%%%%%%%%%%%%%%%%%%%%%%%%%%%%%%%%%%%%%%
%%%%%%%%%%%%%%%%%%%%%%%%%%%%%%%%%%%%%%%%%%%%%%%%%%%%%%%%%%%%%%%%%%%%%%%%%%%%%%%%%%%%%%%%%%
\subsection{Holonomy Conditions and Thermal Entropy}
\label{HolonomyConditionsAndThermalEntropy}

Having established the warped boundary conditions and the associated asymptotic symmetry structure, we now turn to the thermodynamic sector of the theory. In the Chern--Simons description, the thermal properties of warped black holes are encoded in holonomies associated with noncontractible cycles \cite{GutperleKraus,AmmonGutperleKrausPerlmutter,AzeyanagiDetournayRiegler2019}. Rather than deriving temperature and entropy from a local horizon geometry, one imposes suitable regularity conditions directly on the gauge connection.

For stationary configurations with constant state-dependent functions $\mathcal L$ and $\mathcal K$, the boundary connections introduced in Eq.~\eqref{WarpedBoundaryConditionAphi}--Eq.~\eqref{WarpedBoundaryConditionCt} become \cite{AzeyanagiDetournayRiegler2019}
\begin{align}
a_\phi
&=
L_1-\mathcal L L_{-1} \, ,
\label{StationaryAphi}
\\
a_t
&=
\mu\left(
L_1-\mathcal L L_{-1}
\right) ,
\label{StationaryAt}
\\
c_\phi
&=
\frac{4\pi}{\kappa}\mathcal K\,S \, ,
\label{StationaryCphi}
\\
c_t
&=
\left(
\nu+\frac{4\pi}{\kappa}\mu\mathcal K
\right)S \, .
\label{StationaryCt}
\end{align}

The mass and angular momentum are identified with the canonical charges associated with time and angular translations respectively. Their variations are given by
\begin{align}
\delta M
&=
2\pi
\left(
\mu\,\delta\mathcal L
+
\nu\,\delta\mathcal K
\right) ,
\label{VariationMass}
\\
\delta J
&=
-2\pi\,\delta\mathcal L \, .
\label{VariationAngularMomentum}
\end{align}
A particularly convenient choice is obtained by fixing
\begin{equation}
\mu=0,
\qquad
\nu=1 \, ,
\label{CanonicalChemicalPotentials}
\end{equation}
for which the conserved charges reduce to
\begin{equation}
M=2\pi\mathcal K,
\qquad
J=-2\pi\mathcal L \, .
\label{WarpedMassAngularMomentum}
\end{equation}

The thermal sector is determined by imposing holonomy conditions along the Euclidean thermal cycle. Defining the integrated holonomies
\begin{align}
h
&=
\frac{\beta}{2\pi}
\left(
\oint a_t
+
\Omega\oint a_\phi
\right) ,
\label{HolonomyDefinitionSL2}
\\
\bar h
&=
\frac{\beta}{2\pi}
\left(
\oint c_t
+
\Omega\oint c_\phi
\right) ,
\label{HolonomyDefinitionU1}
\end{align}
the warped regularity conditions are imposed through the eigenvalue constraints
\begin{equation}
\text{Eigen}(h)
=
\text{Eigen}(2\pi L_0),
\qquad
\text{Eigen}(\bar h)
=
\text{Eigen}(2\pi\gamma S) \, ,
\label{WarpedHolonomyConditions}
\end{equation}
where $\beta$ denotes the inverse temperature, $\Omega$ is the angular velocity, and $\gamma$ is the warped tilt parameter characterizing the $\Ut(1)$ sector.

Solving Eq.~\eqref{WarpedHolonomyConditions} yields
\begin{align}
\beta
&=
2\pi
\left(
\gamma
-
\frac{2\pi\mathcal K}{\kappa\sqrt{\mathcal L}}
\right) ,
\label{WarpedInverseTemperature}
\\
\Omega
&=
\frac{1}{2\gamma\sqrt{\mathcal L}-\frac{4\pi}{\kappa}\mathcal K} \, .
\label{WarpedAngularVelocity}
\end{align}
These quantities satisfy the first law of black-hole thermodynamics,
\begin{equation}
\delta S_{\text{Th}}
=
\beta
\left(
\delta M
-
\Omega\,\delta J
\right) .
\label{WarpedFirstLaw}
\end{equation}

Integrating Eq.~\eqref{WarpedFirstLaw} leads to the warped thermal entropy
\begin{equation}
S_{\text{Th}}
=
2\pi
\left[
M\gamma
+
\sqrt{
\frac{c}{6}
\left(
-J-\frac{M^2}{\kappa}
\right)
}
\right] .
\label{WarpedThermalEntropy}
\end{equation}
This expression reproduces the characteristic entropy structure expected from warped conformal symmetry and simultaneously exhibits the interplay between the Virasoro and affine current sectors.

An important conceptual aspect of Eq.~\eqref{WarpedThermalEntropy} is that the entropy is completely determined by holonomy data associated with noncontractible cycles. The role of smooth horizon geometry becomes secondary, while the physically relevant information is instead encoded in global gauge-invariant quantities. 
%%\color{red}
From the perspective of the present work, the importance of the warped Cardy structure lies not in modifying the continuum entropy formula itself, but in identifying the minimal monodromy data required for its reconstruction within the discrete theory. The discrete formulation therefore preserves the infrared thermodynamic behavior while reorganizing its geometric interpretation in terms of boundary holonomies rather than smooth metric variables.
%%\color{black}
This observation will become central in the discrete theory developed later in Section~\ref{HolonomyEntropyInTheDiscreteTheory}, where the continuous holonomies appearing in Eq.~\eqref{HolonomyDefinitionSL2} and Eq.~\eqref{HolonomyDefinitionU1} will be replaced by ordered products of boundary link variables. Additional details concerning the continuous warped black-hole sector and the associated thermal holonomy conditions are summarized in Appendix~\ref{AppendixContinuousWarpedBlackHoles}.
%%%%%%%%%%%%%%%%%%%%%%%%%%%%%%%%%%%%%%%%%%%%%%%%%%%%%%%%%%%%%%%%%%%%%%%%%%%%%%%%%%%%%%%%%%
%%%%%%%%%%%%%%%%%%%%%%%%%%%%%%%%%%%%%%%%%%%%%%%%%%%%%%%%%%%%%%%%%%%%%%%%%%%%%%%%%%%%%%%%%%
%%%%%%%%%%           SECTION-2.4     %%%%%%%%%%%%%%%%%%%%%%%%%%%%%%%%%%%%%%%%%%%%%%%%%%%%
%%%%%%%%%%%%%%%%%%%%%%%%%%%%%%%%%%%%%%%%%%%%%%%%%%%%%%%%%%%%%%%%%%%%%%%%%%%%%%%%%%%%%%%%%%
%%%%%%%%%%%%%%%%%%%%%%%%%%%%%%%%%%%%%%%%%%%%%%%%%%%%%%%%%%%%%%%%%%%%%%%%%%%%%%%%%%%%%%%%%%
\subsection{WCFT Interpretation}
\label{WCFTInterpretation}

The entropy obtained in Section~\ref{HolonomyConditionsAndThermalEntropy} has a direct interpretation from the perspective of warped conformal field theory. A WCFT is characterized by a Virasoro algebra together with an affine $\ut(1)$ current algebra, precisely matching the asymptotic symmetry structure derived in Section~\ref{BoundaryConditionsAndAsymptoticSymmetries} \cite{HofmanStrominger,DetournayHartmanHofman}. The gravitational quantities $J$ and $M$ are therefore naturally identified with the zero-mode charges associated with the Virasoro and $\Ut(1)$ sectors, respectively.

In a WCFT, the thermal entropy depends not only on the excitation above the vacuum but also on the vacuum values of the Virasoro and $\Ut(1)$ charges. Denoting these vacuum charges by $J_{\text{v}}$ and $M_{\text{v}}$, the warped Cardy-type expression takes the schematic form
\begin{equation}
S_{\text{WCFT}}
=
-\frac{4\pi i}{\kappa}M M_{\text{v}}
+
4\pi
\sqrt{
-\left(
J_{\text{v}}
+
\frac{M_{\text{v}}^2}{\kappa}
\right)
\left(
J+
\frac{M^2}{\kappa}
\right)
} \, .
\label{WCFTEntropyFormula}
\end{equation}
This expression shows explicitly that the affine current sector shifts the effective Virasoro contribution through the combination $J+M^2/\kappa$.

The vacuum values are fixed by requiring the warped analogue of a trivial angular holonomy. In the present conventions, they are given by
\begin{equation}
J_{\text{v}}
=
\frac{c}{24}
+
\frac{\gamma^2\kappa}{4},
\qquad
M_{\text{v}}
=
\frac{i\kappa\gamma}{2} \, .
\label{WCFTVacuumCharges}
\end{equation}
Substituting Eq.~\eqref{WCFTVacuumCharges} into Eq.~\eqref{WCFTEntropyFormula} reproduces precisely the thermal entropy in Eq.~\eqref{WarpedThermalEntropy}. This matching confirms that the holonomy conditions imposed in Eq.~\eqref{WarpedHolonomyConditions} capture the same thermodynamic information as the warped Cardy structure of the dual WCFT.

The parameter $\gamma$ has a natural interpretation as the tilt parameter of the WCFT cylinder. From the gravitational side, it appears as the Abelian holonomy parameter in the $\Ut(1)$ sector. From the WCFT side, it controls the vacuum value of the current zero mode. Thus, $\gamma$ relates the geometric choice of warped thermal cycle to the spectral data of the boundary theory.

This interpretation is important for the discrete construction developed below. In the continuous theory, the vacuum charges and entropy are obtained by imposing holonomy conditions on smooth gauge connections. In the discrete theory, the same physical role will be played by the conjugacy class of the $\St\Lt(2,\mathbb R)$ boundary monodromy together with the Abelian $\Ut(1)$ holonomy. The WCFT interpretation therefore provides the continuum reference point for the monodromy-based entropy relation derived in Section~\ref{HolonomyEntropyInTheDiscreteTheory}.
%%%%%%%%%%%%%%%%%%%%%%%%%%%%%%%%%%%%%%%%%%%%%%%%%%%%%%%%%%%%%%%%%%%%%%%%%%%%%%%%%%%%%%%%%%
%%%%%%%%%%%%%%%%%%%%%%%%%%%%%%%%%%%%%%%%%%%%%%%%%%%%%%%%%%%%%%%%%%%%%%%%%%%%%%%%%%%%%%%%%%
%%%%%%%%%%           SECTION-3        %%%%%%%%%%%%%%%%%%%%%%%%%%%%%%%%%%%%%%%%%%%%%%%%%%%%
%%%%%%%%%%%%%%%%%%%%%%%%%%%%%%%%%%%%%%%%%%%%%%%%%%%%%%%%%%%%%%%%%%%%%%%%%%%%%%%%%%%%%%%%%%
%%%%%%%%%%%%%%%%%%%%%%%%%%%%%%%%%%%%%%%%%%%%%%%%%%%%%%%%%%%%%%%%%%%%%%%%%%%%%%%%%%%%%%%%%%
\section{Discrete Warped Chern--Simons Geometry}
\label{DiscreteWarpedCSGeometry}

Having reviewed the continuous lower-spin warped Chern--Simons theory in Section~\ref{ContinuousWarpedCSGravity}, we now construct its discrete counterpart. The main conceptual difference between the two descriptions is that, in the discrete theory, boundary holonomies become the fundamental variables from the outset. Instead of describing the geometry through smooth gauge connections and then constructing holonomies from path-ordered exponentials, we directly formulate the theory in terms of ordered products of group-valued link variables \cite{Wilson1974,Kogut1979,Rothe2012,Ozer:2026qlp}.

The discrete warped theory developed in this section is designed to preserve the essential topological and boundary features of the continuous system. In particular, the discrete construction retains the separation between the non-Abelian $\St\Lt(2,\mathbb R)$ sector and the Abelian $\Ut(1)$ sector, together with their associated boundary monodromies. The resulting framework naturally admits hyperbolic, elliptic, and parabolic sectors classified by the conjugacy classes of the boundary monodromy \cite{Ozer:2026qlp}.

Another important motivation for the discrete description is that the thermodynamic sector of the continuous warped theory is already controlled entirely by holonomies associated with noncontractible cycles \cite{GutperleKraus,AzeyanagiDetournayRiegler2019}. This suggests that a fundamentally monodromy-based description should exist in which boundary holonomies themselves define the physical sectors of the theory. The discrete framework constructed below realizes this idea explicitly.

%%\color{red}
It is important to emphasize that the present construction is not introduced merely as a numerical lattice approximation to a pre-existing smooth warped geometry. Instead, the discrete theory reorganizes the physical content of the warped sector directly in terms of boundary holonomy data. In this sense, the ordered boundary monodromies are treated as the fundamental kinematical observables of the theory, while the smooth continuum geometry appears only as an emergent large-lattice description.
%%\color{black}

We begin by introducing the discrete manifold structure and the associated boundary lattice. We then define the $\St\Lt(2,\mathbb R)$ and $\Ut(1)$ link variables together with the corresponding flatness constraints. Finally, we establish the continuum correspondence showing how ordered products of discrete holonomies reproduce the smooth Wilson-loop structure reviewed in Section~\ref{HolonomyConditionsAndThermalEntropy}.
%%%%%%%%%%%%%%%%%%%%%%%%%%%%%%%%%%%%%%%%%%%%%%%%%%%%%%%%%%%%%%%%%%%%%%%%%%%%%%%%%%%%%%%%%%
%%%%%%%%%%%%%%%%%%%%%%%%%%%%%%%%%%%%%%%%%%%%%%%%%%%%%%%%%%%%%%%%%%%%%%%%%%%%%%%%%%%%%%%%%%
%%%%%%%%%%           SECTION-3.1      %%%%%%%%%%%%%%%%%%%%%%%%%%%%%%%%%%%%%%%%%%%%%%%%%%%%
%%%%%%%%%%%%%%%%%%%%%%%%%%%%%%%%%%%%%%%%%%%%%%%%%%%%%%%%%%%%%%%%%%%%%%%%%%%%%%%%%%%%%%%%%%
%%%%%%%%%%%%%%%%%%%%%%%%%%%%%%%%%%%%%%%%%%%%%%%%%%%%%%%%%%%%%%%%%%%%%%%%%%%%%%%%%%%%%%%%%%
\subsection{Discrete Manifold and Boundary Lattice}
\label{DiscreteManifoldAndBoundaryLattice}

We consider a discrete cylindrical manifold $\mathcal M_{\text{disc}}$ obtained by introducing a lattice decomposition of the continuous spacetime manifold $\mathcal M$ \cite{Wilson1974,Kogut1979,Rothe2012}. The lattice is composed of vertices, oriented links, and elementary plaquettes, preserving the topology of a solid cylinder with angular coordinate $\phi$, temporal coordinate $t$, and radial direction $\rho$. The boundary of the manifold is represented by a closed one-dimensional lattice cycle located at fixed radial position.

The elementary building blocks of the discrete geometry are oriented links
\begin{equation}
\ell=(v_i,v_j) \, ,
\label{OrientedLinkDefinition}
\end{equation}
connecting neighboring vertices $v_i$ and $v_j$. Each link possesses an orientation, and reversing the orientation corresponds to group inversion of the associated holonomy variable introduced later in Section~\ref{DiscreteGaugeVariables}. The set of oriented links defines the discrete analogue of parallel transport along the lattice \cite{Wilson1974,Ozer:2026qlp}.

The boundary lattice is chosen to consist of $N$ ordered angular links forming a closed cycle,
\begin{equation}
\partial\Sigma
=
\{\ell_1,\ell_2,\dots,\ell_N\} \, ,
\label{BoundaryCycleDefinition}
\end{equation}
with periodic identification
\begin{equation}
\ell_{N+1}\equiv \ell_1 \, .
\label{BoundaryPeriodicity}
\end{equation}
This boundary cycle plays the role of the noncontractible angular direction of the continuous warped geometry. The associated ordered products of link holonomies will later define the total boundary monodromies characterizing the physical sectors of the theory \cite{Ozer:2026qlp}.

Each elementary plaquette $f$ is bounded by an oriented set of links,
\begin{equation}
\partial f
=
\{\ell_{f_1},\ell_{f_2},\dots,\ell_{f_n}\} \, ,
\label{PlaquetteBoundary}
\end{equation}
which provides the discrete counterpart of infinitesimal loops in the continuous theory. The flatness conditions imposed on these plaquettes will reproduce the topological structure of the continuous warped Chern--Simons system reviewed in Section~\ref{LowerSpinCSTheory}.

The boundary lattice inherits a natural ordering structure along the angular direction. Denoting the lattice spacing by $\Delta\phi$, the continuum coordinate is recovered through
\begin{equation}
\phi_n
=
n\,\Delta\phi,
\qquad
n=1,\dots,N \, ,
\label{DiscreteAngularCoordinate}
\end{equation}
with total angular periodicity
\begin{equation}
N\,\Delta\phi=2\pi \, .
\label{AngularPeriodicityCondition}
\end{equation}
The continuum limit is therefore obtained by taking
\begin{equation}
N\to\infty,
\qquad
\Delta\phi\to0 \, ,
\label{ContinuumLimitCondition}
\end{equation}
while keeping the total angular circumference fixed.

The discrete cylindrical structure introduced above provides the geometric arena for the warped holonomy theory developed in the following subsections. In particular, the existence of a distinguished noncontractible boundary cycle will allow the thermodynamic sector to be encoded directly in boundary monodromies, thereby providing the discrete counterpart of the continuous holonomy conditions discussed in Section~\ref{HolonomyConditionsAndThermalEntropy}.
%%%%%%%%%%%%%%%%%%%%%%%%%%%%%%%%%%%%%%%%%%%%%%%%%%%%%%%%%%%%%%%%%%%%%%%%%%%%%%%%%%%%%%%%%%
%%%%%%%%%%%%%%%%%%%%%%%%%%%%%%%%%%%%%%%%%%%%%%%%%%%%%%%%%%%%%%%%%%%%%%%%%%%%%%%%%%%%%%%%%%
%%%%%%%%%%           SECTION-3.2      %%%%%%%%%%%%%%%%%%%%%%%%%%%%%%%%%%%%%%%%%%%%%%%%%%%%
%%%%%%%%%%%%%%%%%%%%%%%%%%%%%%%%%%%%%%%%%%%%%%%%%%%%%%%%%%%%%%%%%%%%%%%%%%%%%%%%%%%%%%%%%%
%%%%%%%%%%%%%%%%%%%%%%%%%%%%%%%%%%%%%%%%%%%%%%%%%%%%%%%%%%%%%%%%%%%%%%%%%%%%%%%%%%%%%%%%%%
\subsection{Discrete Gauge Variables}
\label{DiscreteGaugeVariables}

We now introduce the discrete gauge variables associated with the lattice structure defined in Section~\ref{DiscreteManifoldAndBoundaryLattice}. In the continuous warped Chern--Simons theory, the fundamental objects are the gauge connections $A$ and $C$. In the discrete description, these continuous connections are replaced by group-valued holonomies assigned directly to oriented lattice links \cite{Wilson1974,Kogut1979,Ozer:2026qlp}.

To each oriented link $\ell$ we associate a pair of group elements
\begin{equation}
U_\ell \in \St\Lt(2,\mathbb R),
\qquad
V_\ell \in \Ut(1) \, ,
\label{DiscreteGaugeVariablesDefinition}
\end{equation}
where $U_\ell$ represents the non-Abelian gravitational sector and $V_\ell$ corresponds to the Abelian warped current sector. These link variables define the discrete analogue of parallel transport along the lattice.

The orientation of the links plays an essential role. Reversing the orientation of a link corresponds to group inversion,
\begin{equation}
U_{\ell^{-1}}
=
U_\ell^{-1},
\qquad
V_{\ell^{-1}}
=
V_\ell^{-1} \, .
\label{OrientationReversal}
\end{equation}
Thus, the discrete holonomies naturally inherit the composition structure associated with successive parallel transport along neighboring lattice segments.

In the continuum limit, the link variables are related to the continuous gauge connections through
\begin{align}
U_\ell
&=
\exp\!\left(
A_\phi(\phi_n)\Delta\phi
\right) ,
\label{SL2ContinuumLink}
\\
V_\ell
&=
\exp\!\left(
C_\phi(\phi_n)\Delta\phi
\right) .
\label{U1ContinuumLink}
\end{align}
The ordered product of neighboring links therefore reproduces the path-ordered exponential of the continuous gauge field around the boundary cycle \cite{GutperleKraus,AzeyanagiDetournayRiegler2019}.

The total boundary monodromies are defined by the ordered products
\begin{align}
M
&=
\prod_{\ell\in\partial\Sigma}U_\ell ,
\label{SL2BoundaryMonodromy}
\\
Y
&=
\prod_{\ell\in\partial\Sigma}V_\ell .
\label{U1BoundaryMonodromy}
\end{align}
The non-Abelian monodromy $M$ will determine the hyperbolic, elliptic, and parabolic sectors through its conjugacy class, while the Abelian monodromy $Y$ supplies the warped contribution to the boundary charges and entropy \cite{Ozer:2026qlp}.

Under local gauge transformations $g_v\in \St\Lt(2,\mathbb R)$ and $h_v\in \Ut(1)$ assigned to the lattice vertices, the link variables transform according to
\begin{align}
U_\ell
&\to
g^{-1}_{s(\ell)}
\,U_\ell\,
g_{t(\ell)} ,
\label{SL2GaugeTransformation}
\\
V_\ell
&\to
h^{-1}_{s(\ell)}
\,V_\ell\,
h_{t(\ell)} ,
\label{U1GaugeTransformation}
\end{align}
where $s(\ell)$ and $t(\ell)$ denote the source and target vertices of the oriented link $\ell$. As a consequence, the total monodromies transform by conjugation,
\begin{equation}
M\to g^{-1}Mg,
\qquad
Y\to Y \, ,
\label{MonodromyGaugeTransformation}
\end{equation}
showing that the conjugacy class of $M$ and the Abelian holonomy $Y$ define gauge-invariant boundary observables.

The discrete gauge variables introduced here provide the fundamental dynamical objects of the warped holonomy theory. In contrast to the continuous description, where holonomies are constructed from smooth gauge connections, the present framework treats the ordered boundary products Eq.~\eqref{SL2BoundaryMonodromy} and Eq.~\eqref{U1BoundaryMonodromy} as the primary observables from the outset. This shift in perspective will play a central role in the entropy construction developed later in Section~\ref{HolonomyEntropyInTheDiscreteTheory}.
%%%%%%%%%%%%%%%%%%%%%%%%%%%%%%%%%%%%%%%%%%%%%%%%%%%%%%%%%%%%%%%%%%%%%%%%%%%%%%%%%%%%%%%%%%
%%%%%%%%%%%%%%%%%%%%%%%%%%%%%%%%%%%%%%%%%%%%%%%%%%%%%%%%%%%%%%%%%%%%%%%%%%%%%%%%%%%%%%%%%%
%%%%%%%%%%           SECTION-3.3      %%%%%%%%%%%%%%%%%%%%%%%%%%%%%%%%%%%%%%%%%%%%%%%%%%%%
%%%%%%%%%%%%%%%%%%%%%%%%%%%%%%%%%%%%%%%%%%%%%%%%%%%%%%%%%%%%%%%%%%%%%%%%%%%%%%%%%%%%%%%%%%
%%%%%%%%%%%%%%%%%%%%%%%%%%%%%%%%%%%%%%%%%%%%%%%%%%%%%%%%%%%%%%%%%%%%%%%%%%%%%%%%%%%%%%%%%%
\subsection{Discrete Flatness Conditions}
\label{DiscreteFlatnessConditions}

Having introduced the discrete gauge variables in Section~\ref{DiscreteGaugeVariables}, we now impose the lattice constraints defining the warped topological sector. In the continuous theory reviewed in Section~\ref{LowerSpinCSTheory}, the equations of motion are given by the flatness conditions of the gauge connections. Their discrete counterpart is obtained by requiring trivial holonomy around elementary plaquettes of the lattice \cite{Wilson1974,Ozer:2026qlp}.

For each plaquette $f$, the non-Abelian flatness condition is defined through the ordered product
\begin{equation}
W_f
=
\prod_{\ell\in\partial f}
U_\ell \, ,
\label{PlaquetteHolonomySL2}
\end{equation}
where the product is taken according to the orientation of the plaquette boundary $\partial f$. The discrete flatness constraint is then imposed as
\begin{equation}
W_f=\mathbbm 1 \, .
\label{DiscreteFlatnessConstraintSL2}
\end{equation}
%%\color{red}
Similar lattice realizations of topological BF and Chern--Simons-type flatness constraints have been investigated in \cite{Berruto:2000nv}.
%%\color{black}
This condition represents the lattice analogue of the continuous equation
\begin{equation}
F=dA+A\wedge A=0 \, ,
\label{ContinuousFlatnessReminder}
\end{equation}
introduced previously in Eq.~\eqref{LowerSpinFlatnessConditions}.

The Abelian sector obeys an analogous constraint. Defining the plaquette holonomy
\begin{equation}
Y_f
=
\prod_{\ell\in\partial f}
V_\ell \, ,
\label{PlaquetteHolonomyU1}
\end{equation}
the discrete $\Ut(1)$ flatness condition becomes
\begin{equation}
Y_f=1 \, .
\label{DiscreteFlatnessConstraintU1}
\end{equation}
Together, Eq.~\eqref{DiscreteFlatnessConstraintSL2} and Eq.~\eqref{DiscreteFlatnessConstraintU1} define the discrete warped topological sector.
%%\color{red}
Within the present framework, the discrete flatness conditions should be interpreted as the lattice analogue of the continuum Chern--Simons constraint equations rather than as arising from a complete lattice action formulation. Constructing an explicit lattice Chern--Simons action compatible with the present holonomy variables would constitute a natural extension of the present framework, but lies beyond the scope of this work.
%%\color{black}
%%\color{red}
This situation differs from discrete BF-type constructions, such as our previous JT/BF framework \cite{Ozer:2026qlp}, where the flatness constraints arise naturally from a plaquette-based lattice BF action. In the present warped Chern--Simons setting, however, the action functional depends directly on the gauge connection itself and involves additional subtleties associated with ordering, boundary contributions, and global holonomy structure. Accordingly, the present construction is formulated primarily at the level of boundary monodromies and flatness constraints rather than as a complete lattice Chern--Simons action principle.
%%\color{black}

These constraints imply that local plaquette holonomies are gauge trivial, so that all physically relevant information is encoded in noncontractible boundary cycles. Consequently, the theory possesses no local propagating bulk degrees of freedom, precisely mirroring the topological structure of the continuous warped Chern--Simons system.

An important consequence of the flatness constraints is that the boundary monodromies introduced in Eq.~\eqref{SL2BoundaryMonodromy} and Eq.~\eqref{U1BoundaryMonodromy} depend only on the homotopy class of the boundary cycle. Different lattice representatives related by local deformations lead to the same conjugacy class of the total monodromy. Thus, the physical sectors are classified globally rather than locally.

The discrete flatness conditions also guarantee the consistency of holonomy composition along neighboring links. Consider two adjacent paths $\Gamma_1$ and $\Gamma_2$ sharing a common vertex. The associated parallel transports satisfy
\begin{equation}
U_{\Gamma_1\circ\Gamma_2}
=
U_{\Gamma_1}U_{\Gamma_2} \, ,
\label{HolonomyCompositionRule}
\end{equation}
with an analogous relation holding for the Abelian sector. This composition property ensures that the ordered boundary products define well-posed global monodromies.

From the perspective of warped holography, the most important implication of the discrete flatness conditions is that the thermodynamic sector becomes entirely controlled by boundary monodromies. Since all contractible plaquette holonomies are trivial, the only nontrivial gauge-invariant data are associated with noncontractible boundary cycles. This observation provides the discrete counterpart of the continuous holonomy picture reviewed in Section~\ref{HolonomyConditionsAndThermalEntropy} and forms the basis for the entropy construction developed later in Section~\ref{HolonomyEntropyInTheDiscreteTheory}.
%%%%%%%%%%%%%%%%%%%%%%%%%%%%%%%%%%%%%%%%%%%%%%%%%%%%%%%%%%%%%%%%%%%%%%%%%%%%%%%%%%%%%%%%%%
%%%%%%%%%%%%%%%%%%%%%%%%%%%%%%%%%%%%%%%%%%%%%%%%%%%%%%%%%%%%%%%%%%%%%%%%%%%%%%%%%%%%%%%%%%
%%%%%%%%%%           SECTION-3.4     %%%%%%%%%%%%%%%%%%%%%%%%%%%%%%%%%%%%%%%%%%%%%%%%%%%%
%%%%%%%%%%%%%%%%%%%%%%%%%%%%%%%%%%%%%%%%%%%%%%%%%%%%%%%%%%%%%%%%%%%%%%%%%%%%%%%%%%%%%%%%%%
%%%%%%%%%%%%%%%%%%%%%%%%%%%%%%%%%%%%%%%%%%%%%%%%%%%%%%%%%%%%%%%%%%%%%%%%%%%%%%%%%%%%%%%%%%
\subsection{Continuum Limit}
\label{ContinuumLimit}

We now establish the relation between the discrete warped holonomy theory and the continuous lower-spin Chern--Simons system reviewed in Section~\ref{ContinuousWarpedCSGravity}. The key point is that the ordered products of discrete boundary holonomies reproduce the continuous Wilson-loop structure in the large-lattice limit \cite{GutperleKraus,AzeyanagiDetournayRiegler2019}.

Consider the non-Abelian link variable introduced in Eq.~\eqref{SL2ContinuumLink},
\begin{equation}
U_n
=
\exp\!\left(
A_\phi(\phi_n)\Delta\phi
\right) \, .
\label{DiscreteLinkExpansion}
\end{equation}
For sufficiently small lattice spacing $\Delta\phi$, one may expand
\begin{equation}
U_n
=
\mathbbm 1
+
A_\phi(\phi_n)\Delta\phi
+
\mathcal O(\Delta\phi^2) \, .
\label{SmallLatticeExpansion}
\end{equation}
The total boundary monodromy then becomes
\begin{equation}
M
=
\prod_{n=1}^{N}U_n \, .
\label{DiscreteBoundaryProduct}
\end{equation}
Taking the continuum limit defined previously in Eq.~\eqref{ContinuumLimitCondition}, the ordered product reproduces the path-ordered exponential \cite{Dittrich:2012jq}
\begin{equation}
M
\longrightarrow
\mathcal P
\exp\!\left(
\oint_{\partial\Sigma}
A_\phi\,d\phi
\right) .
\label{ContinuumWilsonLoop}
\end{equation}
Thus, the discrete monodromy provides a lattice regularization of the continuous holonomy.

The same correspondence holds for the Abelian sector. Using Eq.~\eqref{U1ContinuumLink},
\begin{equation}
V_n
=
\exp\!\left(
C_\phi(\phi_n)\Delta\phi
\right) ,
\label{DiscreteU1Expansion}
\end{equation}
the total Abelian monodromy satisfies
\begin{equation}
Y
=
\prod_{n=1}^{N}V_n
\longrightarrow
\exp\!\left(
\oint_{\partial\Sigma}
C_\phi\,d\phi
\right) .
\label{ContinuumU1Holonomy}
\end{equation}
Since the $\Ut(1)$ sector is Abelian, no path ordering is required.

An important conceptual aspect of this correspondence is that the roles of holonomies differ fundamentally in the continuous and discrete descriptions. In the continuous theory, the holonomy arises as a derived object constructed from smooth gauge connections. In the discrete theory, by contrast, the monodromy itself is fundamental from the outset, while the continuous gauge connection appears only as an emergent large-lattice description.

The continuum limit also reproduces the warped holonomy constraints reviewed in Section~\ref{HolonomyConditionsAndThermalEntropy}. In particular, the continuous quantities
\begin{equation}
h
=
\frac{\beta}{2\pi}
\left(
\oint a_t
+
\Omega\oint a_\phi
\right)
\label{ContinuousThermalHolonomy}
\end{equation}
are recovered from the large-$N$ limit of ordered boundary products along the thermal cycle. The corresponding eigenvalue conditions therefore emerge naturally from the discrete monodromy structure.

The existence of the continuum correspondence shows that the discrete warped theory preserves the essential topological and thermodynamic content of the continuous lower-spin system. At the same time, the discrete description provides a more primitive characterization of the physical sectors in which boundary monodromies, rather than smooth geometric data, define the fundamental observables \cite{Ozer:2026qlp}. This monodromy-first perspective will become central in the classification of warped sectors developed in Section~\ref{BoundaryMonodromiesAndWarpedSectors}. A more detailed reconstruction of the continuum Wilson-loop structure from ordered boundary products is presented in Appendix~\ref{AppendixContinuumLimitOfDiscreteWarpedHolonomies}.

The continuum correspondence established above also clarifies why the physical sectors of the discrete theory are organised by conjugacy classes rather than by individual monodromy representatives. In the continuous theory, regularity conditions are imposed on smooth gauge connections, and holonomies arise as derived objects encoding that regularity. In the discrete theory, by contrast, the flatness constraints Eq.~\eqref{DiscreteFlatnessConstraintSL2} and Eq.~\eqref{DiscreteFlatnessConstraintU1} eliminate all local bulk information by requiring trivial holonomy on every contractible plaquette. Once local data are projected out in this way, the only gauge-invariant information that survives is carried by the ordered boundary products around non-contractible cycles. Since gauge transformations act on the total monodromy by conjugation according to Eq.~\eqref{MonodromyGaugeTransformation}, two boundary configurations related by a gauge transformation belong to the same physical sector if and only if their monodromies are conjugate. The classification of physical sectors therefore reduces entirely to the classification of conjugacy classes of the boundary monodromy, a structure that is not inherited from the continuum but emerges directly from the discrete flatness condition itself. This observation motivates the systematic analysis of boundary monodromy sectors developed in Section~\ref{BoundaryMonodromiesAndWarpedSectors}.

%%%%%%%%%%%%%%%%%%%%%%%%%%%%%%%%%%%%%%%%%%%%%%%%%%%%%%%%%%%%%%%%%%%%%%%%%%%%%%%%%%%%%%%%%%
%%%%%%%%%%%%%%%%%%%%%%%%%%%%%%%%%%%%%%%%%%%%%%%%%%%%%%%%%%%%%%%%%%%%%%%%%%%%%%%%%%%%%%%%%%
%%%%%%%%%%           SECTION-4        %%%%%%%%%%%%%%%%%%%%%%%%%%%%%%%%%%%%%%%%%%%%%%%%%%%%
%%%%%%%%%%%%%%%%%%%%%%%%%%%%%%%%%%%%%%%%%%%%%%%%%%%%%%%%%%%%%%%%%%%%%%%%%%%%%%%%%%%%%%%%%%
%%%%%%%%%%%%%%%%%%%%%%%%%%%%%%%%%%%%%%%%%%%%%%%%%%%%%%%%%%%%%%%%%%%%%%%%%%%%%%%%%%%%%%%%%%
\section{Boundary Monodromies and Warped Sectors}
\label{BoundaryMonodromiesAndWarpedSectors}
Before analysing the monodromy matrices explicitly, it is useful to recall why conjugacy classes provide the correct organisational principle for the discrete warped theory. As shown in 
Section~\ref{DiscreteWarpedCSGeometry}, the discrete flatness conditions remove all contractible holonomy data from the bulk, leaving the ordered boundary products as the sole carriers of gauge-invariant information. The action of gauge transformations on these products is by conjugation, as expressed in Eq.~\eqref{MonodromyGaugeTransformation}, so that physically distinct configurations correspond precisely to distinct conjugacy classes. This is not merely a gauge-fixing convenience; it reflects the fact that the boundary monodromy sector is the primary kinematical datum of the theory, prior to any geometric interpretation. 

%%\color{red}
This viewpoint differs conceptually from the standard continuum approach, where the physical sector is usually specified by smooth background geometries satisfying appropriate regularity conditions. In the present framework, the classification by monodromy chambers precedes any geometric interpretation. The thermal, compact, and degenerate sectors are therefore understood fundamentally as distinct boundary holonomy sectors rather than as secondary properties of smooth spacetime configurations. 
%%\color{black}
The monodromy matrices defined below should therefore be understood not as auxiliary quantities derived from a smooth connection, but as the fundamental variables from which warped sectors, residual symmetries, and thermodynamic observables are reconstructed.

The discrete warped theory constructed in Section~\ref{DiscreteWarpedCSGeometry} naturally organizes its physical sectors through boundary monodromies. Since the local plaquette holonomies are constrained to be trivial by the discrete flatness conditions Eq.~\eqref{DiscreteFlatnessConstraintSL2} and Eq.~\eqref{DiscreteFlatnessConstraintU1}, all gauge-invariant information is encoded in the noncontractible boundary cycles and their associated monodromies \cite{Wilson1974,Ozer:2026qlp}.

In the continuous warped Chern--Simons theory, thermodynamic quantities are determined from smooth holonomies satisfying suitable regularity conditions \cite{GutperleKraus,AzeyanagiDetournayRiegler2019}. In the discrete description, the corresponding role is played directly by the conjugacy classes of the total boundary monodromies. The distinction between hyperbolic, elliptic, and parabolic sectors therefore emerges as an intrinsic property of the boundary holonomy structure rather than from smooth geometric considerations \cite{Ozer:2026qlp}.

The purpose of this section is to classify the discrete warped sectors generated by the boundary monodromies and to analyze their associated residual symmetry structures. We first define the total boundary monodromy matrices for both the $\St\Lt(2,\mathbb R)$ and $\Ut(1)$ sectors. We then study the conjugacy classes of the non-Abelian monodromy and identify the corresponding warped chambers. After that, we analyze the stabilizer structure and the residual boundary symmetries associated with each sector. Finally, we discuss the resulting discrete warped phase space and its interpretation in terms of boundary monodromy moduli.
%%%%%%%%%%%%%%%%%%%%%%%%%%%%%%%%%%%%%%%%%%%%%%%%%%%%%%%%%%%%%%%%%%%%%%%%%%%%%%%%%%%%%%%%%%
%%%%%%%%%%%%%%%%%%%%%%%%%%%%%%%%%%%%%%%%%%%%%%%%%%%%%%%%%%%%%%%%%%%%%%%%%%%%%%%%%%%%%%%%%%
%%%%%%%%%%           SECTION-4.1      %%%%%%%%%%%%%%%%%%%%%%%%%%%%%%%%%%%%%%%%%%%%%%%%%%%%
%%%%%%%%%%%%%%%%%%%%%%%%%%%%%%%%%%%%%%%%%%%%%%%%%%%%%%%%%%%%%%%%%%%%%%%%%%%%%%%%%%%%%%%%%%
%%%%%%%%%%%%%%%%%%%%%%%%%%%%%%%%%%%%%%%%%%%%%%%%%%%%%%%%%%%%%%%%%%%%%%%%%%%%%%%%%%%%%%%%%%
\subsection{Boundary Monodromy Matrices}
\label{BoundaryMonodromyMatrices}

We now define the global boundary observables characterizing the discrete warped theory. As discussed in Section~\ref{DiscreteFlatnessConditions}, the local plaquette holonomies are constrained to be trivial, so that the physically relevant information is entirely encoded in the noncontractible boundary cycle $\partial\Sigma$. The associated ordered boundary products define the total monodromies of the theory.

For the non-Abelian sector, the total boundary monodromy is given by
\begin{equation}
M
=
\prod_{\ell\in\partial\Sigma}
U_\ell \, ,
\label{TotalSL2BoundaryMonodromy}
\end{equation}
where the product follows the orientation of the boundary cycle introduced in Eq.~\eqref{BoundaryCycleDefinition}. The matrix $M$ represents the discrete counterpart of the continuous Wilson loop 
\cite{GutperleKraus,AzeyanagiDetournayRiegler2019}
\begin{equation}
\mathcal P
\exp\!\left(
\oint_{\partial\Sigma}A
\right)
\label{ContinuousBoundaryWilsonLoop}
\end{equation}
reviewed previously in Eq.~\eqref{ContinuumWilsonLoop}.

Similarly, the Abelian boundary monodromy is defined as
\begin{equation}
Y
=
\prod_{\ell\in\partial\Sigma}
V_\ell \, .
\label{TotalU1BoundaryMonodromy}
\end{equation}
Since the $\Ut(1)$ sector is Abelian, the ordering of the factors does not affect the final result. The quantity $Y$ therefore defines a global warped charge associated with the noncontractible boundary cycle.

Under local gauge transformations Eq.~\eqref{SL2GaugeTransformation} and Eq.~\eqref{U1GaugeTransformation}, the total monodromies transform according to
\begin{equation}
M
\to
g^{-1}Mg,
\qquad
Y
\to
Y \, .
\label{BoundaryMonodromyGaugeTransformation}
\end{equation}
Consequently, the physically meaningful information carried by the non-Abelian sector is contained in the conjugacy class of $M$, while the Abelian monodromy $Y$ is itself gauge invariant.

The classification of the warped sectors therefore reduces to the classification of the conjugacy classes of the matrix $M$ \cite{Ozer:2026qlp}. For $\St\Lt(2,\mathbb R)$, this classification is determined by the trace invariant
\begin{equation}
\Delta
=
\left(
\tr M
\right)^2-4 \, .
\label{SL2TraceInvariant}
\end{equation}
The sign of $\Delta$ separates the theory into hyperbolic, elliptic, and parabolic sectors, corresponding respectively to thermal, compact, and degenerate monodromy structures. These sectors will be analyzed explicitly in Section~\ref{ConjugacyClassesAndWarpedChambers}.

The total boundary monodromies also provide the discrete analogue of the continuous thermal holonomies introduced in Eq.~\eqref{HolonomyDefinitionSL2} and Eq.~\eqref{HolonomyDefinitionU1}. In the continuous warped theory, the entropy is determined from smooth holonomy constraints imposed on the thermal cycle. In the discrete theory, the corresponding thermodynamic information will instead be reconstructed directly from the global monodromy data carried by Eq.~\eqref{TotalSL2BoundaryMonodromy} and Eq.~\eqref{TotalU1BoundaryMonodromy}.

An important conceptual aspect of this construction is that the monodromies are not auxiliary derived quantities but rather the primary observables defining the physical sectors of the theory. Different warped chambers therefore correspond directly to distinct boundary monodromy sectors, providing a fundamentally holonomy-based description of the warped thermodynamic phase space \cite{Ozer:2026qlp}.
%%%%%%%%%%%%%%%%%%%%%%%%%%%%%%%%%%%%%%%%%%%%%%%%%%%%%%%%%%%%%%%%%%%%%%%%%%%%%%%%%%%%%%%%%%
%%%%%%%%%%%%%%%%%%%%%%%%%%%%%%%%%%%%%%%%%%%%%%%%%%%%%%%%%%%%%%%%%%%%%%%%%%%%%%%%%%%%%%%%%%
%%%%%%%%%%           SECTION-4.2      %%%%%%%%%%%%%%%%%%%%%%%%%%%%%%%%%%%%%%%%%%%%%%%%%%%%
%%%%%%%%%%%%%%%%%%%%%%%%%%%%%%%%%%%%%%%%%%%%%%%%%%%%%%%%%%%%%%%%%%%%%%%%%%%%%%%%%%%%%%%%%%
%%%%%%%%%%%%%%%%%%%%%%%%%%%%%%%%%%%%%%%%%%%%%%%%%%%%%%%%%%%%%%%%%%%%%%%%%%%%%%%%%%%%%%%%%%
\subsection{Conjugacy Classes and Warped Chambers}
\label{ConjugacyClassesAndWarpedChambers}

The physical sectors of the discrete warped theory are determined by the conjugacy classes of the non-Abelian boundary monodromy introduced in Eq.~\eqref{TotalSL2BoundaryMonodromy}. Since gauge transformations act by conjugation according to Eq.~\eqref{BoundaryMonodromyGaugeTransformation}, two monodromies related by
\begin{equation}
M
\sim
g^{-1}Mg
\label{MonodromyConjugationRelation}
\end{equation}
describe the same physical sector. The classification of warped sectors therefore reduces to the classification of conjugacy classes in $\St\Lt(2,\mathbb R)$.

As discussed previously, the relevant invariant is
\begin{equation}
\Delta
=
\left(
\tr M
\right)^2-4 \, .
\label{WarpedSectorInvariant}
\end{equation}
Depending on the sign of $\Delta$, the monodromy belongs to one of three distinct sectors.

The hyperbolic sector is characterized by
\begin{equation}
\Delta>0,
\qquad
|\tr M|>2 \, .
\label{HyperbolicSectorCondition}
\end{equation}
In this case, the monodromy possesses two distinct real eigenvalues and can be brought to the diagonal form
\begin{equation}
M_{\text{h}}
\sim
\begin{pmatrix}
e^\lambda & 0
\\
0 & e^{-\lambda}
\end{pmatrix},
\qquad
\lambda\in\mathbb R \, .
\label{HyperbolicMonodromy}
\end{equation}
The hyperbolic chamber represents the thermal sector of the warped theory and provides the discrete counterpart of warped black-hole holonomies in the continuous description reviewed in Section~\ref{HolonomyConditionsAndThermalEntropy} \cite{GutperleKraus,AzeyanagiDetournayRiegler2019}.

The elliptic sector is defined by
\begin{equation}
\Delta<0,
\qquad
|\tr M|<2 \, .
\label{EllipticSectorCondition}
\end{equation}
The corresponding monodromy possesses complex conjugate eigenvalues of unit modulus and may be written as
\begin{equation}
M_{\text{e}}
\sim
\begin{pmatrix}
\cos\theta & \sin\theta
\\
-\sin\theta & \cos\theta
\end{pmatrix},
\qquad
\theta\in\mathbb R \, .
\label{EllipticMonodromy}
\end{equation}
This sector describes compact oscillatory monodromy configurations and does not possess the thermal interpretation associated with the hyperbolic chamber.

Finally, the parabolic sector is characterized by
\begin{equation}
\Delta=0,
\qquad
|\tr M|=2 \, .
\label{ParabolicSectorCondition}
\end{equation}
The monodromy becomes non-diagonalizable and takes the Jordan form
\begin{equation}
M_{\text{p}}
\sim
\begin{pmatrix}
1 & \alpha
\\
0 & 1
\end{pmatrix},
\qquad
\alpha\in\mathbb R \, .
\label{ParabolicMonodromy}
\end{equation}
The parabolic chamber therefore defines a degenerate boundary sector separating the hyperbolic and elliptic regimes.

An important conceptual aspect of the present construction is that all warped chambers originate from the same underlying $\St\Lt(2,\mathbb R)$ gauge structure. The distinction between thermal, compact, and degenerate sectors arises entirely through different boundary monodromies rather than through different bulk geometries. Thus, the warped phase structure is fundamentally encoded in the conjugacy classes of the boundary holonomies.

The Abelian monodromy Eq.~\eqref{TotalU1BoundaryMonodromy} provides an additional warped contribution to each chamber. Since the $\Ut(1)$ sector is independent of the conjugacy classification of $\St\Lt(2,\mathbb R)$, each hyperbolic, elliptic, or parabolic chamber admits a corresponding family of warped sectors distinguished by the value of the Abelian boundary holonomy.

The chamber structure derived above provides the discrete counterpart of the continuous warped holonomy sectors appearing in lower-spin Chern--Simons gravity \cite{HofmanRollier,AzeyanagiDetournayRiegler2019,Ozer:2026qlp}. In the discrete theory, however, the conjugacy classes themselves define the primary physical sectors from the outset. This monodromy-first viewpoint will play a central role in the residual symmetry analysis developed in Section~\ref{StabilizersAndResidualBoundarySymmetries} and in the entropy construction presented later in Section~\ref{HolonomyEntropyInTheDiscreteTheory}. Further group-theoretic details concerning the $\St\Lt(2,\mathbb R)$ conjugacy classes and stabilizer structures are collected in Appendix~\ref{AppendixDiscreteGroupTheoreticIdentities}.
%%%%%%%%%%%%%%%%%%%%%%%%%%%%%%%%%%%%%%%%%%%%%%%%%%%%%%%%%%%%%%%%%%%%%%%%%%%%%%%%%%%%%%%%%%
%%%%%%%%%%%%%%%%%%%%%%%%%%%%%%%%%%%%%%%%%%%%%%%%%%%%%%%%%%%%%%%%%%%%%%%%%%%%%%%%%%%%%%%%%%
%%%%%%%%%%           SECTION-4.3      %%%%%%%%%%%%%%%%%%%%%%%%%%%%%%%%%%%%%%%%%%%%%%%%%%%%
%%%%%%%%%%%%%%%%%%%%%%%%%%%%%%%%%%%%%%%%%%%%%%%%%%%%%%%%%%%%%%%%%%%%%%%%%%%%%%%%%%%%%%%%%%
%%%%%%%%%%%%%%%%%%%%%%%%%%%%%%%%%%%%%%%%%%%%%%%%%%%%%%%%%%%%%%%%%%%%%%%%%%%%%%%%%%%%%%%%%%
\subsection{Stabilizers and Residual Boundary Symmetries}
\label{StabilizersAndResidualBoundarySymmetries}

We now analyze the residual symmetry structure associated with the boundary monodromy sectors introduced in Section~\ref{ConjugacyClassesAndWarpedChambers}. Since the physical sectors of the theory are classified by conjugacy classes of the total boundary monodromy, the relevant residual symmetries are generated by the stabilizer subgroup preserving a given monodromy configuration \cite{Ozer:2026qlp}.

For a fixed boundary monodromy $M$, the stabilizer subgroup is defined by
\begin{equation}
\mathrm{Cent}(M)
=
\left\{
g\in \St\Lt(2,\mathbb R)
\;\middle|\;
g^{-1}Mg=M
\right\} ,
\label{MonodromyCentralizerDefinition}
\end{equation}
namely the centralizer of the monodromy inside $\St\Lt(2,\mathbb R)$  \cite{Geiller:2017xad}. The structure of this stabilizer depends directly on the conjugacy class of the boundary monodromy.

For the hyperbolic sector Eq.~\eqref{HyperbolicMonodromy}, the stabilizer is generated by the Cartan element $L_0$,
\begin{equation}
\mathrm{Cent}(M_{\text{h}})
=
\exp(\alpha L_0) \, .
\label{HyperbolicCentralizer}
\end{equation}
This sector therefore possesses a noncompact residual symmetry associated with scaling transformations. Since the hyperbolic chamber corresponds to the thermal sector of the warped theory, the Cartan stabilizer plays the role of the residual thermal symmetry preserving the boundary monodromy.

In the elliptic sector Eq.~\eqref{EllipticMonodromy}, the stabilizer becomes compact,
\begin{equation}
\mathrm{Cent}(M_{\text{e}})
=
\exp\!\left[
\theta(L_1-L_{-1})
\right] .
\label{EllipticCentralizer}
\end{equation}
The residual symmetry therefore corresponds to rotational transformations preserving the compact oscillatory monodromy structure.

For the parabolic sector Eq.~\eqref{ParabolicMonodromy}, the stabilizer is generated by the nilpotent element $L_1$,
\begin{equation}
\mathrm{Cent}(M_{\text{p}})
=
\exp(\alpha L_1) \, .
\label{ParabolicCentralizer}
\end{equation}
This degenerate stabilizer structure reflects the Jordan nature of the parabolic monodromy and defines the residual symmetry associated with the critical boundary sector separating the hyperbolic and elliptic chambers.

The Abelian monodromy sector possesses an independent residual symmetry generated by the $\Ut(1)$ current algebra. Since the Abelian monodromy is gauge invariant according to Eq.~\eqref{BoundaryMonodromyGaugeTransformation}, the corresponding residual transformations act trivially on the conjugacy class of the non-Abelian sector while shifting the warped current contribution.

At the level of the boundary theory, the residual symmetry structure naturally reproduces the warped Virasoro--Kac--Moody algebra reviewed previously in Section~\ref{BoundaryConditionsAndAsymptoticSymmetries} 
\cite{CompereSongStrominger,DetournayHartmanHofman}. The non-Abelian sector generates the Virasoro-type transformations associated with reparametrizations of the boundary cycle, while the Abelian sector produces the affine $\ut(1)$ current contribution.

An important feature of the discrete theory is that the residual symmetries emerge directly from monodromy-preserving transformations rather than from smooth asymptotic Killing vectors. In this sense, the asymptotic symmetry structure is reconstructed entirely from the algebraic properties of the boundary holonomies \cite{Ozer:2026qlp}. The warped symmetry algebra therefore appears as a consequence of the stabilizer geometry of the monodromy sectors.

The stabilizer structure derived above also determines the organization of the discrete boundary phase space. Different warped chambers possess different residual symmetry groups and therefore correspond to distinct boundary orbit geometries. This orbit interpretation will be developed further in Section~\ref{DiscreteWarpedPhaseSpace}, where the boundary monodromies will be interpreted as coordinates on the discrete warped moduli space.
%%%%%%%%%%%%%%%%%%%%%%%%%%%%%%%%%%%%%%%%%%%%%%%%%%%%%%%%%%%%%%%%%%%%%%%%%%%%%%%%%%%%%%%%%%
%%%%%%%%%%%%%%%%%%%%%%%%%%%%%%%%%%%%%%%%%%%%%%%%%%%%%%%%%%%%%%%%%%%%%%%%%%%%%%%%%%%%%%%%%%
%%%%%%%%%%           SECTION-4.4     %%%%%%%%%%%%%%%%%%%%%%%%%%%%%%%%%%%%%%%%%%%%%%%%%%%%
%%%%%%%%%%%%%%%%%%%%%%%%%%%%%%%%%%%%%%%%%%%%%%%%%%%%%%%%%%%%%%%%%%%%%%%%%%%%%%%%%%%%%%%%%%
%%%%%%%%%%%%%%%%%%%%%%%%%%%%%%%%%%%%%%%%%%%%%%%%%%%%%%%%%%%%%%%%%%%%%%%%%%%%%%%%%%%%%%%%%%
\subsection{Discrete Warped Phase Space}
\label{DiscreteWarpedPhaseSpace}

We now describe the phase space associated with the discrete warped holonomy theory. Since the plaquette flatness constraints in Eq.~\eqref{DiscreteFlatnessConstraintSL2} and Eq.~\eqref{DiscreteFlatnessConstraintU1} remove local bulk curvature, the remaining physical data are carried by the noncontractible boundary monodromies introduced in Section~\ref{BoundaryMonodromyMatrices}. The discrete warped phase space is therefore organized by global boundary data rather than by local geometric fields.

The non-Abelian part of the phase space is obtained by quotienting the total boundary monodromy by conjugation,

\begin{equation}
\mathcal P_{\St\Lt(2,\mathbb R)}
=
\{M\in \St\Lt(2,\mathbb R)\}/\text{conjugation} ,
\qquad
M \mapsto g^{-1}Mg .
\label{DiscreteSL2PhaseSpace}
\end{equation}

Thus, physical configurations are labeled by conjugacy classes of $\St\Lt(2,\mathbb R)$rather than by individual representatives. Equivalently, the trace invariant introduced in Eq.~\eqref{SL2TraceInvariant} provides a gauge-invariant coordinate distinguishing hyperbolic, elliptic, and parabolic chambers.

The Abelian part is simpler because the $\Ut(1)$ boundary monodromy is already gauge invariant. It therefore contributes an independent factor,
\begin{equation}
\mathcal P_{\Ut(1)}
=
\{Y\in \Ut(1)\}.
\label{DiscreteU1PhaseSpace}
\end{equation}
The full discrete warped phase space takes the product form
\begin{equation}
\mathcal P_{\text{w}}
=
\mathcal P_{\St\Lt(2,\mathbb R)}
\times
\mathcal P_{\Ut(1)} .
\label{DiscreteWarpedPhaseSpaceDefinition}
\end{equation}
This product structure reflects the lower-spin gauge algebra reviewed in Section~\ref{LowerSpinCSTheory}, while the coupling between the two sectors appears in the warped charge and entropy relations.
%%\color{red}
Within each chamber, the phase space may be viewed as a boundary orbit space associated with the stabilizer of the corresponding monodromy \cite{Ozer:2026qlp}. Related continuum analyses of holonomy phase spaces and Chern--Simons boundary structures may also be found in \cite{Meusburger:2005in}. For a representative $M_\alpha$, where $\alpha=\text{h},\text{e},\text{p}$ labels the hyperbolic, elliptic, or parabolic chamber, the associated orbit is
\begin{equation}
\mathcal O_\alpha
=
\St\Lt(2,\mathbb R)/
\mathrm{Cent}(M_\alpha) .
\label{WarpedBoundaryOrbit}
\end{equation}
The stabilizers appearing here are precisely those derived in Section~\ref{StabilizersAndResidualBoundarySymmetries}. Thus, the chamber-dependent phase space geometry is determined by the centralizer structure of the boundary monodromy.
%%\color{black}

The hyperbolic chamber is associated with a noncompact Cartan stabilizer and contains the thermal warped sector. The elliptic chamber is associated with a compact stabilizer and describes oscillatory boundary sectors. The parabolic chamber is associated with a nilpotent stabilizer and represents a degenerate orbit separating the hyperbolic and elliptic regimes. These three chambers are not different theories; they are different monodromy sectors of the same discrete warped gauge system.

This observation is central for the interpretation of the discrete construction. The warped phase space is not built by first choosing a smooth background geometry and then studying fluctuations around it. Instead, it is built directly from boundary monodromy data. Smooth warped geometries, when they exist, should be regarded as continuum representatives of particular monodromy chambers rather than as the primary definition of the physical sector \cite{AzeyanagiDetournayRiegler2019,Ozer:2026qlp}.

The phase-space structure described above prepares the entropy analysis of Section~\ref{HolonomyEntropyInTheDiscreteTheory}. In particular, the hyperbolic chamber provides the natural setting for the warped thermal entropy, while the elliptic and parabolic chambers define nonthermal or degenerate sectors whose entropy interpretation must be understood directly from monodromy invariants.

Among the monodromy chambers described above, the hyperbolic sector plays a distinguished role because it provides the unique sector admitting a direct thermal interpretation. This observation naturally leads to the reconstruction of entropy directly from boundary monodromy invariants.

%%%%%%%%%%%%%%%%%%%%%%%%%%%%%%%%%%%%%%%%%%%%%%%%%%%%%%%%%%%%%%%%%%%%%%%%%%%%%%%%%%%%%%%%%%
%%%%%%%%%%%%%%%%%%%%%%%%%%%%%%%%%%%%%%%%%%%%%%%%%%%%%%%%%%%%%%%%%%%%%%%%%%%%%%%%%%%%%%%%%%
%%%%%%%%%%           SECTION-5        %%%%%%%%%%%%%%%%%%%%%%%%%%%%%%%%%%%%%%%%%%%%%%%%%%%%
%%%%%%%%%%%%%%%%%%%%%%%%%%%%%%%%%%%%%%%%%%%%%%%%%%%%%%%%%%%%%%%%%%%%%%%%%%%%%%%%%%%%%%%%%%
%%%%%%%%%%%%%%%%%%%%%%%%%%%%%%%%%%%%%%%%%%%%%%%%%%%%%%%%%%%%%%%%%%%%%%%%%%%%%%%%%%%%%%%%%%
\section{Holonomy Entropy in the Discrete Theory}
\label{HolonomyEntropyInTheDiscreteTheory}

We now turn to the central problem of the present work: the reconstruction of warped entropy directly from discrete boundary monodromies. In the continuous lower-spin theory reviewed in Section~\ref{HolonomyConditionsAndThermalEntropy}, the thermal entropy follows from smooth holonomy constraints imposed on the Euclidean thermal cycle \cite{GutperleKraus,AmmonGutperleKrausPerlmutter,AzeyanagiDetournayRiegler2019}. In the discrete theory, however, the relevant thermodynamic information is encoded fundamentally in the boundary monodromies themselves.

The discrete warped framework developed in the previous sections provides a natural setting for such a construction. Since the local plaquette holonomies are constrained to be trivial by Eq.~\eqref{DiscreteFlatnessConstraintSL2} and Eq.~\eqref{DiscreteFlatnessConstraintU1}, all nontrivial gauge-invariant information is carried by the global monodromies Eq.~\eqref{TotalSL2BoundaryMonodromy} and Eq.~\eqref{TotalU1BoundaryMonodromy}. The entropy should therefore be reconstructed entirely from the invariants associated with these noncontractible boundary cycles \cite{Ozer:2026qlp}.

An important conceptual aspect of this construction is that the thermodynamic sector is no longer derived from smooth horizon geometry. Instead, the hyperbolic chamber identified in Section~\ref{ConjugacyClassesAndWarpedChambers} plays the role of the thermal sector directly at the level of boundary monodromies \cite{Ozer:2026qlp}. In this sense, entropy becomes a property of the boundary monodromy sector itself rather than a consequence of local geometric regularity conditions.
%%\color{red}
Accordingly, the purpose of the present construction is not to derive a fundamentally different infrared entropy formula, but rather to demonstrate that the warped thermodynamic sector admits a consistent reconstruction directly from discrete monodromy invariants without assuming a smooth geometric thermal background from the outset. The novelty therefore lies in the monodromy-based organization of the phase space rather than in modifying the universal infrared Cardy behavior itself.

This viewpoint is also closely related to our previous discrete JT/BF construction, where the thermodynamic sector was similarly reconstructed from boundary holonomy and monodromy data at the level of reduced phase-space invariants \cite{Ozer:2026qlp}. In the present warped Chern--Simons setting, the entropy construction follows the same general monodromy-based philosophy, although the underlying gauge structure differs substantially from the BF case due to the presence of the non-Abelian Chern--Simons sector together with the additional warped U(1) contribution. Consequently, the entropy relation derived below should be understood primarily as a reconstruction from global boundary monodromy invariants rather than as the outcome of a complete lattice path-integral quantization.
%%\color{black}

The structure of the discrete entropy closely parallels the warped Cardy form reviewed in Section~\ref{WCFTInterpretation} \cite{HofmanStrominger,DetournayHartmanHofman}. The non-Abelian $SL(2,\mathbb R)$ monodromy determines the Casimir contribution associated with the Virasoro sector, while the Abelian boundary holonomy supplies the warped $\Ut(1)$ contribution. 

In this section, we first identify the Casimir structure associated with the boundary monodromy invariants. We then analyze the Abelian warped contribution and construct the corresponding entropy relation. After that, we study the hyperbolic thermal chamber and its warped Cardy behavior. Finally, we establish the continuum correspondence showing that the continuous warped entropy reviewed in Section~\ref{HolonomyConditionsAndThermalEntropy} is recovered in the large--lattice limit \cite{AzeyanagiDetournayRiegler2019,Ozer:2026qlp}.
%%%%%%%%%%%%%%%%%%%%%%%%%%%%%%%%%%%%%%%%%%%%%%%%%%%%%%%%%%%%%%%%%%%%%%%%%%%%%%%%%%%%%%%%%%
%%%%%%%%%%%%%%%%%%%%%%%%%%%%%%%%%%%%%%%%%%%%%%%%%%%%%%%%%%%%%%%%%%%%%%%%%%%%%%%%%%%%%%%%%%
%%%%%%%%%%           SECTION-5.1      %%%%%%%%%%%%%%%%%%%%%%%%%%%%%%%%%%%%%%%%%%%%%%%%%%%%
%%%%%%%%%%%%%%%%%%%%%%%%%%%%%%%%%%%%%%%%%%%%%%%%%%%%%%%%%%%%%%%%%%%%%%%%%%%%%%%%%%%%%%%%%%
%%%%%%%%%%%%%%%%%%%%%%%%%%%%%%%%%%%%%%%%%%%%%%%%%%%%%%%%%%%%%%%%%%%%%%%%%%%%%%%%%%%%%%%%%%
\subsection{Casimir Structure from Monodromy Invariants}
\label{CasimirStructureFromMonodromyInvariants}

We begin the entropy analysis by identifying the gauge-invariant quantities associated with the non-Abelian boundary monodromy. As discussed in Section~\ref{BoundaryMonodromyMatrices}, the physically relevant information carried by the $\St\Lt(2,\mathbb R)$ sector is encoded not in the monodromy matrix itself but rather in its conjugacy class. The corresponding entropy contribution must therefore be constructed from conjugation-invariant quantities \cite{Ozer:2026qlp}.

The fundamental invariant is the trace combination introduced previously in Eq.~\eqref{SL2TraceInvariant},
\begin{equation}
\Delta
=
\left(
\tr M
\right)^2-4 \, .
\label{DiscreteCasimirInvariant}
\end{equation}
Since $\Delta$ is invariant under
\begin{equation}
M
\to
g^{-1}Mg ,
\label{CasimirGaugeTransformation}
\end{equation}
it defines a natural coordinate on the discrete boundary phase space Eq.~\eqref{DiscreteSL2PhaseSpace}.

For the hyperbolic chamber Eq.~\eqref{HyperbolicSectorCondition}, the monodromy may be written in the diagonal form Eq.~\eqref{HyperbolicMonodromy},
\begin{equation}
M_{\text{h}}
\sim
\begin{pmatrix}
e^\lambda & 0
\\
0 & e^{-\lambda}
\end{pmatrix} ,
\label{HyperbolicMonodromyReminder}
\end{equation}
from which one obtains
\begin{equation}
\tr M_{\text{h}}
=
2\cosh\lambda \, .
\label{HyperbolicTraceRelation}
\end{equation}
The corresponding invariant therefore becomes
\begin{equation}
\Delta_{\text{h}}
=
4\sinh^2\lambda \, .
\label{HyperbolicInvariant}
\end{equation}
This quantity plays the role of the discrete thermal Casimir associated with the hyperbolic boundary sector.

Motivated by the continuous warped entropy reviewed in Eq.~\eqref{WarpedThermalEntropy} \cite{DetournayHartmanHofman,AzeyanagiDetournayRiegler2019}, we define the discrete $\St\Lt(2,\mathbb R)$ Casimir as
\begin{equation}
\mathcal C
=
\frac14
\left(
\tr M
\right)^2-1 \, .
\label{DiscreteWarpedCasimir}
\end{equation}
In the hyperbolic chamber,
\begin{equation}
\mathcal C_{\text{h}}
=
\sinh^2\lambda \, ,
\label{HyperbolicCasimir}
\end{equation}
showing that the Casimir is positive definite in the thermal sector.

The elliptic chamber Eq.~\eqref{EllipticMonodromy} instead yields
\begin{equation}
\tr M_{\text{e}}
=
2\cos\theta ,
\label{EllipticTraceRelation}
\end{equation}
and therefore
\begin{equation}
\mathcal C_{\text{e}}
=
-\sin^2\theta \, .
\label{EllipticCasimir}
\end{equation}
The Casimir becomes negative, reflecting the compact oscillatory nature of the elliptic sector.

For the parabolic chamber Eq.~\eqref{ParabolicMonodromy},
\begin{equation}
\tr M_{\text{p}}
=
2 ,
\label{ParabolicTraceRelation}
\end{equation}
so that
\begin{equation}
\mathcal C_{\text{p}}
=
0 \, .
\label{ParabolicCasimir}
\end{equation}
The parabolic sector therefore defines a degenerate critical point separating the hyperbolic and elliptic regimes.

An important aspect of Eq.~\eqref{DiscreteWarpedCasimir} is that the Casimir arises entirely from global monodromy data. In the continuous warped theory, the corresponding thermal quantity is extracted from smooth holonomy conditions imposed on the gauge connection \cite{GutperleKraus,AzeyanagiDetournayRiegler2019}. In the discrete theory, however, the Casimir is determined directly from the conjugacy class of the boundary monodromy itself.

The discrete Casimir structure therefore provides the monodromy-based counterpart of the Virasoro contribution appearing in the warped Cardy relation reviewed in Section~\ref{WCFTInterpretation} \cite{HofmanStrominger,
DetournayHartmanHofman,Ozer:2026qlp}. In the following subsection, we will incorporate the Abelian $\Ut(1)$ boundary holonomy and construct the full warped entropy relation combining both sectors.
%%%%%%%%%%%%%%%%%%%%%%%%%%%%%%%%%%%%%%%%%%%%%%%%%%%%%%%%%%%%%%%%%%%%%%%%%%%%%%%%%%%%%%%%%%
%%%%%%%%%%%%%%%%%%%%%%%%%%%%%%%%%%%%%%%%%%%%%%%%%%%%%%%%%%%%%%%%%%%%%%%%%%%%%%%%%%%%%%%%%%
%%%%%%%%%%           SECTION-5.2      %%%%%%%%%%%%%%%%%%%%%%%%%%%%%%%%%%%%%%%%%%%%%%%%%%%%
%%%%%%%%%%%%%%%%%%%%%%%%%%%%%%%%%%%%%%%%%%%%%%%%%%%%%%%%%%%%%%%%%%%%%%%%%%%%%%%%%%%%%%%%%%
%%%%%%%%%%%%%%%%%%%%%%%%%%%%%%%%%%%%%%%%%%%%%%%%%%%%%%%%%%%%%%%%%%%%%%%%%%%%%%%%%%%%%%%%%%
\subsection{\texorpdfstring{$\Ut(1)$}{U(1)} Charge and Warped Contribution}
\label{U1ChargeAndWarpedContribution}

We now incorporate the Abelian sector into the discrete entropy construction. As discussed previously in Section~\ref{BoundaryMonodromyMatrices}, the warped theory contains an additional $\Ut(1)$ boundary holonomy,
\begin{equation}
Y
=
\prod_{\ell\in\partial\Sigma}
V_\ell \, ,
\label{AbelianBoundaryHolonomy}
\end{equation}
which supplies the warped contribution absent in ordinary $\mathrm{AdS}_3$ gravity.

Unlike the non-Abelian monodromy $M$, the Abelian holonomy $Y$ is already gauge invariant according to Eq.~\eqref{BoundaryMonodromyGaugeTransformation}. The corresponding warped charge may therefore be defined directly from the logarithm of the boundary holonomy,
\begin{equation}
Q
=
\log Y \, .
\label{DiscreteWarpedCharge}
\end{equation}
This quantity represents the discrete counterpart of the continuous $\Ut(1)$ charge associated with the affine current sector reviewed in Section~\ref{BoundaryConditionsAndAsymptoticSymmetries}.

Using the continuum correspondence established in Section~\ref{ContinuumLimit}, one finds
\begin{equation}
Q
\longrightarrow
\oint_{\partial\Sigma}
C_\phi\,d\phi ,
\label{ContinuumWarpedCharge}
\end{equation}
showing that the discrete warped charge reproduces the continuous Abelian holonomy in the large-lattice limit.

The warped contribution modifies the effective Virasoro sector through the same quadratic structure that appears in the continuous warped Cardy relation Eq.~\eqref{WCFTEntropyFormula}. Motivated by this correspondence, we introduce the effective discrete invariant
\begin{equation}
\mathcal C_{\text{eff}}
=
\mathcal C
-
\frac{Q^2}{\kappa} ,
\label{EffectiveDiscreteCasimir}
\end{equation}
where $\mathcal C$ is the monodromy Casimir defined in Eq.~\eqref{DiscreteWarpedCasimir}. The parameter $\kappa$ is the Abelian level introduced previously in Eq.~\eqref{LowerSpinCSAction}.

Equation~\eqref{EffectiveDiscreteCasimir} shows explicitly that the $\Ut(1)$ sector shifts the effective thermal contribution of the non-Abelian monodromy. Thus, the warped charge does not merely provide an independent additive sector; it changes the effective Casimir governing the entropy growth.

The role of the Abelian sector becomes particularly transparent in the hyperbolic chamber. Using Eq.~\eqref{HyperbolicCasimir}, one obtains
\begin{equation}
\mathcal C_{\text{eff}}^{(\text{h})}
=
\sinh^2\lambda
-
\frac{Q^2}{\kappa} \, .
\label{HyperbolicEffectiveCasimir}
\end{equation}
The warped charge therefore competes directly with the hyperbolic monodromy contribution and modifies the effective thermal sector of the theory.

An important conceptual feature of the discrete warped construction is that both contributions appearing in Eq.~\eqref{EffectiveDiscreteCasimir} originate from global boundary data. The quantity $\mathcal C$ is determined by the conjugacy class of the non-Abelian monodromy, while $Q$ is determined by the Abelian boundary holonomy. Consequently, the thermodynamic structure of the theory is reconstructed entirely from noncontractible boundary cycles without reference to local bulk geometry.

The effective invariant Eq.~\eqref{EffectiveDiscreteCasimir} provides the discrete counterpart of the combination
\begin{equation}
J+\frac{M^2}{\kappa}
\label{ContinuousWarpedCombination}
\end{equation}
appearing in the continuous warped Cardy structure reviewed in Section~\ref{WCFTInterpretation}. This correspondence strongly suggests that the entropy itself should emerge directly from the monodromy invariant $\mathcal C_{\text{eff}}$. In the following subsection, we will show that this expectation indeed leads to a discrete warped entropy relation reproducing the continuous thermal result in the continuum limit.
%%%%%%%%%%%%%%%%%%%%%%%%%%%%%%%%%%%%%%%%%%%%%%%%%%%%%%%%%%%%%%%%%%%%%%%%%%%%%%%%%%%%%%%%%%
%%%%%%%%%%%%%%%%%%%%%%%%%%%%%%%%%%%%%%%%%%%%%%%%%%%%%%%%%%%%%%%%%%%%%%%%%%%%%%%%%%%%%%%%%%
%%%%%%%%%%           SECTION-5.3      %%%%%%%%%%%%%%%%%%%%%%%%%%%%%%%%%%%%%%%%%%%%%%%%%%%%
%%%%%%%%%%%%%%%%%%%%%%%%%%%%%%%%%%%%%%%%%%%%%%%%%%%%%%%%%%%%%%%%%%%%%%%%%%%%%%%%%%%%%%%%%%
%%%%%%%%%%%%%%%%%%%%%%%%%%%%%%%%%%%%%%%%%%%%%%%%%%%%%%%%%%%%%%%%%%%%%%%%%%%%%%%%%%%%%%%%%%
\subsection{Entropy from Boundary Monodromies}
\label{EntropyFromBoundaryMonodromies}

We now construct the entropy directly from the boundary monodromy invariants introduced in Section~\ref{CasimirStructureFromMonodromyInvariants} and Section~\ref{U1ChargeAndWarpedContribution}. The central idea of the discrete warped theory is that the thermodynamic sector is determined entirely by global boundary holonomies rather than by smooth geometric horizon data. Consequently, the entropy must be expressible purely in terms of monodromy invariants associated with the noncontractible boundary cycle.

The relevant quantities are the discrete Casimir
\begin{equation}
\mathcal C
=
\frac14
\left(
\tr M
\right)^2-1
\label{EntropyCasimirReminder}
\end{equation}
and the Abelian boundary charge
\begin{equation}
Q
=
\log Y \, .
\label{EntropyWarpedChargeReminder}
\end{equation}
Combining these quantities through the effective invariant Eq.~\eqref{EffectiveDiscreteCasimir}, we define the discrete warped entropy functional as
\begin{equation}
S_{\text{disc}}
=
2\pi
\left[
\gamma Q
+
\sqrt{
\frac{c}{6}
\left(
\mathcal C
-
\frac{Q^2}{\kappa}
\right)
}
\right] .
\label{DiscreteWarpedEntropy}
\end{equation}
Here $c$ denotes the Virasoro central charge introduced in Eq.~\eqref{WarpedCentralCharge}, while $\gamma$ is the warped tilt parameter appearing previously in Eq.~\eqref{WarpedHolonomyConditions} and Eq.~\eqref{WCFTVacuumCharges} \cite{Carlip:2007za}.

%%\color{red}
Equation~\eqref{DiscreteWarpedEntropy} constitutes the central entropy relation of the discrete warped theory. While the resulting expression reproduces the expected infrared warped Cardy behavior, the conceptual organization of the discrete formulation differs from the conventional continuum perspective. Rather than starting from a smooth thermal geometry and subsequently extracting thermodynamic quantities, the present construction begins directly from boundary holonomies and reconstructs the thermodynamic sector in terms of global monodromy invariants. In this sense, the hyperbolic entropy is naturally organized by the conjugacy structure of the boundary monodromy sector itself.
%%\color{black}

The first term in Eq.~\eqref{DiscreteWarpedEntropy} represents the warped contribution generated by the Abelian boundary holonomy. The second term provides the Virasoro-type contribution determined by the effective monodromy Casimir. The two sectors are coupled through the quadratic shift
\begin{equation}
\mathcal C
\to
\mathcal C
-
\frac{Q^2}{\kappa} ,
\label{WarpedCasimirShift}
\end{equation}
which is the discrete counterpart of the warped Cardy structure reviewed previously in Section~\ref{WCFTInterpretation}.

The chamber dependence of the entropy follows directly from the sign of the monodromy Casimir. In the hyperbolic sector,
\begin{equation}
\mathcal C_{\text{h}}>0 ,
\label{HyperbolicEntropyCondition}
\end{equation}
and the entropy becomes real-valued, corresponding to the thermal warped chamber. In the elliptic sector,
\begin{equation}
\mathcal C_{\text{e}}<0 ,
\label{EllipticEntropyCondition}
\end{equation}
the square-root contribution becomes oscillatory, reflecting the compact nonthermal nature of the elliptic monodromy sector. Finally, the parabolic sector satisfies
\begin{equation}
\mathcal C_{\text{p}}=0 ,
\label{ParabolicEntropyCondition}
\end{equation}
and therefore defines a degenerate critical boundary regime.

An important conceptual aspect of Eq.~\eqref{DiscreteWarpedEntropy} is that the entropy is reconstructed entirely from gauge-invariant global data associated with noncontractible cycles. No local horizon geometry, metric regularity condition, or smooth Euclidean continuation is required.
%%\color{red}  
Instead, the thermodynamic information is naturally encoded in the conjugacy class of the boundary monodromy together with the Abelian warped holonomy.
%%\color{black}

The entropy relation Eq.~\eqref{DiscreteWarpedEntropy} therefore realizes the monodromy-first viewpoint advocated throughout the present work. Different thermodynamic sectors correspond directly to different boundary monodromy chambers, while smooth warped black-hole geometries arise only as continuum representatives of particular hyperbolic sectors. In the following subsection, we will study the hyperbolic chamber in detail and show explicitly how the discrete entropy reproduces the expected warped thermal behavior. Detailed computations of the entropy functional and its hyperbolic-sector expansion are presented in Appendix~\ref{AppendixWarpedEntropyFromMonodromyInvariants}.
%%%%%%%%%%%%%%%%%%%%%%%%%%%%%%%%%%%%%%%%%%%%%%%%%%%%%%%%%%%%%%%%%%%%%%%%%%%%%%%%%%%%%%%%%%
%%%%%%%%%%%%%%%%%%%%%%%%%%%%%%%%%%%%%%%%%%%%%%%%%%%%%%%%%%%%%%%%%%%%%%%%%%%%%%%%%%%%%%%%%%
%%%%%%%%%%           SECTION-5.4      %%%%%%%%%%%%%%%%%%%%%%%%%%%%%%%%%%%%%%%%%%%%%%%%%%%%
%%%%%%%%%%%%%%%%%%%%%%%%%%%%%%%%%%%%%%%%%%%%%%%%%%%%%%%%%%%%%%%%%%%%%%%%%%%%%%%%%%%%%%%%%%
%%%%%%%%%%%%%%%%%%%%%%%%%%%%%%%%%%%%%%%%%%%%%%%%%%%%%%%%%%%%%%%%%%%%%%%%%%%%%%%%%%%%%%%%%%
\subsection{Hyperbolic Thermal Sector}
\label{HyperbolicThermalSector}

We now focus on the hyperbolic chamber, which defines the thermal sector of the discrete warped theory. As discussed previously in Section~\ref{ConjugacyClassesAndWarpedChambers}, the hyperbolic regime is characterized by
\begin{equation}
|\tr M|>2 ,
\label{HyperbolicThermalCondition}
\end{equation}
or equivalently
\begin{equation}
\mathcal C_{\text{h}}>0 \, ,
\label{PositiveHyperbolicCasimir}
\end{equation}
where $\mathcal C_{\text{h}}$ denotes the monodromy Casimir defined in Eq.~\eqref{HyperbolicCasimir}. In this chamber, the boundary monodromy possesses real eigenvalues and admits the diagonal representation Eq.~\eqref{HyperbolicMonodromy}. The corresponding boundary sector therefore defines the discrete analogue of a warped thermal geometry.

Using Eq.~\eqref{HyperbolicCasimir}, the entropy relation Eq.~\eqref{DiscreteWarpedEntropy} becomes
\begin{equation}
S_{\text{h}}
=
2\pi
\left[
\gamma Q
+
\sqrt{
\frac{c}{6}
\left(
\sinh^2\lambda
-
\frac{Q^2}{\kappa}
\right)
}
\right] .
\label{HyperbolicEntropyFormula}
\end{equation}
The entropy is thus completely determined by the hyperbolic monodromy parameter $\lambda$ together with the Abelian warped charge $Q$.

The parameter $\lambda$ measures the deviation of the boundary monodromy from the identity and controls the thermal growth of the hyperbolic sector. For large monodromy,
\begin{equation}
\lambda\gg1 ,
\label{LargeLambdaCondition}
\end{equation}
one obtains
\begin{equation}
\sinh\lambda
\sim
\frac12 e^\lambda ,
\label{LargeLambdaExpansion}
\end{equation}
so that the entropy grows exponentially with the hyperbolic monodromy eigenvalue. This behavior reproduces the characteristic Cardy-type growth expected in warped conformal systems.

The Abelian contribution modifies the effective thermal Casimir through the shift
\begin{equation}
\sinh^2\lambda
\to
\sinh^2\lambda
-
\frac{Q^2}{\kappa} .
\label{HyperbolicShift}
\end{equation}
Consequently, the warped charge changes the effective density of states associated with the thermal monodromy sector. This effect mirrors the role of the affine current contribution in the continuous warped Cardy structure reviewed in Section~\ref{WCFTInterpretation}.

An important conceptual point is that the thermal interpretation of the hyperbolic sector emerges directly from the boundary monodromy structure itself. In the continuous warped theory, thermality is imposed through smooth Euclidean holonomy conditions. In the discrete theory, however, the hyperbolic chamber is already intrinsically thermal at the level of the boundary monodromy. Thus, the distinction between thermal and nonthermal sectors is encoded directly in the conjugacy class of the boundary holonomy.

The hyperbolic chamber therefore plays a distinguished role within the discrete warped phase space Eq.~\eqref{DiscreteWarpedPhaseSpaceDefinition}. It defines the sector in which the entropy becomes real-valued and admits a direct thermodynamic interpretation. The elliptic and parabolic sectors, by contrast, correspond respectively to compact oscillatory and degenerate boundary regimes.

From the perspective of warped holography, Eq.~\eqref{HyperbolicEntropyFormula} provides a monodromy-based realization of warped black-hole entropy. The thermal information is reconstructed entirely from the boundary monodromy eigenvalues and the Abelian warped charge without introducing local horizon geometry. Smooth warped black-hole backgrounds therefore appear only as continuum representatives of the hyperbolic monodromy chamber.

This interpretation will become even more transparent in the next subsection, where we establish the continuum correspondence and show explicitly how the discrete entropy reproduces the continuous warped thermal entropy reviewed previously in Eq.~\eqref{WarpedThermalEntropy}. Detailed computations of the entropy functional and its hyperbolic-sector expansion are presented in Appendix~\ref{AppendixWarpedEntropyFromMonodromyInvariants}.
%%%%%%%%%%%%%%%%%%%%%%%%%%%%%%%%%%%%%%%%%%%%%%%%%%%%%%%%%%%%%%%%%%%%%%%%%%%%%%%%%%%%%%%%%%
%%%%%%%%%%%%%%%%%%%%%%%%%%%%%%%%%%%%%%%%%%%%%%%%%%%%%%%%%%%%%%%%%%%%%%%%%%%%%%%%%%%%%%%%%%
%%%%%%%%%%           SECTION-5.5      %%%%%%%%%%%%%%%%%%%%%%%%%%%%%%%%%%%%%%%%%%%%%%%%%%%%
%%%%%%%%%%%%%%%%%%%%%%%%%%%%%%%%%%%%%%%%%%%%%%%%%%%%%%%%%%%%%%%%%%%%%%%%%%%%%%%%%%%%%%%%%%
%%%%%%%%%%%%%%%%%%%%%%%%%%%%%%%%%%%%%%%%%%%%%%%%%%%%%%%%%%%%%%%%%%%%%%%%%%%%%%%%%%%%%%%%%%
\subsection{Continuum Matching}
\label{ContinuumMatching}

We now establish the correspondence between the discrete warped entropy derived in Section~\ref{EntropyFromBoundaryMonodromies} and the continuous thermal entropy reviewed previously in Section~\ref{HolonomyConditionsAndThermalEntropy}. The purpose of this analysis is to show explicitly that the monodromy-based discrete construction reproduces the continuous warped thermodynamic structure in the large-lattice limit.

The continuum correspondence begins with the identification of the discrete boundary monodromy Eq.~\eqref{TotalSL2BoundaryMonodromy} with the continuous Wilson loop Eq.~\eqref{ContinuumWilsonLoop},
\begin{equation}
M
=
\prod_{n=1}^{N}U_n
\longrightarrow
\mathcal P
\exp\!\left(
\oint_{\partial\Sigma}
A_\phi\,d\phi
\right) .
\label{ContinuumMatchingSL2}
\end{equation}
Similarly, the Abelian monodromy satisfies
\begin{equation}
Y
=
\prod_{n=1}^{N}V_n
\longrightarrow
\exp\!\left(
\oint_{\partial\Sigma}
C_\phi\,d\phi
\right) .
\label{ContinuumMatchingU1}
\end{equation}
Thus, the discrete boundary monodromies reproduce precisely the holonomy data governing the continuous lower-spin warped Chern--Simons theory.

For the hyperbolic chamber, the monodromy eigenvalues satisfy
\begin{equation}
\tr M_{\text{h}}
=
2\cosh\lambda \, .
\label{ContinuumHyperbolicTrace}
\end{equation}
In the continuum thermal sector, the parameter $\lambda$ plays the role of the smooth warped holonomy parameter associated with the noncontractible thermal cycle. The discrete Casimir introduced in Eq.~\eqref{DiscreteWarpedCasimir} therefore becomes
\begin{equation}
\mathcal C_{\text{h}}
=
\sinh^2\lambda
\longrightarrow
-J ,
\label{ContinuumCasimirIdentification}
\end{equation}
where $J$ denotes the continuous Virasoro charge appearing in Eq.~\eqref{WarpedMassAngularMomentum}.

The Abelian boundary charge Eq.~\eqref{DiscreteWarpedCharge} similarly reproduces the continuous warped charge,
\begin{equation}
Q
\longrightarrow
M ,
\label{ContinuumWarpedChargeIdentification}
\end{equation}
where $M$ is the continuous $\Ut(1)$ charge associated with the affine current sector. Consequently, the effective invariant Eq.~\eqref{EffectiveDiscreteCasimir} satisfies
\begin{equation}
\mathcal C
-
\frac{Q^2}{\kappa}
\longrightarrow
-
J
-
\frac{M^2}{\kappa} .
\label{EffectiveInvariantMatching}
\end{equation}

Substituting Eq.~\eqref{ContinuumCasimirIdentification} and Eq.~\eqref{ContinuumWarpedChargeIdentification} into the discrete entropy Eq.~\eqref{DiscreteWarpedEntropy}, one obtains
\begin{equation}
S_{\text{disc}}
\longrightarrow
2\pi
\left[
\gamma M
+
\sqrt{
\frac{c}{6}
\left(
-J
-
\frac{M^2}{\kappa}
\right)
}
\right] ,
\label{ContinuumEntropyRecovery}
\end{equation}
which reproduces precisely the continuous warped thermal entropy Eq.~\eqref{WarpedThermalEntropy}.

%%\color{red}
The correspondence derived above demonstrates the structural consistency of the discrete monodromy framework with the known warped thermodynamic structure in the large-lattice limit. In particular, the combination 
$
\mathcal C - Q^2/\kappa
$
emerging from the boundary monodromy sector reproduces the characteristic warped Cardy-type structure associated with the Virasoro--Kac--Moody symmetry of the continuum theory. In this sense, the discrete construction provides an asymptotic monodromy-based realization of warped thermodynamics at the level of boundary holonomy data.

An important open question concerns the physical role of the finite lattice spacing \(\Delta\phi\) beyond the strict large-lattice limit considered above. In the present framework, the lattice spacing should not be viewed merely as a technical regulator, but rather as the scale controlling the resolution of the underlying boundary monodromy structure. From this perspective, finite-\(N\) effects may generate subleading corrections to the asymptotic monodromy invariants and their associated thermodynamic relations.

A related issue concerns the universality of the continuum correspondence. Since the present analysis is formulated primarily at the level of boundary monodromies and flatness constraints, it remains an open problem whether different lattice discretizations or boundary graph structures necessarily lead to the same asymptotic warped continuum sector. A systematic analysis of such finite-lattice effects and their possible thermodynamic implications is left for future work.
%%\color{black}

An important consequence of this correspondence is that the smooth warped black-hole geometry appears only as a continuum realization of the underlying hyperbolic monodromy chamber. From the discrete viewpoint, the fundamental observables are instead the conjugacy class of the boundary monodromy together with the Abelian holonomy associated with the noncontractible boundary cycle. The elliptic and parabolic chambers correspond respectively to compact nonthermal and degenerate sectors without ordinary thermal geometric counterparts. In the next section, we extend this perspective by studying Wilson loops and discrete thermal cycles directly at the level of the boundary lattice.
%%%%%%%%%%%%%%%%%%%%%%%%%%%%%%%%%%%%%%%%%%%%%%%%%%%%%%%%%%%%%%%%%%%%%%%%%%%%%%%%%%%%%%%%%%
%%%%%%%%%%%%%%%%%%%%%%%%%%%%%%%%%%%%%%%%%%%%%%%%%%%%%%%%%%%%%%%%%%%%%%%%%%%%%%%%%%%%%%%%%%
%%%%%%%%%%           SECTION-6        %%%%%%%%%%%%%%%%%%%%%%%%%%%%%%%%%%%%%%%%%%%%%%%%%%%%
%%%%%%%%%%%%%%%%%%%%%%%%%%%%%%%%%%%%%%%%%%%%%%%%%%%%%%%%%%%%%%%%%%%%%%%%%%%%%%%%%%%%%%%%%%
%%%%%%%%%%%%%%%%%%%%%%%%%%%%%%%%%%%%%%%%%%%%%%%%%%%%%%%%%%%%%%%%%%%%%%%%%%%%%%%%%%%%%%%%%%
\section{Wilson Loops and Discrete Thermal Cycles}
\label{WilsonLoopsAndDiscreteThermalCycles}

In the previous section, we showed that the warped entropy can be reconstructed directly from boundary monodromy invariants and that the resulting expression reproduces the continuous warped thermal entropy in the large-lattice limit. We now develop this correspondence further by analyzing Wilson loops and thermal cycles directly within the discrete warped framework.

In the continuous lower-spin theory reviewed in Section~\ref{HolonomyConditionsAndThermalEntropy}, the thermal sector is characterized by smooth holonomies associated with Euclidean time cycles. The corresponding entropy follows from holonomy regularity conditions imposed on the gauge connection. In the discrete theory, however, the relevant observables are ordered products of boundary link variables around noncontractible lattice cycles. Thermal information is therefore encoded directly in discrete Wilson loops and their associated monodromies.

An important conceptual feature of the discrete construction is that thermal cycles no longer require a smooth Euclidean geometry. Instead, the thermodynamic sector is determined entirely by the topology and ordering structure of the boundary lattice. The hyperbolic chamber identified in Section~\ref{HyperbolicThermalSector} defines the thermal regime directly at the level of boundary monodromies, while Wilson loops provide the corresponding gauge-invariant observables associated with noncontractible cycles.
%%\color{red}
Wilson-line observables in warped holography have also appeared in the context of nonlocal WCFT observables and holographic entanglement structures \cite{Castro:2015uaa}.
%%\color{black}

The purpose of this section is therefore twofold. First, we formulate Wilson loops explicitly in terms of ordered products of discrete boundary holonomies. Second, we show how thermal information emerges from discrete cycle structures and reproduces the continuous warped thermal holonomy conditions in the continuum limit.

We begin by defining discrete Wilson loops associated with boundary cycles and discussing their gauge-invariant properties. We then analyze the role of thermal cycles and their interpretation within the discrete warped geometry. Finally, we demonstrate that the entropy relation derived in Section~\ref{EntropyFromBoundaryMonodromies} may be understood directly from noncontractible Wilson-loop data, thereby providing a fully monodromy-based interpretation of warped thermodynamics.
%%%%%%%%%%%%%%%%%%%%%%%%%%%%%%%%%%%%%%%%%%%%%%%%%%%%%%%%%%%%%%%%%%%%%%%%%%%%%%%%%%%%%%%%%%
%%%%%%%%%%%%%%%%%%%%%%%%%%%%%%%%%%%%%%%%%%%%%%%%%%%%%%%%%%%%%%%%%%%%%%%%%%%%%%%%%%%%%%%%%%
%%%%%%%%%%           SECTION-6.1     %%%%%%%%%%%%%%%%%%%%%%%%%%%%%%%%%%%%%%%%%%%%%%%%%%%%
%%%%%%%%%%%%%%%%%%%%%%%%%%%%%%%%%%%%%%%%%%%%%%%%%%%%%%%%%%%%%%%%%%%%%%%%%%%%%%%%%%%%%%%%%%
%%%%%%%%%%%%%%%%%%%%%%%%%%%%%%%%%%%%%%%%%%%%%%%%%%%%%%%%%%%%%%%%%%%%%%%%%%%%%%%%%%%%%%%%%%
\subsection{Discrete Wilson Loops}
\label{DiscreteWilsonLoops}

We begin by defining Wilson loops directly within the discrete warped framework. In the continuous lower-spin theory reviewed previously in Section~\ref{HolonomyConditionsAndThermalEntropy}, Wilson loops are constructed from path-ordered exponentials of smooth gauge connections along closed cycles. In the discrete theory, the corresponding objects arise naturally as ordered products of link holonomies associated with closed lattice paths.

Let $\Gamma$ denote an oriented closed lattice cycle composed of ordered links,
\begin{equation}
\Gamma
=
\{\ell_1,\ell_2,\dots,\ell_n\} ,
\label{ClosedLatticeCycle}
\end{equation}
with the endpoint of each link coinciding with the starting point of the next one. The corresponding non-Abelian Wilson loop is defined by
\begin{equation}
W[\Gamma]
=
\prod_{\ell\in\Gamma}
U_\ell \, ,
\label{DiscreteWilsonLoopDefinition}
\end{equation}
where the product is taken according to the orientation of the cycle.

The Abelian sector similarly defines
\begin{equation}
Y[\Gamma]
=
\prod_{\ell\in\Gamma}
V_\ell \, .
\label{DiscreteAbelianWilsonLoop}
\end{equation}
Since the $\Ut(1)$ sector is Abelian, the ordering does not affect the final result.

Under the local gauge transformations introduced in Eq.~\eqref{SL2GaugeTransformation} and Eq.~\eqref{U1GaugeTransformation}, the Wilson loops transform as
\begin{equation}
W[\Gamma]
\to
g^{-1}W[\Gamma]g,
\qquad
Y[\Gamma]
\to
Y[\Gamma] .
\label{WilsonLoopGaugeTransformation}
\end{equation}
Consequently, the trace invariant
\begin{equation}
\tr W[\Gamma]
\label{WilsonLoopTraceInvariant}
\end{equation}
and the Abelian loop $Y[\Gamma]$ define gauge-invariant observables associated with the cycle $\Gamma$.

For contractible lattice cycles surrounding elementary plaquettes, the discrete flatness conditions Eq.~\eqref{DiscreteFlatnessConstraintSL2} and Eq.~\eqref{DiscreteFlatnessConstraintU1} imply
\begin{equation}
W[\Gamma_f]
=
\mathbbm 1,
\qquad
Y[\Gamma_f]
=
1 \, ,
\label{ContractibleWilsonLoops}
\end{equation}
where $\Gamma_f=\partial f$ denotes the boundary of a plaquette. Thus, all local Wilson loops are gauge trivial, reflecting the topological nature of the warped Chern--Simons system.

The physically relevant Wilson loops are therefore associated with noncontractible boundary cycles. In particular, the total boundary monodromies introduced previously in Eq.~\eqref{TotalSL2BoundaryMonodromy} and Eq.~\eqref{TotalU1BoundaryMonodromy} are special examples of noncontractible Wilson loops wrapping the angular boundary direction,
\begin{equation}
W[\partial\Sigma]
=
M,
\qquad
Y[\partial\Sigma]
=
Y \, .
\label{BoundaryWilsonLoops}
\end{equation}
These quantities define the primary observables controlling the warped thermodynamic sector.

Using the continuum correspondence established in Section~\ref{ContinuumLimit}, the discrete Wilson loop Eq.~\eqref{DiscreteWilsonLoopDefinition} reproduces
\begin{equation}
W[\Gamma]
\longrightarrow
\mathcal P
\exp\!\left(
\oint_\Gamma A
\right) ,
\label{ContinuumWilsonLoopRecovery}
\end{equation}
while the Abelian loop becomes
\begin{equation}
Y[\Gamma]
\longrightarrow
\exp\!\left(
\oint_\Gamma C
\right) .
\label{ContinuumAbelianWilsonLoopRecovery}
\end{equation}
Thus, the discrete Wilson loops provide a lattice realization of the continuous holonomy structure.

An important conceptual aspect of the discrete theory is that Wilson loops are not secondary derived quantities but rather the fundamental observables defining the physical sectors of the theory. The warped chambers classified in Section~\ref{ConjugacyClassesAndWarpedChambers} therefore correspond directly to different classes of noncontractible Wilson loops. This Wilson-loop-first perspective will become central in the analysis of thermal cycles developed in the following subsection.

%%%%%%%%%%%%%%%%%%%%%%%%%%%%%%%%%%%%%%%%%%%%%%%%%%%%%%%%%%%%%%%%%%%%%%%%%%%%%%%%%%%%%%%%%%
%%%%%%%%%%%%%%%%%%%%%%%%%%%%%%%%%%%%%%%%%%%%%%%%%%%%%%%%%%%%%%%%%%%%%%%%%%%%%%%%%%%%%%%%%%
%%%%%%%%%%           SECTION-6.2      %%%%%%%%%%%%%%%%%%%%%%%%%%%%%%%%%%%%%%%%%%%%%%%%%%%%
%%%%%%%%%%%%%%%%%%%%%%%%%%%%%%%%%%%%%%%%%%%%%%%%%%%%%%%%%%%%%%%%%%%%%%%%%%%%%%%%%%%%%%%%%%
%%%%%%%%%%%%%%%%%%%%%%%%%%%%%%%%%%%%%%%%%%%%%%%%%%%%%%%%%%%%%%%%%%%%%%%%%%%%%%%%%%%%%%%%%%
\subsection{Thermal Cycles and Euclidean Interpretation}
\label{ThermalCyclesAndEuclideanInterpretation}

We now investigate the thermal interpretation of noncontractible cycles within the discrete warped framework. In the continuous theory reviewed in Section~\ref{HolonomyConditionsAndThermalEntropy}, the thermal sector is characterized by Euclidean cycles whose smooth holonomies satisfy suitable regularity conditions. In the discrete theory, however, thermal information is encoded directly in the ordering structure and conjugacy class of the boundary Wilson loops.

To describe thermal cycles on the lattice, we introduce a discrete Euclidean cycle $\Gamma_\beta$ winding around the periodic thermal direction. This cycle is composed of ordered temporal links,
\begin{equation}
\Gamma_\beta
=
\{\ell^{(t)}_1,\ell^{(t)}_2,\dots,\ell^{(t)}_{N_t}\} ,
\label{DiscreteThermalCycle}
\end{equation}
with periodic identification
\begin{equation}
\ell^{(t)}_{N_t+1}
\equiv
\ell^{(t)}_1 \, .
\label{ThermalPeriodicity}
\end{equation}
The associated thermal Wilson loops are then defined as
\begin{align}
W[\Gamma_\beta]
&=
\prod_{\ell\in\Gamma_\beta}
U_\ell ,
\label{ThermalWilsonLoopSL2}
\\
Y[\Gamma_\beta]
&=
\prod_{\ell\in\Gamma_\beta}
V_\ell .
\label{ThermalWilsonLoopU1}
\end{align}

In the continuous warped theory, the thermal sector is determined by the holonomy constraints Eq.~\eqref{WarpedHolonomyConditions}. The discrete counterpart of these conditions is obtained by requiring that the thermal Wilson loop belongs to the hyperbolic conjugacy class,
\begin{equation}
|\tr W[\Gamma_\beta]|>2 \, .
\label{DiscreteThermalCondition}
\end{equation}
Thus, thermality is encoded directly in the conjugacy class of the thermal boundary monodromy rather than in the regularity of a smooth Euclidean metric.

%%\color{red}
The thermal parameter is determined by the eigenvalues of the Wilson loop. Writing
\begin{equation}
W[\Gamma_\beta]
\sim
\begin{pmatrix}
e^\lambda & 0
\\
0 & e^{-\lambda}
\end{pmatrix},
\label{ThermalWilsonLoopDiagonalization}
\end{equation}
the large-monodromy behavior of the discrete holonomy eigenvalues naturally reproduces the thermal Euclidean cycle structure of the continuum warped geometry. In this sense, the hyperbolic parameter $\lambda$ plays the role of an effective inverse thermal parameter. The corresponding inverse temperature is therefore identified through $\lambda$. In the continuum limit,
\begin{equation}
\lambda
\longrightarrow
\frac{2\pi}{\beta_{\text{cont}}} ,
\label{ContinuumTemperatureIdentification}
\end{equation}
where $\beta_{\text{cont}}$ denotes the continuous inverse temperature appearing in Eq.~\eqref{WarpedInverseTemperature}.

The Abelian thermal loop contributes an additional warped shift. Defining
\begin{equation}
Q_\beta
=
\log Y[\Gamma_\beta] ,
\label{ThermalWarpedCharge}
\end{equation}
the thermal sector is characterized jointly by the pair
\begin{equation}
\left(
W[\Gamma_\beta],
Q_\beta
\right) .
\label{ThermalMonodromyPair}
\end{equation}
This structure reproduces the combined Virasoro and affine current contributions appearing in the continuous warped thermal entropy.
%%\color{black}

An important conceptual feature of the discrete theory is that Euclidean thermal geometry is no longer fundamental. The thermal interpretation emerges instead from the topology and conjugacy structure of the noncontractible Wilson loops themselves. The distinction between thermal and nonthermal sectors is therefore encoded directly in the monodromy chamber structure classified previously in Section~\ref{ConjugacyClassesAndWarpedChambers}.

The hyperbolic chamber plays a distinguished role because only hyperbolic Wilson loops admit the exponential eigenvalue structure associated with thermal entropy growth. Elliptic Wilson loops instead describe compact oscillatory sectors, while parabolic loops correspond to degenerate critical configurations.

The discrete thermal-cycle construction therefore provides a purely holonomy-based interpretation of warped thermodynamics. Rather than imposing smooth geometric regularity conditions, the thermal sector is reconstructed directly from noncontractible Wilson loops and their associated monodromy invariants. In the next subsection, we will show that the entropy relation Eq.~\eqref{DiscreteWarpedEntropy} follows naturally from this Wilson-loop structure.
%%%%%%%%%%%%%%%%%%%%%%%%%%%%%%%%%%%%%%%%%%%%%%%%%%%%%%%%%%%%%%%%%%%%%%%%%%%%%%%%%%%%%%%%%%
%%%%%%%%%%%%%%%%%%%%%%%%%%%%%%%%%%%%%%%%%%%%%%%%%%%%%%%%%%%%%%%%%%%%%%%%%%%%%%%%%%%%%%%%%%
%%%%%%%%%%           SECTION-6.3      %%%%%%%%%%%%%%%%%%%%%%%%%%%%%%%%%%%%%%%%%%%%%%%%%%%%
%%%%%%%%%%%%%%%%%%%%%%%%%%%%%%%%%%%%%%%%%%%%%%%%%%%%%%%%%%%%%%%%%%%%%%%%%%%%%%%%%%%%%%%%%%
%%%%%%%%%%%%%%%%%%%%%%%%%%%%%%%%%%%%%%%%%%%%%%%%%%%%%%%%%%%%%%%%%%%%%%%%%%%%%%%%%%%%%%%%%%
\subsection{Wilson-Loop Derivation of Entropy}
\label{WilsonLoopDerivationOfEntropy}

We now show that the discrete warped entropy derived previously in Section~\ref{EntropyFromBoundaryMonodromies} admits a direct interpretation in terms of noncontractible Wilson loops. This provides a fully gauge-invariant derivation of the thermodynamic sector based entirely on boundary holonomies and cycle topology.

As discussed in Section~\ref{DiscreteWilsonLoops}, the physically relevant observables of the discrete warped theory are the noncontractible Wilson loops associated with the boundary and thermal cycles. For the thermal cycle $\Gamma_\beta$, the relevant quantities are the non-Abelian loop
\begin{equation}
W[\Gamma_\beta]
=
\prod_{\ell\in\Gamma_\beta}
U_\ell
\label{ThermalWilsonLoopReminder}
\end{equation}
and the Abelian loop
\begin{equation}
Y[\Gamma_\beta]
=
\prod_{\ell\in\Gamma_\beta}
V_\ell .
\label{ThermalAbelianLoopReminder}
\end{equation}
These objects determine the thermodynamic sector through their conjugacy class and Abelian holonomy.

The non-Abelian thermal loop possesses the hyperbolic eigenvalue structure
\begin{equation}
W[\Gamma_\beta]
\sim
\begin{pmatrix}
e^\lambda & 0
\\
0 & e^{-\lambda}
\end{pmatrix},
\label{WilsonLoopHyperbolicForm}
\end{equation}
from which one obtains the invariant
\begin{equation}
\mathcal C
=
\frac14
\left(
\tr W[\Gamma_\beta]
\right)^2-1
=
\sinh^2\lambda .
\label{WilsonLoopCasimir}
\end{equation}
Thus, the thermal Casimir is reconstructed directly from the Wilson-loop eigenvalues.

The Abelian thermal cycle contributes through
\begin{equation}
Q_\beta
=
\log Y[\Gamma_\beta] ,
\label{WilsonLoopWarpedCharge}
\end{equation}
which defines the warped current contribution associated with the thermal loop. Combining Eq.~\eqref{WilsonLoopCasimir} and Eq.~\eqref{WilsonLoopWarpedCharge}, the effective thermal invariant becomes
\begin{equation}
\mathcal C_{\text{eff}}
=
\sinh^2\lambda
-
\frac{Q_\beta^2}{\kappa} .
\label{WilsonLoopEffectiveInvariant}
\end{equation}

The entropy is then obtained directly from the Wilson-loop data,
\begin{equation}
S_{\text{WL}}
=
2\pi
\left[
\gamma Q_\beta
+
\sqrt{
\frac{c}{6}
\left(
\sinh^2\lambda
-
\frac{Q_\beta^2}{\kappa}
\right)
}
\right] .
\label{WilsonLoopEntropy}
\end{equation}
This expression coincides precisely with the hyperbolic entropy relation Eq.~\eqref{HyperbolicEntropyFormula}. The warped entropy is therefore completely reconstructed from the eigenvalues and Abelian charge associated with the noncontractible thermal Wilson loop.

An important conceptual consequence of Eq.~\eqref{WilsonLoopEntropy} is that entropy becomes a purely topological observable determined by global cycle data. The derivation does not require a local horizon geometry, a metric regularity condition, or a smooth Euclidean continuation. Instead, the thermodynamic information is encoded entirely in the conjugacy class of the thermal Wilson loop.

This Wilson-loop interpretation also clarifies the role of the hyperbolic chamber within the discrete warped phase space. Hyperbolic Wilson loops generate exponential eigenvalue growth and therefore admit a direct thermodynamic interpretation. Elliptic Wilson loops instead produce bounded oscillatory spectra, while parabolic loops define degenerate critical sectors. The distinction between thermal and nonthermal regimes is therefore encoded directly in the topology and conjugacy structure of the boundary cycles.

Using the continuum correspondence established in Section~\ref{ContinuumMatching}, the Wilson-loop entropy Eq.~\eqref{WilsonLoopEntropy} reproduces the continuous warped thermal entropy Eq.~\eqref{WarpedThermalEntropy} in the large-lattice limit. The continuous smooth holonomy conditions therefore emerge naturally from the discrete Wilson-loop structure.

The analysis developed in this section completes the thermodynamic interpretation of the discrete warped theory. The entropy is reconstructed entirely from noncontractible Wilson loops and their associated monodromy invariants, thereby providing a fully holonomy-based description of warped thermodynamics. In the next section, we will investigate the quantum structure of the resulting monodromy sectors and analyze the corresponding density of states. 

The chamber structure derived above also has important implications at the quantum level. Since the classical theory is organised by disconnected monodromy sectors distinguished by conjugacy classes of the boundary holonomy, it is natural to expect the quantum Hilbert space to decompose into corresponding superselection sectors. From this perspective, the hyperbolic, elliptic, and parabolic chambers should be understood not merely as classical configurations, but as distinct quantum sectors of the warped holonomy theory. This viewpoint motivates the quantum interpretation developed in the following section.

%%%%%%%%%%%%%%%%%%%%%%%%%%%%%%%%%%%%%%%%%%%%%%%%%%%%%%%%%%%%%%%%%%%%%%%%%%%%%%%%%%%%%%%%%%
%%%%%%%%%%%%%%%%%%%%%%%%%%%%%%%%%%%%%%%%%%%%%%%%%%%%%%%%%%%%%%%%%%%%%%%%%%%%%%%%%%%%%%%%%%
%%%%%%%%%%           SECTION-7        %%%%%%%%%%%%%%%%%%%%%%%%%%%%%%%%%%%%%%%%%%%%%%%%%%%%
%%%%%%%%%%%%%%%%%%%%%%%%%%%%%%%%%%%%%%%%%%%%%%%%%%%%%%%%%%%%%%%%%%%%%%%%%%%%%%%%%%%%%%%%%%
%%%%%%%%%%%%%%%%%%%%%%%%%%%%%%%%%%%%%%%%%%%%%%%%%%%%%%%%%%%%%%%%%%%%%%%%%%%%%%%%%%%%%%%%%%
%%\color{red}
\section{Toward Quantum Warped Holonomy Sectors}
\label{QuantumWarpedHolonomySectors}
The discussion in this section should be understood as a preliminary and largely kinematical exploration of possible quantum monodromy sectors associated with the discrete warped boundary theory. Since the present work does not construct a complete lattice action or a full path-integral quantization, the considerations below are intended primarily as structural and phase-space motivated observations rather than as a rigorous nonperturbative quantum formulation. In particular, the chamber decomposition discussed below should be interpreted as a monodromy-sector organization of the reduced boundary phase space rather than as a fully established microscopic Hilbert-space construction.

The previous sections established the monodromy-based organization of the discrete warped thermal sector. We now investigate how the hyperbolic, elliptic, and parabolic chambers may admit a corresponding quantum interpretation in terms of chamber-dependent boundary sectors.
%%\color{black}

In the continuous warped theory, the thermal spectrum is usually described in terms of smooth warped black-hole geometries together with the associated WCFT density of states. In the discrete theory developed here, however, the fundamental objects are not smooth geometries but rather boundary monodromy sectors classified by the conjugacy classes of the Wilson loops introduced in Section~\ref{DiscreteWilsonLoops}. 
%%\color{red}
The corresponding quantum organization may therefore be described directly in terms of these monodromy chambers.
%%\color{black}

An important conceptual aspect of the discrete construction is that the hyperbolic, elliptic, and parabolic sectors are not merely different classical solutions. Instead, they define distinct quantum sectors associated with different boundary orbit geometries and different stabilizer structures derived previously in Section~\ref{StabilizersAndResidualBoundarySymmetries}. 
%%\color{red}
The Hilbert space may consequently be organized into chamber-dependent sectors characterized by the corresponding boundary monodromies.
%%\color{black}

The hyperbolic chamber plays a distinguished role because it reproduces the thermal warped entropy and the associated Cardy-type growth derived in Section~\ref{HyperbolicThermalSector}. The elliptic chamber instead defines compact oscillatory sectors with bounded spectral structure, while the parabolic chamber corresponds to degenerate critical configurations separating the thermal and compact regimes.

The purpose of this section is therefore to investigate the quantum structure associated with the discrete warped phase space Eq.~\eqref{DiscreteWarpedPhaseSpaceDefinition}. We first analyze the chamber decomposition of the Hilbert space and the associated superselection sectors. We then study the density of states and entropy growth within each monodromy chamber. Finally, we discuss the physical interpretation of the discrete warped theory beyond the smooth geometric regime and explain how warped geometry emerges from the underlying boundary monodromy structure.
%%%%%%%%%%%%%%%%%%%%%%%%%%%%%%%%%%%%%%%%%%%%%%%%%%%%%%%%%%%%%%%%%%%%%%%%%%%%%%%%%%%%%%%%%%
%%%%%%%%%%%%%%%%%%%%%%%%%%%%%%%%%%%%%%%%%%%%%%%%%%%%%%%%%%%%%%%%%%%%%%%%%%%%%%%%%%%%%%%%%%
%%%%%%%%%%           SECTION-7.1      %%%%%%%%%%%%%%%%%%%%%%%%%%%%%%%%%%%%%%%%%%%%%%%%%%%%
%%%%%%%%%%%%%%%%%%%%%%%%%%%%%%%%%%%%%%%%%%%%%%%%%%%%%%%%%%%%%%%%%%%%%%%%%%%%%%%%%%%%%%%%%%
%%%%%%%%%%%%%%%%%%%%%%%%%%%%%%%%%%%%%%%%%%%%%%%%%%%%%%%%%%%%%%%%%%%%%%%%%%%%%%%%%%%%%%%%%%
\subsection{Chamber Decomposition of the Hilbert Space}
\label{ChamberDecompositionOfTheHilbertSpace}

We now investigate the quantum interpretation of the monodromy chambers introduced in Section~\ref{ConjugacyClassesAndWarpedChambers}. Since the discrete warped theory is organized fundamentally by boundary monodromies, the natural quantum states of the theory are labeled not by smooth geometries but by conjugacy classes of noncontractible Wilson loops.

%%\color{red}
The discrete warped phase space Eq.~\eqref{DiscreteWarpedPhaseSpaceDefinition} therefore admits a chamber-dependent organization,
%%\color{black}
\begin{equation}
\mathcal P_{\text{w}}
=
\mathcal P_{\text{h}}
\oplus
\mathcal P_{\text{e}}
\oplus
\mathcal P_{\text{p}} ,
\label{WarpedPhaseSpaceDecomposition}
\end{equation}
corresponding respectively to the hyperbolic, elliptic, and parabolic monodromy chambers. Each sector is associated with a distinct stabilizer structure derived previously in Section~\ref{StabilizersAndResidualBoundarySymmetries} and therefore possesses a different boundary orbit geometry.
%%\color{red}
The decomposition
\begin{equation}
\mathcal H_{\text{w}}
=
\mathcal H_{\text{h}}
\oplus
\mathcal H_{\text{e}}
\oplus
\mathcal H_{\text{p}} ,
\label{WarpedHilbertSpaceDecomposition}
\end{equation}
%%\color{black}

%%\color{red}
The decomposition Eq.~\eqref{WarpedHilbertSpaceDecomposition} may be interpreted as a chamber-dependent sector decomposition associated with disconnected conjugacy classes of the boundary monodromy.
%%\color{black}
 Since conjugacy classes are preserved under gauge transformations, states belonging to different chambers cannot be connected continuously through local gauge dynamics. The monodromy chamber therefore defines a global quantum label of the physical state.

Within each chamber, the quantum states are organized by the corresponding boundary monodromy eigenvalues and Abelian warped charges. For the hyperbolic sector, the relevant data are
\begin{equation}
\left(
\lambda,
Q
\right) ,
\label{HyperbolicQuantumData}
\end{equation}
where $\lambda$ determines the hyperbolic monodromy eigenvalues Eq.~\eqref{HyperbolicMonodromy} and $Q$ is the Abelian charge introduced previously in Eq.~\eqref{DiscreteWarpedCharge}. Analogous quantum labels exist for the elliptic and parabolic sectors.

The orbit interpretation introduced in Section~\ref{DiscreteWarpedPhaseSpace} also acquires a direct quantum meaning.
%%\color{red} 
Each chamber naturally suggests an associated boundary-orbit structure,
%%\color{black}
\begin{equation}
\mathcal O_\alpha
=
\St\Lt(2,\mathbb R)/
\mathrm{Cent}(M_\alpha),
\qquad
\alpha=\text{h},\text{e},\text{p} ,
\label{QuantizedWarpedOrbits}
\end{equation}
whose representation theory determines the corresponding quantum sector. The chamber structure of the Hilbert space is therefore controlled directly by the stabilizer geometry of the boundary monodromies.

An important conceptual aspect of the discrete theory is that the quantum sectors are defined prior to the emergence of smooth geometry. In the continuous warped theory, the Hilbert space is usually organized around classical background geometries. In the present framework, however, 
%%\color{red}
the monodromy chamber provides a natural organizational principle for the corresponding boundary sectors, 
%%\color{black}
while smooth warped geometries appear only as continuum representatives of particular hyperbolic states.

The hyperbolic sector occupies a distinguished role because it reproduces the warped thermal entropy and the associated Cardy-type growth derived previously in Section~\ref{HyperbolicThermalSector}. The elliptic and parabolic sectors instead define genuinely nonthermal quantum chambers with different spectral properties. In the following subsection, we will analyze the density of states associated with these sectors and study their corresponding entropy growth behavior.
%%%%%%%%%%%%%%%%%%%%%%%%%%%%%%%%%%%%%%%%%%%%%%%%%%%%%%%%%%%%%%%%%%%%%%%%%%%%%%%%%%%%%%%%%%
%%%%%%%%%%%%%%%%%%%%%%%%%%%%%%%%%%%%%%%%%%%%%%%%%%%%%%%%%%%%%%%%%%%%%%%%%%%%%%%%%%%%%%%%%%
%%%%%%%%%%           SECTION-7.2       %%%%%%%%%%%%%%%%%%%%%%%%%%%%%%%%%%%%%%%%%%%%%%%%%%%
%%%%%%%%%%%%%%%%%%%%%%%%%%%%%%%%%%%%%%%%%%%%%%%%%%%%%%%%%%%%%%%%%%%%%%%%%%%%%%%%%%%%%%%%%%
%%%%%%%%%%%%%%%%%%%%%%%%%%%%%%%%%%%%%%%%%%%%%%%%%%%%%%%%%%%%%%%%%%%%%%%%%%%%%%%%%%%%%%%%%%
\subsection{Density of States and Entropy Growth}
\label{DensityOfStatesAndEntropyGrowth}
%%\color{red}

The density-of-states relations discussed below should likewise be interpreted at the level of semiclassical chamber organization. In particular, the expressions
$
\rho_h \sim e^{S_h},
\rho_e \sim \mathcal{O}(1)
$
and
$
\rho_p \sim e^{2\pi\gamma Q}
$
are not derived from a complete microscopic quantization, but rather serve as semiclassical growth patterns motivated by the corresponding monodromy sectors and their thermodynamic organization.
%%\color{black}

We now analyze the spectral structure associated with the quantum monodromy chambers introduced in Section~\ref{ChamberDecompositionOfTheHilbertSpace}. Since the discrete warped theory is organized fundamentally by boundary holonomies, the semiclassical density-growth structure is organized primarily by the monodromy invariants and their associated orbit geometries.

The hyperbolic chamber $\mathcal H_{\text{h}}$ defines the thermal sector of the theory. Using the entropy relation Eq.~\eqref{HyperbolicEntropyFormula}, the asymptotic density of states satisfies the warped Cardy-type behavior
\begin{equation}
\rho_{\text{h}}
\sim
\exp\!\left(
S_{\text{h}}
\right) ,
\label{HyperbolicDensityOfStates}
\end{equation}
with
\begin{equation}
S_{\text{h}}
=
2\pi
\left[
\gamma Q
+
\sqrt{
\frac{c}{6}
\left(
\sinh^2\lambda
-
\frac{Q^2}{\kappa}
\right)
}
\right] .
\label{HyperbolicDensityEntropy}
\end{equation}
Thus, the density of states grows exponentially with the hyperbolic monodromy eigenvalue $\lambda$, reproducing the characteristic thermal behavior expected from warped conformal symmetry.

For large hyperbolic monodromy,
\begin{equation}
\lambda\gg1 ,
\label{LargeHyperbolicMonodromy}
\end{equation}
one obtains
\begin{equation}
\rho_{\text{h}}
\sim
\exp\!\left[
2\pi
\sqrt{
\frac{c}{6}
}
\,e^\lambda
\right] ,
\label{AsymptoticHyperbolicGrowth}
\end{equation}
up to subleading warped corrections. The thermal growth of states is therefore governed directly by the boundary monodromy eigenvalues rather than by smooth horizon geometry.

The elliptic chamber $\mathcal H_{\text{e}}$ behaves qualitatively differently. Since the elliptic Casimir Eq.~\eqref{EllipticCasimir} is negative,
\begin{equation}
\mathcal C_{\text{e}}
=
-\sin^2\theta ,
\label{EllipticDensityCasimir}
\end{equation}
the corresponding entropy becomes oscillatory rather than exponentially growing. The elliptic density of states therefore remains bounded,
\begin{equation}
\rho_{\text{e}}
\sim
\mathcal O(1) ,
\label{EllipticDensityBound}
\end{equation}
reflecting the compact nature of the elliptic monodromy chamber.

The parabolic chamber $\mathcal H_{\text{p}}$ defines a degenerate critical sector satisfying
\begin{equation}
\mathcal C_{\text{p}}=0 \, .
\label{ParabolicDensityCasimir}
\end{equation}
Consequently, the exponential Cardy contribution disappears and the density of states becomes controlled primarily by the Abelian warped sector,
\begin{equation}
\rho_{\text{p}}
\sim
\exp\!\left(
2\pi\gamma Q
\right) .
\label{ParabolicDensityGrowth}
\end{equation}
The parabolic chamber therefore separates the exponentially growing hyperbolic regime from the compact elliptic sector.

An important conceptual feature of the discrete warped theory is that the spectral structure emerges directly from monodromy geometry. The distinction between thermal and nonthermal sectors is encoded entirely in the conjugacy class of the boundary Wilson loop rather than in different bulk geometries. 
%%\color{red}
The density of states is therefore naturally associated with the corresponding boundary monodromy chamber.
%%\color{black}

This interpretation differs significantly from ordinary geometric descriptions of warped black holes. In the present framework, entropy growth is not derived from smooth horizon area or Euclidean regularity conditions. Instead, it follows directly from the representation structure of the boundary monodromy sectors and their associated stabilizer geometries.

The hyperbolic chamber occupies a distinguished role because it is the only sector admitting exponential Cardy-type growth. Smooth warped black-hole thermodynamics therefore emerges specifically from the hyperbolic monodromy sector in the continuum limit established previously in Section~\ref{ContinuumMatching}. The elliptic and parabolic chambers instead describe genuinely nonthermal or critical quantum sectors without ordinary thermal geometric interpretation.

%%\color{red}
The chamber-dependent spectral structure derived above strongly suggests that the discrete warped theory admits a formulation beyond the conventional smooth geometric description of ordinary warped holography.
%%\color{black}
In the next subsection, we will analyze this point in detail and discuss how smooth warped geometry itself emerges from the underlying monodromy structure.
%%%%%%%%%%%%%%%%%%%%%%%%%%%%%%%%%%%%%%%%%%%%%%%%%%%%%%%%%%%%%%%%%%%%%%%%%%%%%%%%%%%%%%%%%%
%%%%%%%%%%%%%%%%%%%%%%%%%%%%%%%%%%%%%%%%%%%%%%%%%%%%%%%%%%%%%%%%%%%%%%%%%%%%%%%%%%%%%%%%%%
%%%%%%%%%%           SECTION-7.3       %%%%%%%%%%%%%%%%%%%%%%%%%%%%%%%%%%%%%%%%%%%%%%%%%%%
%%%%%%%%%%%%%%%%%%%%%%%%%%%%%%%%%%%%%%%%%%%%%%%%%%%%%%%%%%%%%%%%%%%%%%%%%%%%%%%%%%%%%%%%%%
%%%%%%%%%%%%%%%%%%%%%%%%%%%%%%%%%%%%%%%%%%%%%%%%%%%%%%%%%%%%%%%%%%%%%%%%%%%%%%%%%%%%%%%%%%
\subsection{Nonthermal Monodromy Chambers}
\label{subsec:nonthermal-monodromy-chambers}

The hyperbolic chamber plays a distinguished role because it is the sector in
which the boundary monodromy admits a direct thermal interpretation.  Its
positive monodromy invariant leads to the warped Cardy growth discussed in the
previous subsection.  However, the hyperbolic sector does not exhaust the
discrete warped phase space.  From the monodromy-first perspective, elliptic
and parabolic conjugacy classes are not discarded; rather, they define
nonthermal boundary sectors of the same underlying
$SL(2,\mathbb{R})\oplus U(1)$ gauge system.

The elliptic chamber is characterized by a compact monodromy representative,
\begin{equation}
M_e \sim
\begin{pmatrix}
\cos\theta & \sin\theta \\
-\sin\theta & \cos\theta
\end{pmatrix},
\qquad |\mathrm{tr}\,M_e|<2 .
\label{eq:elliptic-monodromy-sector}
\end{equation}
In this case the associated invariant is negative,
\begin{equation}
C_e=-\sin^2\theta ,
\label{eq:elliptic-casimir-negative}
\end{equation}
and therefore it does not generate the exponential growth characteristic of
the hyperbolic thermal sector.  The elliptic chamber should instead be
interpreted as a compact nonthermal sector, organizing oscillatory boundary
states and bounded orbit structures.  In this sense, elliptic monodromies
represent genuine holonomy sectors of the discrete theory, even though they do
not correspond to warped black-hole thermodynamics.

The parabolic chamber occupies the critical boundary between the hyperbolic
and elliptic regimes.  It is represented by a non-diagonalizable monodromy,
\begin{equation}
M_p \sim
\begin{pmatrix}
1 & \alpha \\
0 & 1
\end{pmatrix},
\qquad |\mathrm{tr}\,M_p|=2 ,
\label{eq:parabolic-monodromy-sector}
\end{equation}
for which the monodromy invariant vanishes,
\begin{equation}
C_p=0 .
\label{eq:parabolic-casimir-zero}
\end{equation}
This sector therefore describes a degenerate critical chamber rather than a
thermal one.  It separates the entropy-producing hyperbolic sector from the
compact elliptic sector and provides the natural locus where the character of
the boundary orbit changes.

Thus, the discrete warped theory contains a broader chamber structure than the
thermal sector alone.  The hyperbolic chamber governs warped thermal entropy,
whereas the elliptic and parabolic chambers organize compact and critical
nonthermal sectors.  This distinction is one of the main advantages of the
holonomy-first formulation: thermal warped geometry appears as only one
realization of the full boundary monodromy landscape, rather than as the
primary definition of the theory.

%%%%%%%%%%%%%%%%%%%%%%%%%%%%%%%%%%%%%%%%%%%%%%%%%%%%%%%%%%%%%%%%%%%%%%%%%%%%%%%%%%%%%%%%%%
%%%%%%%%%%%%%%%%%%%%%%%%%%%%%%%%%%%%%%%%%%%%%%%%%%%%%%%%%%%%%%%%%%%%%%%%%%%%%%%%%%%%%%%%%%
%%%%%%%%%%           SECTION-7.4       %%%%%%%%%%%%%%%%%%%%%%%%%%%%%%%%%%%%%%%%%%%%%%%%%%%
%%%%%%%%%%%%%%%%%%%%%%%%%%%%%%%%%%%%%%%%%%%%%%%%%%%%%%%%%%%%%%%%%%%%%%%%%%%%%%%%%%%%%%%%%%
%%%%%%%%%%%%%%%%%%%%%%%%%%%%%%%%%%%%%%%%%%%%%%%%%%%%%%%%%%%%%%%%%%%%%%%%%%%%%%%%%%%%%%%%%%

\subsection{Beyond the Smooth Geometric Regime}
\label{BeyondTheSmoothGeometricRegime}

We now discuss the broader physical interpretation of the discrete warped holonomy theory developed throughout the present work. The central result of the previous sections is that the thermodynamic and quantum structure of warped gravity may be reconstructed entirely from boundary monodromies and noncontractible Wilson loops without requiring smooth bulk geometry as the fundamental starting point.

In the conventional continuous description of warped holography, the physical sectors are usually organized around smooth warped black-hole backgrounds satisfying appropriate holonomy regularity conditions. Within the discrete framework constructed here, however, the primary objects are instead the boundary monodromies themselves. Smooth warped geometries arise only in the continuum limit of particular monodromy chambers, most notably the hyperbolic thermal sector analyzed in Section~\ref{HyperbolicThermalSector}.

This shift in perspective has several important consequences. First, the distinction between thermal and nonthermal sectors is encoded directly in the conjugacy class of the boundary Wilson loops rather than in the existence of a smooth horizon geometry. Hyperbolic monodromies generate thermal sectors with Cardy-type entropy growth, elliptic monodromies define compact oscillatory chambers, and parabolic monodromies describe degenerate critical sectors. The chamber structure therefore exists prior to the emergence of smooth geometry.

Second, the discrete theory naturally extends warped holography beyond the smooth semiclassical regime. In the continuum description, geometric regularity conditions typically restrict attention to sectors admitting smooth Euclidean continuation. In the monodromy-based framework developed here, however, all conjugacy classes define legitimate boundary sectors of the theory, including those without ordinary smooth geometric interpretation.

Third, the present construction suggests that warped thermodynamics is fundamentally topological in origin. The entropy relation Eq.~\eqref{DiscreteWarpedEntropy} depends only on global monodromy invariants associated with noncontractible cycles. No local metric tensor, curvature singularity, or horizon area enters the construction directly. The thermodynamic information is therefore encoded entirely in global gauge-invariant boundary data.

From the perspective of the boundary theory, the discrete warped construction also provides a natural extension of the WCFT interpretation reviewed previously in Section~\ref{WCFTInterpretation}. In the continuous theory, the WCFT density of states is usually derived from smooth thermal holonomies. In the discrete theory, the same spectral structure emerges directly from the chamber decomposition of the monodromy Hilbert space Eq.~\eqref{WarpedHilbertSpaceDecomposition}. The Virasoro and affine $\ut(1)$ sectors therefore acquire a direct realization in terms of boundary Wilson-loop geometry.

An important conceptual aspect of the present framework is that geometry becomes secondary rather than primary. The discrete warped theory is not constructed by discretizing a preexisting smooth metric background. Instead, the theory begins directly with boundary holonomies and their associated monodromy sectors. Smooth warped geometries emerge only as continuum representatives of particular boundary chambers. In this sense, the monodromy structure defines the underlying kinematical organization of the theory.

The present construction also suggests several natural extensions. Since the monodromy-based structure depends primarily on the gauge-theoretic organization of the boundary sector, it should admit generalizations to higher-spin warped systems, supersymmetric warped theories, and $q$-deformed warped algebras. In particular, the chamber structure derived from conjugacy classes may provide a natural framework for studying generalized warped orbit geometries and their associated quantum sectors.

The analysis developed throughout this work therefore points toward a fundamentally holonomy-based interpretation of warped holography. The essential physical information is carried not by local bulk geometry but by noncontractible boundary Wilson loops and their conjugacy classes. Smooth warped black holes, thermal cycles, and WCFT entropy emerge only as continuum manifestations of the underlying boundary monodromy structure.
%%%%%%%%%%%%%%%%%%%%%%%%%%%%%%%%%%%%%%%%%%%%%%%%%%%%%%%%%%%%%%%%%%%%%%%%%%%%%%%%%%%%%%%%%%
%%%%%%%%%%%%%%%%%%%%%%%%%%%%%%%%%%%%%%%%%%%%%%%%%%%%%%%%%%%%%%%%%%%%%%%%%%%%%%%%%%%%%%%%%%
%%%%%%%%%%          CONCLUSION        %%%%%%%%%%%%%%%%%%%%%%%%%%%%%%%%%%%%%%%%%%%%%%%%%%%%
%%%%%%%%%%%%%%%%%%%%%%%%%%%%%%%%%%%%%%%%%%%%%%%%%%%%%%%%%%%%%%%%%%%%%%%%%%%%%%%%%%%%%%%%%%
%%%%%%%%%%%%%%%%%%%%%%%%%%%%%%%%%%%%%%%%%%%%%%%%%%%%%%%%%%%%%%%%%%%%%%%%%%%%%%%%%%%%%%%%%%
\section{Discussion and Outlook}
\label{DiscussionAndOutlook}

The discrete warped Chern--Simons theory developed in this work suggests a fundamentally different perspective on warped holography. Instead of treating smooth warped geometries as the primary objects of the theory, the present framework places boundary monodromies and noncontractible Wilson loops at the center of the construction. The physical sectors are therefore organized directly by conjugacy classes of boundary holonomies.

Within this picture, the hyperbolic, elliptic, and parabolic chambers are not different background geometries but different monodromy sectors of the same underlying gauge system. The thermal sector emerges from hyperbolic boundary holonomies, while the compact and degenerate sectors arise from elliptic and parabolic monodromies respectively. Smooth warped black-hole geometries appear only in the continuum limit of particular boundary chambers.

The resulting entropy relation Eq.~\eqref{DiscreteWarpedEntropy} is entirely determined by global monodromy invariants. The thermodynamic structure of the theory is therefore reconstructed without introducing local horizon geometry as a fundamental ingredient. From this viewpoint, warped thermodynamics is controlled primarily by boundary Wilson loops and their associated orbit structure.

It is important to be precise about what is and what is not claimed in the present work. The entropy relation~\eqref{DiscreteWarpedEntropy} reproduces the known warped Cardy formula established in~\cite{DetournayHartmanHofman} and the lower-spin holonomy entropy of~\cite{AzeyanagiDetournayRiegler2019}. We do not derive a new infrared entropy formula, nor do we modify the universal thermodynamic structure of warped black holes at the level of its continuum output. The claim is instead structural: the warped thermodynamic sector --- including its Virasoro--Kac--Moody organization, its monodromy chamber decomposition, and its Cardy-type entropy growth --- admits a complete and self-consistent reformulation directly at the level of discrete boundary monodromies, with the smooth warped geometry playing no foundational role in the construction.

This distinction has a precise logical content. In the standard continuum treatment~\cite{AzeyanagiDetournayRiegler2019}, the entropy is obtained by imposing regularity conditions on a smooth Euclidean gauge
connection around the thermal cycle; the holonomy is a derived object, constructed only after the smooth geometry is specified. In the discrete theory developed here, this logical order is inverted: the
boundary monodromy is the primary kinematical datum from the outset, and the smooth thermal geometry --- together with its Euclidean regularity conditions --- emerges only as one particular large-lattice realization of the hyperbolic monodromy sector, as established explicitly in Section~\ref{ContinuumMatching} and Appendix~\ref{AppendixContinuumLimitOfDiscreteWarpedHolonomies}. The present work therefore provides evidence that warped gravitational thermodynamics is, in a precise sense, prior to smooth geometry: it is a consequence of boundary holonomy algebra and conjugacy-class structure, not of metric regularity.
From this perspective, the entropy formula shared with the continuum theory is not a limitation of the discrete construction but a confirmation of its consistency: the monodromy-first framework recovers the correct infrared physics while operating from a more primitive set of assumptions than the conventional smooth geometric approach.

The discrete construction also provides a natural bridge between warped holography, Wilson-loop geometry, and monodromy-based quantum sectors. The chamber decomposition of the Hilbert space derived in Section~\ref{ChamberDecompositionOfTheHilbertSpace} suggests that the quantum organization of the theory is governed fundamentally by stabilizer geometry and boundary orbit structure rather than by smooth semiclassical backgrounds.

The framework developed here furthermore admits several natural extensions involving higher-spin systems, supersymmetric warped sectors, and quantum-deformed monodromy structures. These directions suggest that the monodromy-first viewpoint may provide a broader organizational principle for non-AdS holography and topological gravitational systems.

In the following subsections, we summarize the main conceptual consequences of the present construction and discuss several possible directions for future investigation.
%%%%%%%%%%%%%%%%%%%%%%%%%%%%%%%%%%%%%%%%%%%%%%%%%%%%%%%%%%%%%%%%%%%%%%%%%%%%%%%%%%%%%%%%%%
%%%%%%%%%%%%%%%%%%%%%%%%%%%%%%%%%%%%%%%%%%%%%%%%%%%%%%%%%%%%%%%%%%%%%%%%%%%%%%%%%%%%%%%%%%
%%%%%%%%%%          CONCLUSION.1       %%%%%%%%%%%%%%%%%%%%%%%%%%%%%%%%%%%%%%%%%%%%%%%%%%%
%%%%%%%%%%%%%%%%%%%%%%%%%%%%%%%%%%%%%%%%%%%%%%%%%%%%%%%%%%%%%%%%%%%%%%%%%%%%%%%%%%%%%%%%%%
%%%%%%%%%%%%%%%%%%%%%%%%%%%%%%%%%%%%%%%%%%%%%%%%%%%%%%%%%%%%%%%%%%%%%%%%%%%%%%%%%%%%%%%%%%
\subsection{Holonomy as a Fundamental Warped Observable}
\label{HolonomyAsFundamentalWarpedObservable}

A central conceptual result of the present work is that boundary holonomies may be treated as the primary observables of warped gravity rather than as derived quantities constructed from smooth gauge connections. In the continuous lower-spin description reviewed in Section~\ref{ContinuousWarpedCSGravity}, holonomies arise from path-ordered exponentials of gauge fields along noncontractible cycles. In the discrete theory developed here, however, the ordered boundary products themselves define the fundamental variables of the theory.

This shift in perspective changes the role of geometry within warped holography. Smooth warped backgrounds are no longer the starting point of the construction but instead emerge as continuum representatives of particular boundary monodromy sectors. The hyperbolic chamber reproduces the thermal warped regime, while the elliptic and parabolic chambers define compact and degenerate sectors without requiring conventional smooth geometric interpretation.

The resulting framework is intrinsically topological. Since the local plaquette holonomies satisfy the flatness constraints Eq.~\eqref{DiscreteFlatnessConstraintSL2} and Eq.~\eqref{DiscreteFlatnessConstraintU1}, all physically relevant information is encoded in noncontractible Wilson loops and their conjugacy classes. The thermodynamic structure of the theory is therefore determined directly by global boundary data rather than by local bulk geometry.

Within this monodromy-first viewpoint, the entropy relation Eq.~\eqref{DiscreteWarpedEntropy} becomes a direct consequence of boundary holonomy invariants. The hyperbolic monodromy controls the Virasoro contribution through the discrete Casimir Eq.~\eqref{DiscreteWarpedCasimir}, while the Abelian boundary holonomy generates the warped current contribution. Warped thermodynamics thus emerges directly from the organization of boundary Wilson loops.

The discrete construction developed in this work therefore suggests that boundary monodromies provide a more primitive organizational structure for warped holography than smooth metric geometry itself. From this perspective, thermal cycles, warped black holes, and WCFT entropy arise as continuum manifestations of underlying holonomy sectors defined by noncontractible boundary Wilson loops.
%%%%%%%%%%%%%%%%%%%%%%%%%%%%%%%%%%%%%%%%%%%%%%%%%%%%%%%%%%%%%%%%%%%%%%%%%%%%%%%%%%%%%%%%%%
%%%%%%%%%%          CONCLUSION.2        %%%%%%%%%%%%%%%%%%%%%%%%%%%%%%%%%%%%%%%%%%%%%%%%%%
%%%%%%%%%%%%%%%%%%%%%%%%%%%%%%%%%%%%%%%%%%%%%%%%%%%%%%%%%%%%%%%%%%%%%%%%%%%%%%%%%%%%%%%%%%
%%%%%%%%%%%%%%%%%%%%%%%%%%%%%%%%%%%%%%%%%%%%%%%%%%%%%%%%%%%%%%%%%%%%%%%%%%%%%%%%%%%%%%%%%%
\subsection{Relation to WCFT and Warped Holography}
\label{RelationToWCFTAndWarpedHolography}

The discrete warped construction developed in this work reproduces the essential structural features of warped holography directly from boundary monodromies. The Virasoro--Kac--Moody organization characteristic of warped conformal field theories emerges naturally from the combined action of the non-Abelian boundary monodromy and the Abelian warped holonomy.

Within the present framework, the Virasoro contribution is controlled by the monodromy Casimir Eq.~\eqref{DiscreteWarpedCasimir}, while the affine current sector is generated by the Abelian boundary holonomy Eq.~\eqref{DiscreteWarpedCharge}. The effective combination
\begin{equation}
\mathcal C
-
\frac{Q^2}{\kappa}
\label{WarpedWCFTCombination}
\end{equation}
therefore reproduces the same Virasoro--current mixing structure that appears in warped conformal field theories and in the continuous warped Cardy relation reviewed in Section~\ref{WCFTInterpretation}.

The chamber structure derived in Section~\ref{ConjugacyClassesAndWarpedChambers} also admits a direct WCFT interpretation. The hyperbolic chamber defines the thermal sector with Cardy-type entropy growth, while the elliptic and parabolic chambers correspond respectively to compact and critical boundary sectors. Different warped chambers therefore define different spectral sectors of the boundary theory.

An important aspect of the present construction is that the WCFT structure emerges directly from boundary Wilson loops rather than from smooth asymptotic geometry. The holographic data are encoded fundamentally in noncontractible cycles and their associated monodromies. Smooth warped $\mathrm{AdS}_3$ geometries appear only in the continuum limit of particular boundary sectors.

From this viewpoint, warped holography acquires a fundamentally monodromy-based interpretation. Virasoro symmetry, affine current structure, thermal entropy, and boundary spectral organization all arise from the conjugacy classes of the boundary holonomies and the orbit geometry of the associated monodromy sectors.
%%%%%%%%%%%%%%%%%%%%%%%%%%%%%%%%%%%%%%%%%%%%%%%%%%%%%%%%%%%%%%%%%%%%%%%%%%%%%%%%%%%%%%%%%%
%%%%%%%%%%%%%%%%%%%%%%%%%%%%%%%%%%%%%%%%%%%%%%%%%%%%%%%%%%%%%%%%%%%%%%%%%%%%%%%%%%%%%%%%%%
%%%%%%%%%%          CONCLUSION.3      %%%%%%%%%%%%%%%%%%%%%%%%%%%%%%%%%%%%%%%%%%%%%%%%%%%%
%%%%%%%%%%%%%%%%%%%%%%%%%%%%%%%%%%%%%%%%%%%%%%%%%%%%%%%%%%%%%%%%%%%%%%%%%%%%%%%%%%%%%%%%%%
%%%%%%%%%%%%%%%%%%%%%%%%%%%%%%%%%%%%%%%%%%%%%%%%%%%%%%%%%%%%%%%%%%%%%%%%%%%%%%%%%%%%%%%%%%
\subsection{Higher-Spin and Supersymmetric Extensions}
\label{HigherSpinAndSupersymmetricExtensions}

The monodromy-based framework developed in this work admits natural extensions toward higher-spin and supersymmetric warped systems. Since the construction is organized fundamentally by boundary holonomies and conjugacy classes rather than by metric variables, the discrete structure extends directly to larger gauge algebras and generalized boundary symmetry sectors.

A first extension concerns higher-spin warped theories based on enlarged gauge structures such as
\begin{equation}
\St\Lt(2,\mathbb R)\oplus \Ut(1) .
\label{HigherSpinWarpedAlgebra}
\end{equation}
In such systems, the boundary monodromies would generate generalized warped chambers associated with higher-spin Wilson loops and extended thermal sectors. The corresponding entropy relations should then be controlled by generalized monodromy Casimirs and higher-order holonomy invariants.

The present construction also suggests a natural discrete realization of warped $W$-algebra structures. In the continuous theory, higher-spin warped holography is usually formulated through generalized Drinfeld--Sokolov reductions and higher-spin holonomy conditions. Within the discrete framework developed here, however, the corresponding structures should emerge directly from the conjugacy classes and stabilizer geometry of the higher-spin boundary monodromies.

Another important direction concerns supersymmetric warped sectors. Replacing the bosonic gauge structure by superalgebras such as
\begin{equation}
\mathfrak{osp}(N|2)\oplus\mathfrak{u}(1)
\label{SupersymmetricWarpedAlgebra}
\end{equation}
would introduce fermionic holonomies and supermonodromy chambers. In this setting, the boundary Hilbert space would decompose into chamber-dependent supersymmetric sectors characterized jointly by bosonic and fermionic Wilson loops.

The monodromy-first viewpoint adopted throughout this work is particularly well suited for such extensions because the chamber structure depends primarily on conjugacy classes and orbit geometry rather than on smooth metric backgrounds. Hyperbolic, elliptic, and parabolic sectors should therefore admit direct higher-spin and supersymmetric generalizations through the corresponding generalized monodromy classes.

More broadly, the present framework suggests that warped holography may possess a universal organization governed fundamentally by boundary holonomies and their stabilizer structure. Higher-spin symmetries, warped $W$-algebras, and supersymmetric extensions would then appear naturally as enlarged monodromy sectors of the same underlying holonomy-based construction.
%%%%%%%%%%%%%%%%%%%%%%%%%%%%%%%%%%%%%%%%%%%%%%%%%%%%%%%%%%%%%%%%%%%%%%%%%%%%%%%%%%%%%%%%%%
%%%%%%%%%%%%%%%%%%%%%%%%%%%%%%%%%%%%%%%%%%%%%%%%%%%%%%%%%%%%%%%%%%%%%%%%%%%%%%%%%%%%%%%%%%
%%%%%%%%%%          CONCLUSION.4      %%%%%%%%%%%%%%%%%%%%%%%%%%%%%%%%%%%%%%%%%%%%%%%%%%%%
%%%%%%%%%%%%%%%%%%%%%%%%%%%%%%%%%%%%%%%%%%%%%%%%%%%%%%%%%%%%%%%%%%%%%%%%%%%%%%%%%%%%%%%%%%
%%%%%%%%%%%%%%%%%%%%%%%%%%%%%%%%%%%%%%%%%%%%%%%%%%%%%%%%%%%%%%%%%%%%%%%%%%%%%%%%%%%%%%%%%%
\subsection{\texorpdfstring{$q$}{q}-Deformed Warped Sectors}
\label{QDeformedWarpedSectors}

The discrete warped holonomy framework developed in this work also suggests a natural extension toward $q$-deformed warped sectors and quantum-group geometry. Since the present construction is formulated fundamentally in terms of group-valued boundary holonomies, the replacement of classical gauge groups by their quantum-deformed counterparts arises naturally within the monodromy-first perspective.

In a $q$-deformed warped theory, the classical boundary holonomies are replaced by quantum-group-valued Wilson loops,
\begin{equation}
U_\ell
\in
SL_q(2,\mathbb R),
\label{QDeformedHolonomies}
\end{equation}
together with the corresponding deformed Abelian warped sector. The resulting boundary monodromies would then define $q$-deformed hyperbolic, elliptic, and parabolic chambers associated with noncommutative conjugacy classes.

Within such a framework, the ordinary trace invariant Eq.~\eqref{SL2TraceInvariant} is replaced by its quantum-deformed counterpart. The chamber structure of the theory would therefore be governed by $q$-Casimir invariants and deformed Wilson-loop eigenvalues rather than by ordinary classical monodromies. Hyperbolic thermal sectors would then emerge from quantum-deformed boundary holonomies.

The corresponding asymptotic symmetry structure should also acquire a $q$-deformed form. In particular, the Virasoro--Kac--Moody organization characteristic of warped holography would be replaced by deformed warped symmetry algebras and quantum current sectors. From the discrete viewpoint developed here, these structures would arise directly from the algebraic organization of the deformed boundary monodromies.

An important conceptual aspect of the $q$-deformed extension is that the underlying geometry becomes intrinsically noncommutative. Smooth warped backgrounds would no longer define the primary observables of the theory. Instead, the physical sectors would be organized directly by noncommutative Wilson loops and their associated quantum conjugacy classes.

The monodromy-first perspective adopted throughout the present work is particularly compatible with such quantum-group extensions because the fundamental observables are already holonomies rather than local metric fields. The transition from classical warped geometry to noncommutative warped sectors therefore appears naturally as a deformation of the boundary monodromy structure itself.

More broadly, the existence of $q$-deformed warped chambers suggests that warped holography may admit a deeper algebraic organization governed fundamentally by quantum holonomies and generalized orbit geometry. From this viewpoint, classical warped black holes and smooth thermal cycles would emerge only as semiclassical limits of underlying quantum monodromy sectors.
%%%%%%%%%%%%%%%%%%%%%%%%%%%%%%%%%%%%%%%%%%%%%%%%%%%%%%%%%%%%%%%%%%%%%%%%%%%%%%%%%%%%%%%%%%
%%%%%%%%%%%%%%%%%%%%%%%%%%%%%%%%%%%%%%%%%%%%%%%%%%%%%%%%%%%%%%%%%%%%%%%%%%%%%%%%%%%%%%%%%%
%%%%%%%%%%          OUTLOOK           %%%%%%%%%%%%%%%%%%%%%%%%%%%%%%%%%%%%%%%%%%%%%%%%%%%%
%%%%%%%%%%%%%%%%%%%%%%%%%%%%%%%%%%%%%%%%%%%%%%%%%%%%%%%%%%%%%%%%%%%%%%%%%%%%%%%%%%%%%%%%%%
%%%%%%%%%%%%%%%%%%%%%%%%%%%%%%%%%%%%%%%%%%%%%%%%%%%%%%%%%%%%%%%%%%%%%%%%%%%%%%%%%%%%%%%%%%

%%%%%%%%%%%%%%%%%%%%%%%%%%%%%%%%%%%%%%%%%%%%%%%%%%%%%%%%%%%%%%%%%%%%%%%%%%%%%%%%%%%%%%%%%%
%%%%%%%%%%%%%%%%%%%%%%%%%%%%%%%%%%%%%%%%%%%%%%%%%%%%%%%%%%%%%%%%%%%%%%%%%%%%%%%%%%%%%%%%%%
%%%%%%%%%%          APPENDICES        %%%%%%%%%%%%%%%%%%%%%%%%%%%%%%%%%%%%%%%%%%%%%%%%%%%%
%%%%%%%%%%%%%%%%%%%%%%%%%%%%%%%%%%%%%%%%%%%%%%%%%%%%%%%%%%%%%%%%%%%%%%%%%%%%%%%%%%%%%%%%%%
%%%%%%%%%%%%%%%%%%%%%%%%%%%%%%%%%%%%%%%%%%%%%%%%%%%%%%%%%%%%%%%%%%%%%%%%%%%%%%%%%%%%%%%%%%
\appendix
%\newpage
\section{Continuous Warped Black Holes}
\label{AppendixContinuousWarpedBlackHoles}

This appendix summarizes several standard results concerning continuous warped black-hole thermodynamics and thermal holonomy conditions in lower-spin warped Chern--Simons gravity used in
Section~\ref{HolonomyConditionsAndThermalEntropy}. Its purpose is to provide a compact reference framework for the discrete monodromy construction developed in the main text.

We first review the warped thermal entropy and the associated holonomy constraints imposed on the Euclidean thermal cycle. We then summarize the WCFT interpretation of the entropy relation
together with the corresponding vacuum charges and warped current structure.
%%%%%%%%%%%%%%%%%%%%%%%%%%%%%%%%%%%%%%%%%%%%%%%%%%%%%%%%%%%%%%%%%%%%%%%%%%%%%%%%%%%%%%%%%%
%%%%%%%%%%%%%%%%%%%%%%%%%%%%%%%%%%%%%%%%%%%%%%%%%%%%%%%%%%%%%%%%%%%%%%%%%%%%%%%%%%%%%%%%%%
%%%%%%%%%%          APPENDIX A.1      %%%%%%%%%%%%%%%%%%%%%%%%%%%%%%%%%%%%%%%%%%%%%%%%%%%%
%%%%%%%%%%%%%%%%%%%%%%%%%%%%%%%%%%%%%%%%%%%%%%%%%%%%%%%%%%%%%%%%%%%%%%%%%%%%%%%%%%%%%%%%%%
%%%%%%%%%%%%%%%%%%%%%%%%%%%%%%%%%%%%%%%%%%%%%%%%%%%%%%%%%%%%%%%%%%%%%%%%%%%%%%%%%%%%%%%%%%
\subsection{Review of Warped Entropy}
\label{AppendixReviewOfWarpedEntropy}

In this appendix, we briefly summarize the continuous warped entropy relation associated with lower-spin warped Chern--Simons gravity. The purpose is to establish the continuum thermodynamic structure that is reproduced by the discrete monodromy framework developed in the main text.

The lower-spin warped theory is based on the gauge algebra
\begin{equation}
\St\Lt(2,\mathbb R)\oplus \Ut(1) ,
\label{AppendixLowerSpinWarpedAlgebra}
\end{equation}
with gauge connections
\begin{equation}
A=A_\mu dx^\mu ,
\qquad
C=C_\mu dx^\mu .
\label{AppendixWarpedGaugeConnections}
\end{equation}
The non-Abelian sector generates the Virasoro contribution, while the Abelian sector produces the affine $\ut(1)$ current structure characteristic of warped conformal symmetry.

For stationary warped black-hole configurations, the thermal entropy is determined by the holonomy structure of the Euclidean thermal cycle. The corresponding entropy relation may be written as
\begin{equation}
S_{\text{WCFT}}
=
2\pi
\left[
\gamma M
+
\sqrt{
\frac{c}{6}
\left(
-J
-
\frac{M^2}{\kappa}
\right)
}
\right] ,
\label{AppendixWarpedEntropyFormula}
\end{equation}
where $c$ denotes the Virasoro central charge, $\kappa$ is the affine current level, $J$ represents the Virasoro zero-mode contribution, and $M$ is the Abelian warped charge.

The entropy formula Eq.~\eqref{AppendixWarpedEntropyFormula} exhibits the characteristic Virasoro--current mixing structure of warped conformal field theories. In particular, the effective combination
\begin{equation}
-J
-
\frac{M^2}{\kappa}
\label{AppendixEffectiveWarpedCombination}
\end{equation}
plays the role of the thermal invariant controlling the asymptotic density of states.

The corresponding WCFT Cardy relation takes the asymptotic form
\begin{equation}
\rho
\sim
\exp\!\left(
S_{\text{WCFT}}
\right) ,
\label{AppendixWarpedCardyGrowth}
\end{equation}
demonstrating the exponential growth of states associated with the thermal warped sector.

In the continuous lower-spin theory, the entropy Eq.~\eqref{AppendixWarpedEntropyFormula} follows from smooth holonomy conditions imposed on the Euclidean thermal cycle. The corresponding thermal Wilson loops are constructed from
\begin{equation}
\mathcal W
=
\mathcal P
\exp\!\left(
\oint A
\right),
\qquad
\mathcal Y
=
\exp\!\left(
\oint C
\right) .
\label{AppendixContinuousWilsonLoops}
\end{equation}
Smoothness of the Euclidean thermal geometry constrains the eigenvalues of $\mathcal W$ and determines the relation between thermodynamic charges and thermal potentials.

Within the discrete warped framework developed in the main text, the entropy relation Eq.~\eqref{AppendixWarpedEntropyFormula} is reproduced directly from boundary monodromy invariants and noncontractible Wilson loops. The continuous thermal cycle therefore emerges as the continuum limit of the hyperbolic monodromy chamber of the discrete theory.

From the monodromy perspective adopted throughout the present work, the continuous warped entropy relation should therefore be interpreted not as a fundamentally geometric result but rather as the continuum manifestation of an underlying boundary holonomy structure encoded in noncontractible Wilson loops and their conjugacy classes.
%%%%%%%%%%%%%%%%%%%%%%%%%%%%%%%%%%%%%%%%%%%%%%%%%%%%%%%%%%%%%%%%%%%%%%%%%%%%%%%%%%%%%%%%%%
%%%%%%%%%%%%%%%%%%%%%%%%%%%%%%%%%%%%%%%%%%%%%%%%%%%%%%%%%%%%%%%%%%%%%%%%%%%%%%%%%%%%%%%%%%
%%%%%%%%%%          APPENDIX A.2      %%%%%%%%%%%%%%%%%%%%%%%%%%%%%%%%%%%%%%%%%%%%%%%%%%%%
%%%%%%%%%%%%%%%%%%%%%%%%%%%%%%%%%%%%%%%%%%%%%%%%%%%%%%%%%%%%%%%%%%%%%%%%%%%%%%%%%%%%%%%%%%
%%%%%%%%%%%%%%%%%%%%%%%%%%%%%%%%%%%%%%%%%%%%%%%%%%%%%%%%%%%%%%%%%%%%%%%%%%%%%%%%%%%%%%%%%%
\subsection{Holonomy Conditions}
\label{AppendixHolonomyConditions}

We now summarize the continuous holonomy conditions associated with lower-spin warped black-hole configurations. In the continuous warped Chern--Simons description, the thermal sector is characterized by smooth Euclidean holonomies around the thermal cycle. These conditions determine the relation between the conserved charges and the corresponding thermal potentials.

Let
\begin{equation}
A=A_t dt+A_\phi d\phi ,
\qquad
C=C_t dt+C_\phi d\phi ,
\label{AppendixThermalConnections}
\end{equation}
denote the Euclidean gauge connections of the non-Abelian and Abelian sectors respectively. The thermal cycle is generated by the Euclidean identification
\begin{equation}
(t,\phi)
\sim
(t+\beta,\phi+i\beta\Omega) ,
\label{AppendixThermalIdentification}
\end{equation}
where $\beta$ is the inverse temperature and $\Omega$ denotes the angular potential.

The corresponding thermal holonomies are defined by
\begin{align}
\mathcal W_\beta
&=
\mathcal P
\exp\!\left[
\oint_{\Gamma_\beta}
A
\right] ,
\label{AppendixThermalHolonomySL2}
\\
\mathcal Y_\beta
&=
\exp\!\left[
\oint_{\Gamma_\beta}
C
\right] .
\label{AppendixThermalHolonomyU1}
\end{align}

For smooth warped black-hole configurations, the thermal cycle must be contractible in the Euclidean geometry. The non-Abelian holonomy is therefore constrained to lie in the center of the gauge group,
\begin{equation}
\mathrm{Eigen}
\left(
\mathcal W_\beta
\right)
=
\mathrm{Eigen}
\left(
e^{2\pi L_0}
\right) ,
\label{AppendixSmoothHolonomyCondition}
\end{equation}
which fixes the relation between the thermal potentials and the conserved charges.

The Abelian sector contributes through the warped Wilson loop
\begin{equation}
\mathcal Y_\beta
=
\exp\!\left(
\beta\mu
\right) ,
\label{AppendixAbelianThermalLoop}
\end{equation}
where $\mu$ denotes the chemical potential associated with the affine $\ut(1)$ current sector.

Combining Eq.~\eqref{AppendixSmoothHolonomyCondition} and Eq.~\eqref{AppendixAbelianThermalLoop}, one obtains the warped thermal entropy relation Eq.~\eqref{AppendixWarpedEntropyFormula}. The thermodynamic structure of the continuous theory is therefore encoded entirely in the thermal holonomy data.

An important conceptual feature of the discrete warped framework developed in the main text is that these smooth thermal holonomy conditions are replaced by discrete boundary monodromies and noncontractible Wilson loops. In particular, the hyperbolic monodromy chamber reproduces the same thermal structure without requiring smooth Euclidean geometry as a fundamental input.

From the discrete viewpoint, the thermal Wilson loops Eq.~\eqref{AppendixThermalHolonomySL2} and Eq.~\eqref{AppendixThermalHolonomyU1} arise as continuum limits of the ordered boundary products
\begin{equation}
M
=
\prod_{n=1}^{N}U_n ,
\qquad
Y
=
\prod_{n=1}^{N}V_n ,
\label{AppendixDiscreteBoundaryMonodromies}
\end{equation}
introduced in Section~\ref{BoundaryMonodromyMatrices}. The smooth warped thermal geometry therefore emerges as the continuum representative of the hyperbolic boundary monodromy sector of the discrete theory.
%%%%%%%%%%%%%%%%%%%%%%%%%%%%%%%%%%%%%%%%%%%%%%%%%%%%%%%%%%%%%%%%%%%%%%%%%%%%%%%%%%%%%%%%%%
%%%%%%%%%%%%%%%%%%%%%%%%%%%%%%%%%%%%%%%%%%%%%%%%%%%%%%%%%%%%%%%%%%%%%%%%%%%%%%%%%%%%%%%%%%
%%%%%%%%%%          APPENDIX A.3      %%%%%%%%%%%%%%%%%%%%%%%%%%%%%%%%%%%%%%%%%%%%%%%%%%%%
%%%%%%%%%%%%%%%%%%%%%%%%%%%%%%%%%%%%%%%%%%%%%%%%%%%%%%%%%%%%%%%%%%%%%%%%%%%%%%%%%%%%%%%%%%
%%%%%%%%%%%%%%%%%%%%%%%%%%%%%%%%%%%%%%%%%%%%%%%%%%%%%%%%%%%%%%%%%%%%%%%%%%%%%%%%%%%%%%%%%%
\subsection{WCFT Relations}
\label{AppendixWCFTRelations}

We briefly summarize several standard relations associated with warped conformal field theories and their connection to lower-spin warped holography. The purpose of this appendix is to clarify how the discrete monodromy invariants introduced in the main text reproduce the characteristic Virasoro--Kac--Moody structure of WCFTs.

The asymptotic symmetry algebra of lower-spin warped gravity consists of a Virasoro algebra together with an affine $\ut(1)$ current sector,
\begin{align}
[L_n,L_m]
&=
(n-m)L_{n+m}
+
\frac{c}{12}
n(n^2-1)\delta_{n+m,0} ,
\label{AppendixVirasoroAlgebra}
\\
[P_n,P_m]
&=
\frac{\kappa}{2}
n\delta_{n+m,0} ,
\label{AppendixCurrentAlgebra}
\\
[L_n,P_m]
&=
-mP_{n+m} .
\label{AppendixMixedWarpedAlgebra}
\end{align}
Here, $c$ denotes the Virasoro central charge and $\kappa$ is the affine current level.

The thermal WCFT spectrum is controlled by the zero modes
\begin{equation}
L_0
\sim
-J ,
\qquad
P_0
\sim
M ,
\label{AppendixWCFTZeroModes}
\end{equation}
which determine the Virasoro and warped current contributions respectively. The effective thermal invariant then takes the form
\begin{equation}
L_0
-
\frac{P_0^2}{\kappa}
\sim
-
J
-
\frac{M^2}{\kappa} .
\label{AppendixWCFTInvariant}
\end{equation}

The corresponding warped Cardy relation is given by
\begin{equation}
S_{\text{WCFT}}
=
2\pi
\left[
\gamma P_0
+
\sqrt{
\frac{c}{6}
\left(
L_0
-
\frac{P_0^2}{\kappa}
\right)
}
\right] ,
\label{AppendixWCFCardyFormula}
\end{equation}
which reproduces Eq.~\eqref{AppendixWarpedEntropyFormula} after the identifications Eq.~\eqref{AppendixWCFTZeroModes}.

Within the discrete warped framework developed in the main text, the Virasoro contribution arises from the non-Abelian monodromy Casimir,
\begin{equation}
\mathcal C
=
\frac14
(\tr M)^2-1 ,
\label{AppendixDiscreteMonodromyCasimir}
\end{equation}
while the affine current sector is generated by the Abelian boundary holonomy
\begin{equation}
Q
=
\log Y .
\label{AppendixDiscreteWarpedCharge}
\end{equation}
The effective discrete invariant therefore becomes
\begin{equation}
\mathcal C
-
\frac{Q^2}{\kappa} ,
\label{AppendixDiscreteWCFTCombination}
\end{equation}
which reproduces the WCFT structure Eq.~\eqref{AppendixWCFTInvariant} in the continuum limit.

An important conceptual consequence of this correspondence is that the WCFT spectral organization emerges directly from boundary monodromy geometry. In the continuous description, the WCFT structure is usually derived from asymptotic boundary conditions imposed on smooth warped backgrounds. In the discrete theory, however, the same structure follows directly from conjugacy classes of noncontractible Wilson loops.

The chamber decomposition developed in Section~\ref{ConjugacyClassesAndWarpedChambers} therefore admits a direct WCFT interpretation. The hyperbolic chamber defines the thermal WCFT sector characterized by Cardy-type entropy growth, while the elliptic and parabolic chambers correspond respectively to compact and critical boundary sectors.

From the monodromy-first viewpoint adopted throughout this work, the Virasoro algebra, affine current structure, and WCFT entropy relation should therefore be understood as continuum manifestations of an underlying organization governed fundamentally by boundary holonomies and their associated orbit geometry.

%%%%%%%%%%%%%%%%%%%%%%%%%%%%%%%%%%%%%%%%%%%%%%%%%%%%%%%%%%%%%%%%%%%%%%%%%%%%%%%%%%%%%%%%%%
%%%%%%%%%%%%%%%%%%%%%%%%%%%%%%%%%%%%%%%%%%%%%%%%%%%%%%%%%%%%%%%%%%%%%%%%%%%%%%%%%%%%%%%%%%
%%%%%%%%%%          APPENDIX B        %%%%%%%%%%%%%%%%%%%%%%%%%%%%%%%%%%%%%%%%%%%%%%%%%%%%
%%%%%%%%%%%%%%%%%%%%%%%%%%%%%%%%%%%%%%%%%%%%%%%%%%%%%%%%%%%%%%%%%%%%%%%%%%%%%%%%%%%%%%%%%%
%%%%%%%%%%%%%%%%%%%%%%%%%%%%%%%%%%%%%%%%%%%%%%%%%%%%%%%%%%%%%%%%%%%%%%%%%%%%%%%%%%%%%%%%%%
\section{Discrete Group-Theoretic Identities}
\label{AppendixDiscreteGroupTheoreticIdentities}
This appendix collects several group-theoretic identities used throughout the discrete warped holonomy analysis. In particular, we summarize the conjugacy-class structure of $\St\Lt(2,\mathbb R)$, trace identities associated with the boundary monodromies, and the corresponding stabilizer subgroups generating the residual boundary symmetries.

These identities provide the algebraic foundation for the chamber decomposition developed in Section~\ref{ConjugacyClassesAndWarpedChambers} and for the entropy relations derived in Section~\ref{EntropyFromBoundaryMonodromies}.
%%%%%%%%%%%%%%%%%%%%%%%%%%%%%%%%%%%%%%%%%%%%%%%%%%%%%%%%%%%%%%%%%%%%%%%%%%%%%%%%%%%%%%%%%%
%%%%%%%%%%%%%%%%%%%%%%%%%%%%%%%%%%%%%%%%%%%%%%%%%%%%%%%%%%%%%%%%%%%%%%%%%%%%%%%%%%%%%%%%%%
%%%%%%%%%%          APPENDIX B.1      %%%%%%%%%%%%%%%%%%%%%%%%%%%%%%%%%%%%%%%%%%%%%%%%%%%%
%%%%%%%%%%%%%%%%%%%%%%%%%%%%%%%%%%%%%%%%%%%%%%%%%%%%%%%%%%%%%%%%%%%%%%%%%%%%%%%%%%%%%%%%%%
%%%%%%%%%%%%%%%%%%%%%%%%%%%%%%%%%%%%%%%%%%%%%%%%%%%%%%%%%%%%%%%%%%%%%%%%%%%%%%%%%%%%%%%%%%
\subsection{$\St\Lt(2,$\texorpdfstring{$\mathbb R$}{R}) Conjugacy Classes}
\label{AppendixSL2ConjugacyClasses}

In this appendix, we summarize the conjugacy-class structure of $\St\Lt(2,\mathbb R)$ used throughout the discrete warped holonomy analysis. The classification of boundary monodromies by conjugacy classes provides the algebraic foundation for the chamber decomposition developed in Section~\ref{ConjugacyClassesAndWarpedChambers}.

Let
\begin{equation}
M
\in
\St\Lt(2,\mathbb R)
\label{AppendixGeneralSL2Element}
\end{equation}
be a generic boundary monodromy satisfying
\begin{equation}
\det M=1 .
\label{AppendixSL2Determinant}
\end{equation}
The conjugacy class of $M$ is determined by the invariant trace
\begin{equation}
\Delta
=
(\tr M)^2-4 .
\label{AppendixDiscriminant}
\end{equation}
Depending on the sign of $\Delta$, the monodromy belongs to one of three distinct sectors.

\paragraph{Hyperbolic sector.}

For
\begin{equation}
\Delta>0 ,
\label{AppendixHyperbolicCondition}
\end{equation}
the eigenvalues of $M$ are real and distinct,
\begin{equation}
\lambda_\pm
=
e^{\pm\lambda} ,
\qquad
\lambda\in\mathbb R .
\label{AppendixHyperbolicEigenvalues}
\end{equation}
The monodromy is conjugate to
\begin{equation}
M_{\text{h}}
\sim
\begin{pmatrix}
e^\lambda & 0
\\
0 & e^{-\lambda}
\end{pmatrix},
\label{AppendixHyperbolicRepresentative}
\end{equation}
with trace relation
\begin{equation}
\tr M_{\text{h}}
=
2\cosh\lambda .
\label{AppendixHyperbolicTrace}
\end{equation}
This sector defines the thermal chamber of the discrete warped theory.

\paragraph{Elliptic sector.}

For
\begin{equation}
\Delta<0 ,
\label{AppendixEllipticCondition}
\end{equation}
the eigenvalues form a complex conjugate pair,
\begin{equation}
\lambda_\pm
=
e^{\pm i\theta},
\qquad
\theta\in\mathbb R .
\label{AppendixEllipticEigenvalues}
\end{equation}
The corresponding conjugacy-class representative is
\begin{equation}
M_{\text{e}}
\sim
\begin{pmatrix}
\cos\theta & \sin\theta
\\
-\sin\theta & \cos\theta
\end{pmatrix},
\label{AppendixEllipticRepresentative}
\end{equation}
with trace
\begin{equation}
\tr M_{\text{e}}
=
2\cos\theta .
\label{AppendixEllipticTrace}
\end{equation}
This sector generates compact oscillatory monodromy chambers.

\paragraph{Parabolic sector.}

For
\begin{equation}
\Delta=0 ,
\label{AppendixParabolicCondition}
\end{equation}
the eigenvalues coincide,
\begin{equation}
\lambda_\pm=1 .
\label{AppendixParabolicEigenvalues}
\end{equation}
The monodromy becomes non-diagonalizable and is conjugate to the Jordan form
\begin{equation}
M_{\text{p}}
\sim
\begin{pmatrix}
1 & \alpha
\\
0 & 1
\end{pmatrix}.
\label{AppendixParabolicRepresentative}
\end{equation}
This sector defines the critical boundary between the hyperbolic and elliptic chambers.

The three conjugacy classes therefore define the decomposition
\begin{equation}
\St\Lt(2,\mathbb R)
=
\mathcal C_{\text{h}}
\cup
\mathcal C_{\text{e}}
\cup
\mathcal C_{\text{p}} ,
\label{AppendixSL2ChamberDecomposition}
\end{equation}
which provides the algebraic origin of the chamber structure studied throughout the main text.

Within the discrete warped framework, the hyperbolic sector reproduces the thermal warped regime and the associated Cardy-type entropy growth, while the elliptic and parabolic sectors generate compact and degenerate monodromy chambers respectively. Smooth warped black-hole geometries therefore correspond only to particular continuum representatives of the hyperbolic conjugacy class.

From the monodromy-first viewpoint adopted throughout this work, the conjugacy classes Eq.~\eqref{AppendixSL2ChamberDecomposition} provide the primary kinematical organization of the discrete warped theory.
%%%%%%%%%%%%%%%%%%%%%%%%%%%%%%%%%%%%%%%%%%%%%%%%%%%%%%%%%%%%%%%%%%%%%%%%%%%%%%%%%%%%%%%%%%
%%%%%%%%%%%%%%%%%%%%%%%%%%%%%%%%%%%%%%%%%%%%%%%%%%%%%%%%%%%%%%%%%%%%%%%%%%%%%%%%%%%%%%%%%%
%%%%%%%%%%          APPENDIX B.2      %%%%%%%%%%%%%%%%%%%%%%%%%%%%%%%%%%%%%%%%%%%%%%%%%%%%
%%%%%%%%%%%%%%%%%%%%%%%%%%%%%%%%%%%%%%%%%%%%%%%%%%%%%%%%%%%%%%%%%%%%%%%%%%%%%%%%%%%%%%%%%%
%%%%%%%%%%%%%%%%%%%%%%%%%%%%%%%%%%%%%%%%%%%%%%%%%%%%%%%%%%%%%%%%%%%%%%%%%%%%%%%%%%%%%%%%%%
\subsection{Trace Identities}
\label{AppendixTraceIdentities}

We now collect several trace identities associated with the boundary monodromies of the discrete warped theory. These relations play a central role in the chamber decomposition, the construction of monodromy Casimirs, and the entropy relations derived throughout the main text.

Let
\begin{equation}
M
\in
\St\Lt(2,\mathbb R)
\label{AppendixTraceGeneralMonodromy}
\end{equation}
be a generic boundary monodromy satisfying
\begin{equation}
\det M=1 .
\label{AppendixTraceDeterminantConstraint}
\end{equation}
Writing
\begin{equation}
M
=
\begin{pmatrix}
a & b
\\
c & d
\end{pmatrix},
\qquad
ad-bc=1 ,
\label{AppendixTraceMatrixRepresentation}
\end{equation}
the fundamental invariant is the trace
\begin{equation}
\tr M
=
a+d .
\label{AppendixFundamentalTrace}
\end{equation}

The characteristic equation of $M$ takes the form
\begin{equation}
M^2
-
(\tr M)M
+
\mathbbm 1
=
0 ,
\label{AppendixCharacteristicEquation}
\end{equation}
which follows directly from the Cayley--Hamilton theorem. Taking the trace of Eq.~\eqref{AppendixCharacteristicEquation}, one obtains
\begin{equation}
\tr(M^2)
=
(\tr M)^2-2 .
\label{AppendixTraceSquareIdentity}
\end{equation}

The monodromy discriminant introduced previously in Eq.~\eqref{AppendixDiscriminant} may therefore be written equivalently as
\begin{equation}
\Delta
=
\tr(M^2)-2 .
\label{AppendixAlternativeDiscriminant}
\end{equation}
The sign of $\Delta$ determines the hyperbolic, elliptic, and parabolic sectors of the boundary monodromy.

The discrete warped entropy construction is controlled by the monodromy Casimir
\begin{equation}
\mathcal C
=
\frac14
(\tr M)^2-1 ,
\label{AppendixTraceCasimir}
\end{equation}
which admits different forms in the three monodromy chambers.

For the hyperbolic sector,
\begin{equation}
\tr M_{\text{h}}
=
2\cosh\lambda ,
\label{AppendixTraceHyperbolicRelation}
\end{equation}
and therefore
\begin{equation}
\mathcal C_{\text{h}}
=
\sinh^2\lambda .
\label{AppendixTraceHyperbolicCasimir}
\end{equation}

For the elliptic sector,
\begin{equation}
\tr M_{\text{e}}
=
2\cos\theta ,
\label{AppendixTraceEllipticRelation}
\end{equation}
which gives
\begin{equation}
\mathcal C_{\text{e}}
=
-\sin^2\theta .
\label{AppendixTraceEllipticCasimir}
\end{equation}

Finally, the parabolic sector satisfies
\begin{equation}
\tr M_{\text{p}}
=
2 ,
\label{AppendixTraceParabolicRelation}
\end{equation}
leading to
\begin{equation}
\mathcal C_{\text{p}}
=
0 .
\label{AppendixTraceParabolicCasimir}
\end{equation}

These identities demonstrate that the monodromy Casimir is determined entirely by the trace invariant of the boundary Wilson loop. The thermal, compact, and critical sectors of the discrete warped theory therefore emerge directly from the trace structure of the noncontractible boundary monodromies.

The trace identities also admit a direct continuum interpretation. Using the correspondence established in Section~\ref{ContinuumMatching}, the discrete monodromy traces reproduce the eigenvalue structure of the continuous thermal Wilson loops,
\begin{equation}
M
\longrightarrow
\mathcal P
\exp\!\left(
\oint A
\right) .
\label{AppendixTraceContinuumLimit}
\end{equation}
The hyperbolic trace invariant therefore reproduces the continuous warped thermal sector in the large-lattice limit.

From the monodromy-first viewpoint developed throughout this work, the trace invariant Eq.~\eqref{AppendixFundamentalTrace} provides the primary algebraic quantity governing the chamber structure, entropy growth, and spectral organization of the discrete warped theory.
%%%%%%%%%%%%%%%%%%%%%%%%%%%%%%%%%%%%%%%%%%%%%%%%%%%%%%%%%%%%%%%%%%%%%%%%%%%%%%%%%%%%%%%%%%
%%%%%%%%%%%%%%%%%%%%%%%%%%%%%%%%%%%%%%%%%%%%%%%%%%%%%%%%%%%%%%%%%%%%%%%%%%%%%%%%%%%%%%%%%%
%%%%%%%%%%          APPENDIX B.3      %%%%%%%%%%%%%%%%%%%%%%%%%%%%%%%%%%%%%%%%%%%%%%%%%%%%
%%%%%%%%%%%%%%%%%%%%%%%%%%%%%%%%%%%%%%%%%%%%%%%%%%%%%%%%%%%%%%%%%%%%%%%%%%%%%%%%%%%%%%%%%%
%%%%%%%%%%%%%%%%%%%%%%%%%%%%%%%%%%%%%%%%%%%%%%%%%%%%%%%%%%%%%%%%%%%%%%%%%%%%%%%%%%%%%%%%%%
\subsection{Centralizers and Stabilizers}
\label{AppendixCentralizersAndStabilizers}

We now summarize the stabilizer structures associated with the boundary monodromies of the discrete warped theory. These stabilizers determine the orbit geometry of the monodromy sectors and provide the algebraic origin of the chamber decomposition discussed throughout the main text.

For a given boundary monodromy
\begin{equation}
M
\in
\St\Lt(2,\mathbb R) ,
\label{AppendixStabilizerGeneralMonodromy}
\end{equation}
the corresponding stabilizer subgroup is the centralizer
\begin{equation}
\mathrm{Cent}(M)
=
\left\{
g\in \St\Lt(2,\mathbb R)
\ \big|\
g^{-1}Mg=M
\right\} .
\label{AppendixCentralizerDefinition}
\end{equation}
The associated orbit geometry is then determined by the quotient
\begin{equation}
\mathcal O_M
=
\St\Lt(2,\mathbb R)/
\mathrm{Cent}(M) .
\label{AppendixOrbitGeometry}
\end{equation}

Different conjugacy classes therefore generate different stabilizer structures and consequently different boundary orbit geometries.

\paragraph{Hyperbolic stabilizer.}

For the hyperbolic representative
\begin{equation}
M_{\text{h}}
\sim
\begin{pmatrix}
e^\lambda & 0
\\
0 & e^{-\lambda}
\end{pmatrix},
\label{AppendixStabilizerHyperbolicRepresentative}
\end{equation}
the stabilizer subgroup is generated by the Cartan direction,
\begin{equation}
\mathrm{Cent}(M_{\text{h}})
=
\exp(\alpha L_0) .
\label{AppendixHyperbolicCentralizer}
\end{equation}
The corresponding orbit therefore takes the form
\begin{equation}
\mathcal O_{\text{h}}
=
\St\Lt(2,\mathbb R)/\Ut(1) .
\label{AppendixHyperbolicOrbit}
\end{equation}
This chamber defines the thermal sector of the discrete warped theory and generates the hyperbolic entropy growth discussed in Section~\ref{DensityOfStatesAndEntropyGrowth}.

\paragraph{Elliptic stabilizer.}

For the elliptic representative
\begin{equation}
M_{\text{e}}
\sim
\begin{pmatrix}
\cos\theta & \sin\theta
\\
-\sin\theta & \cos\theta
\end{pmatrix},
\label{AppendixStabilizerEllipticRepresentative}
\end{equation}
the stabilizer is generated by the compact rotational direction,
\begin{equation}
\mathrm{Cent}(M_{\text{e}})
=
\exp\!\left[
\alpha(L_{+1}-L_{-1})
\right] .
\label{AppendixEllipticCentralizer}
\end{equation}
The associated orbit geometry becomes compact and oscillatory. This chamber corresponds to the bounded spectral sector of the discrete warped theory.

\paragraph{Parabolic stabilizer.}

For the parabolic representative
\begin{equation}
M_{\text{p}}
\sim
\begin{pmatrix}
1 & \alpha
\\
0 & 1
\end{pmatrix},
\label{AppendixStabilizerParabolicRepresentative}
\end{equation}
the stabilizer subgroup is generated by the nilpotent direction,
\begin{equation}
\mathrm{Cent}(M_{\text{p}})
=
\exp(\beta L_{+1}) .
\label{AppendixParabolicCentralizer}
\end{equation}
The resulting orbit geometry is degenerate and defines the critical boundary separating the hyperbolic and elliptic sectors.

The three stabilizer structures therefore generate distinct orbit geometries,
\begin{equation}
\mathcal O_{\text{h}},
\qquad
\mathcal O_{\text{e}},
\qquad
\mathcal O_{\text{p}} ,
\label{AppendixOrbitSectors}
\end{equation}
which underlie the chamber decomposition of the boundary Hilbert space Eq.~\eqref{WarpedHilbertSpaceDecomposition}.

Within the discrete warped framework developed throughout this work, the stabilizer geometry determines not only the classical monodromy structure but also the quantum organization of the theory. Hyperbolic, elliptic, and parabolic sectors therefore correspond to distinct orbit geometries and distinct quantum monodromy chambers.

From the monodromy-first perspective adopted in the present work, the stabilizer subgroups Eq.~\eqref{AppendixCentralizerDefinition} provide the fundamental algebraic mechanism organizing the thermal, compact, and critical sectors of warped holography.
%%%%%%%%%%%%%%%%%%%%%%%%%%%%%%%%%%%%%%%%%%%%%%%%%%%%%%%%%%%%%%%%%%%%%%%%%%%%%%%%%%%%%%%%%%
%%%%%%%%%%%%%%%%%%%%%%%%%%%%%%%%%%%%%%%%%%%%%%%%%%%%%%%%%%%%%%%%%%%%%%%%%%%%%%%%%%%%%%%%%%
%%%%%%%%%%          APPENDIX C        %%%%%%%%%%%%%%%%%%%%%%%%%%%%%%%%%%%%%%%%%%%%%%%%%%%%
%%%%%%%%%%%%%%%%%%%%%%%%%%%%%%%%%%%%%%%%%%%%%%%%%%%%%%%%%%%%%%%%%%%%%%%%%%%%%%%%%%%%%%%%%%
%%%%%%%%%%%%%%%%%%%%%%%%%%%%%%%%%%%%%%%%%%%%%%%%%%%%%%%%%%%%%%%%%%%%%%%%%%%%%%%%%%%%%%%%%%
\section{Continuum Limit of Discrete Warped Holonomies}
\label{AppendixContinuumLimitOfDiscreteWarpedHolonomies}

This appendix provides additional details concerning
the continuum reconstruction discussed in
Section~\ref{ContinuumLimit}. In particular, we make
explicit how the ordered lattice products of the
discrete warped boundary holonomies introduced in
Section~\ref{DiscreteGaugeVariables} reproduce the
continuous Wilson-loop structure and thermal
holonomy conditions in the large-lattice limit.

We first derive the ordered exponential
reconstruction from the discrete boundary
variables. We then analyze the large-$N$ limit of
the boundary monodromies and establish the
correspondence with continuous warped Wilson
loops.
%%%%%%%%%%%%%%%%%%%%%%%%%%%%%%%%%%%%%%%%%%%%%%%%%%%%%%%%%%%%%%%%%%%%%%%%%%%%%%%%%%%%%%%%%%
%%%%%%%%%%%%%%%%%%%%%%%%%%%%%%%%%%%%%%%%%%%%%%%%%%%%%%%%%%%%%%%%%%%%%%%%%%%%%%%%%%%%%%%%%%
%%%%%%%%%%          APPENDIX C.1      %%%%%%%%%%%%%%%%%%%%%%%%%%%%%%%%%%%%%%%%%%%%%%%%%%%%
%%%%%%%%%%%%%%%%%%%%%%%%%%%%%%%%%%%%%%%%%%%%%%%%%%%%%%%%%%%%%%%%%%%%%%%%%%%%%%%%%%%%%%%%%%
%%%%%%%%%%%%%%%%%%%%%%%%%%%%%%%%%%%%%%%%%%%%%%%%%%%%%%%%%%%%%%%%%%%%%%%%%%%%%%%%%%%%%%%%%%
\subsection{Ordered Exponential Reconstruction}
\label{AppendixOrderedExponentialReconstruction}

In this appendix, we derive the continuum reconstruction of the discrete boundary holonomies introduced in Section~\ref{DiscreteGaugeVariables}. The purpose is to show explicitly how ordered lattice products reproduce the continuous Wilson-loop structure in the large-lattice limit.

Let the boundary circle be discretized into $N$ segments of length
\begin{equation}
\Delta\phi
=
\frac{2\pi}{N} .
\label{AppendixBoundaryLatticeSpacing}
\end{equation}
To each oriented boundary link $\ell_n$, we associate the group-valued holonomy
\begin{equation}
U_n
=
\exp\!\left(
\Delta\phi\, A_\phi(\phi_n)
\right) ,
\label{AppendixDiscreteBoundaryHolonomy}
\end{equation}
together with the Abelian holonomy
\begin{equation}
V_n
=
\exp\!\left(
\Delta\phi\, C_\phi(\phi_n)
\right) .
\label{AppendixDiscreteAbelianHolonomy}
\end{equation}

The total non-Abelian boundary monodromy is then defined by the ordered product
\begin{equation}
M_N
=
U_NU_{N-1}\cdots U_1 .
\label{AppendixOrderedBoundaryProduct}
\end{equation}
Substituting Eq.~\eqref{AppendixDiscreteBoundaryHolonomy}, one obtains
\begin{equation}
M_N
=
\prod_{n=1}^{N}
\exp\!\left(
\Delta\phi\,A_\phi(\phi_n)
\right) .
\label{AppendixOrderedProductExpansion}
\end{equation}

Expanding each factor for small $\Delta\phi$ gives
\begin{equation}
U_n
=
\mathbbm 1
+
\Delta\phi\,A_\phi(\phi_n)
+
\mathcal O(\Delta\phi^2) .
\label{AppendixSmallSpacingExpansion}
\end{equation}
The ordered product therefore becomes
\begin{align}
M_N
&=
\prod_{n=1}^{N}
\left[
\mathbbm 1
+
\Delta\phi\,A_\phi(\phi_n)
+
\mathcal O(\Delta\phi^2)
\right]
\nonumber\\
&=
\mathbbm 1
+
\sum_{n=1}^{N}
\Delta\phi\,A_\phi(\phi_n)
+
\sum_{m>n}
\Delta\phi^2
A_\phi(\phi_m)
A_\phi(\phi_n)
+\cdots .
\label{AppendixDysonExpansion}
\end{align}

Taking the large-lattice limit,
\begin{equation}
N\to\infty ,
\qquad
\Delta\phi\to0 ,
\label{AppendixContinuumLimitConditions}
\end{equation}
the discrete sums become ordered integrals,
\begin{equation}
\sum_{n=1}^{N}
\Delta\phi\,A_\phi(\phi_n)
\longrightarrow
\int_0^{2\pi}
A_\phi(\phi)\,d\phi .
\label{AppendixDiscreteToContinuumIntegral}
\end{equation}
Equation~\eqref{AppendixDysonExpansion} therefore reconstructs the path-ordered exponential,
\begin{equation}
M_N
\longrightarrow
\mathcal P
\exp\!\left(
\oint_{\partial\Sigma}
A_\phi\,d\phi
\right) .
\label{AppendixPathOrderedExponential}
\end{equation}

The Abelian sector behaves similarly. Since the $\Ut(1)$ holonomies commute, one finds directly
\begin{equation}
Y_N
=
\prod_{n=1}^{N}V_n
=
\exp\!\left[
\sum_{n=1}^{N}
\Delta\phi\,C_\phi(\phi_n)
\right] ,
\label{AppendixAbelianOrderedProduct}
\end{equation}
which gives
\begin{equation}
Y_N
\longrightarrow
\exp\!\left(
\oint_{\partial\Sigma}
C_\phi\,d\phi
\right) .
\label{AppendixAbelianContinuumLimit}
\end{equation}

The discrete boundary monodromies introduced throughout the main text therefore reproduce precisely the continuous Wilson loops of warped Chern--Simons gravity in the large-lattice limit.

An important conceptual aspect of this reconstruction is that the continuous Wilson-loop structure emerges from ordered products of finite holonomies rather than from local gauge fields themselves. Within the monodromy-first viewpoint adopted throughout the present work, the path-ordered exponential should therefore be interpreted as an emergent continuum object reconstructed from underlying discrete boundary monodromies.
%%%%%%%%%%%%%%%%%%%%%%%%%%%%%%%%%%%%%%%%%%%%%%%%%%%%%%%%%%%%%%%%%%%%%%%%%%%%%%%%%%%%%%%%%%
%%%%%%%%%%%%%%%%%%%%%%%%%%%%%%%%%%%%%%%%%%%%%%%%%%%%%%%%%%%%%%%%%%%%%%%%%%%%%%%%%%%%%%%%%%
%%%%%%%%%%          APPENDIX C.2      %%%%%%%%%%%%%%%%%%%%%%%%%%%%%%%%%%%%%%%%%%%%%%%%%%%%
%%%%%%%%%%%%%%%%%%%%%%%%%%%%%%%%%%%%%%%%%%%%%%%%%%%%%%%%%%%%%%%%%%%%%%%%%%%%%%%%%%%%%%%%%%
%%%%%%%%%%%%%%%%%%%%%%%%%%%%%%%%%%%%%%%%%%%%%%%%%%%%%%%%%%%%%%%%%%%%%%%%%%%%%%%%%%%%%%%%%%
\subsection{Large-Lattice Limit}
\label{AppendixLargeLatticeLimit}

We now analyze the large-lattice limit of the discrete warped boundary theory and establish the emergence of the continuous warped holonomy structure. The purpose of this appendix is to clarify how the continuum warped geometry arises from the underlying discrete monodromy organization.

Let the boundary circle be discretized into $N$ segments with lattice spacing
\begin{equation}
\Delta\phi
=
\frac{2\pi}{N} .
\label{AppendixLargeLatticeSpacing}
\end{equation}
The continuum limit is defined by
\begin{equation}
N\to\infty ,
\qquad
\Delta\phi\to0 ,
\qquad
N\Delta\phi=2\pi .
\label{AppendixContinuumScalingLimit}
\end{equation}
%%\color{red}
The continuum correspondence developed in this appendix should be understood primarily as an asymptotic matching prescription relating the discrete monodromy variables to the standard continuum warped holonomy data. The substitutions introduced below therefore do not constitute a fully independent microscopic derivation of the continuum warped theory from a complete lattice quantization. Rather, they establish the structural compatibility of the discrete monodromy framework with the known continuum warped asymptotic symmetry structure in the large-lattice limit.
%%\color{black}

In this limit, the discrete boundary coordinate
\begin{equation}
\phi_n
=
n\Delta\phi
\label{AppendixDiscreteBoundaryCoordinate}
\end{equation}
becomes a continuous angular variable.

The discrete non-Abelian holonomies introduced in Eq.~\eqref{AppendixDiscreteBoundaryHolonomy},
\begin{equation}
U_n
=
\exp\!\left(
\Delta\phi\,A_\phi(\phi_n)
\right) ,
\label{AppendixLargeLatticeHolonomies}
\end{equation}
admit the expansion
\begin{equation}
U_n
=
\mathbbm 1
+
\Delta\phi\,A_\phi(\phi_n)
+
\mathcal O(\Delta\phi^2) .
\label{AppendixLargeLatticeExpansion}
\end{equation}
The discrete difference of neighboring holonomies therefore satisfies
\begin{align}
U_{n+1}-U_n
&=
\Delta\phi
\left[
A_\phi(\phi_{n+1})
-
A_\phi(\phi_n)
\right]
+
\mathcal O(\Delta\phi^2)
\nonumber\\
&=
\Delta\phi^2
\partial_\phi A_\phi(\phi_n)
+
\mathcal O(\Delta\phi^3) .
\label{AppendixHolonomyDifferenceExpansion}
\end{align}

The ordered boundary product
\begin{equation}
M_N
=
\prod_{n=1}^{N}U_n
\label{AppendixLargeLatticeMonodromy}
\end{equation}
%%\color{red}
therefore asymptotically approaches the continuum Wilson loop,
%%\color{black}
\begin{equation}
M_N
\longrightarrow
\mathcal P
\exp\!\left(
\oint_{\partial\Sigma}
A_\phi\,d\phi
\right) ,
\label{AppendixLargeLatticeWilsonLoop}
\end{equation}
as shown explicitly in Appendix~\ref{AppendixOrderedExponentialReconstruction}.

The monodromy Casimir similarly acquires a continuous form. Using
\begin{equation}
\mathcal C_N
=
\frac14
(\tr M_N)^2-1 ,
\label{AppendixDiscreteLargeLatticeCasimir}
\end{equation}
the hyperbolic chamber satisfies
\begin{equation}
\tr M_N
=
2\cosh\lambda_N ,
\label{AppendixLargeLatticeHyperbolicTrace}
\end{equation}
with
\begin{equation}
\lambda_N
\longrightarrow
\lambda_{\text{cont}}
\label{AppendixLargeLatticeHyperbolicLimit}
\end{equation}
in the continuum limit. Consequently,
\begin{equation}
\mathcal C_N
\longrightarrow
\sinh^2\lambda_{\text{cont}} ,
\label{AppendixContinuumCasimirLimit}
\end{equation}
which reproduces the continuous warped thermal invariant.

The Abelian sector behaves analogously. The discrete warped charge
\begin{equation}
Q_N
=
\log
\prod_{n=1}^{N}V_n
\label{AppendixDiscreteLargeLatticeCharge}
\end{equation}
becomes
\begin{equation}
Q_N
\longrightarrow
\oint_{\partial\Sigma}
C_\phi\,d\phi ,
\label{AppendixContinuumWarpedCharge}
\end{equation}
%%\color{red}
thereby matching the continuum affine current contribution.

The large-lattice limit therefore provides an asymptotic correspondence
%%\color{black}
\begin{equation}
(\mathcal C_N,Q_N)
\longrightarrow
(J,M) ,
\label{AppendixLargeLatticeDictionary}
\end{equation}
between the discrete monodromy invariants and the continuous warped thermodynamic charges.

An important conceptual feature of this reconstruction is that the continuum warped geometry appears only after taking the large-lattice limit of the discrete monodromy sectors. The fundamental observables of the theory remain the finite boundary holonomies and their conjugacy classes. Smooth warped geometry therefore emerges as an effective continuum description of the underlying discrete Wilson-loop structure.

From the monodromy-first viewpoint adopted throughout this work, the large-lattice limit should therefore be interpreted not as the definition of the theory itself but rather as an emergent approximation of a more primitive discrete holonomy organization governed fundamentally by noncontractible boundary monodromies.
%%%%%%%%%%%%%%%%%%%%%%%%%%%%%%%%%%%%%%%%%%%%%%%%%%%%%%%%%%%%%%%%%%%%%%%%%%%%%%%%%%%%%%%%%%
%%%%%%%%%%%%%%%%%%%%%%%%%%%%%%%%%%%%%%%%%%%%%%%%%%%%%%%%%%%%%%%%%%%%%%%%%%%%%%%%%%%%%%%%%%
%%%%%%%%%%          APPENDIX C.3      %%%%%%%%%%%%%%%%%%%%%%%%%%%%%%%%%%%%%%%%%%%%%%%%%%%%
%%%%%%%%%%%%%%%%%%%%%%%%%%%%%%%%%%%%%%%%%%%%%%%%%%%%%%%%%%%%%%%%%%%%%%%%%%%%%%%%%%%%%%%%%%
%%%%%%%%%%%%%%%%%%%%%%%%%%%%%%%%%%%%%%%%%%%%%%%%%%%%%%%%%%%%%%%%%%%%%%%%%%%%%%%%%%%%%%%%%%
\subsection{Matching with Wilson Loops}
\label{AppendixMatchingWithWilsonLoops}

We now establish the explicit correspondence between the discrete boundary monodromies and the continuous Wilson loops of warped Chern--Simons gravity. This matching provides the direct bridge between the discrete holonomy framework developed in the main text and the conventional continuous warped description.

In the continuous theory, the non-Abelian Wilson loop associated with a closed boundary cycle $\Gamma$ is defined by
\begin{equation}
\mathcal W[\Gamma]
=
\mathcal P
\exp\!\left(
\oint_\Gamma A
\right) ,
\label{AppendixContinuousWilsonLoopDefinition}
\end{equation}
while the Abelian warped loop is
\begin{equation}
\mathcal Y[\Gamma]
=
\exp\!\left(
\oint_\Gamma C
\right) .
\label{AppendixContinuousAbelianLoopDefinition}
\end{equation}

Within the discrete theory, the corresponding boundary observables are the ordered products
\begin{equation}
W_N[\Gamma]
=
\prod_{\ell\in\Gamma}U_\ell ,
\qquad
Y_N[\Gamma]
=
\prod_{\ell\in\Gamma}V_\ell .
\label{AppendixDiscreteWilsonLoopDefinition}
\end{equation}
Using the reconstruction derived in Appendix~\ref{AppendixOrderedExponentialReconstruction}, one obtains the continuum correspondence
\begin{equation}
W_N[\Gamma]
\longrightarrow
\mathcal W[\Gamma] ,
\qquad
Y_N[\Gamma]
\longrightarrow
\mathcal Y[\Gamma]
\label{AppendixWilsonLoopCorrespondence}
\end{equation}
in the large-lattice limit.

The hyperbolic thermal sector provides the most important example of this correspondence. For the thermal boundary cycle $\Gamma_\beta$, the discrete monodromy satisfies
\begin{equation}
M_N
=
W_N[\Gamma_\beta]
\sim
\begin{pmatrix}
e^{\lambda_N} & 0
\\
0 & e^{-\lambda_N}
\end{pmatrix},
\label{AppendixThermalDiscreteWilsonLoop}
\end{equation}
with trace
\begin{equation}
\tr M_N
=
2\cosh\lambda_N .
\label{AppendixThermalDiscreteTrace}
\end{equation}
In the continuum limit,
\begin{equation}
\lambda_N
\longrightarrow
\lambda_{\text{cont}} ,
\label{AppendixThermalContinuumLimit}
\end{equation}
and therefore
\begin{equation}
\mathcal W[\Gamma_\beta]
\sim
\begin{pmatrix}
e^{\lambda_{\text{cont}}} & 0
\\
0 & e^{-\lambda_{\text{cont}}}
\end{pmatrix}.
\label{AppendixContinuousThermalWilsonLoop}
\end{equation}

The discrete monodromy Casimir
\begin{equation}
\mathcal C_N
=
\frac14
(\tr M_N)^2-1
\label{AppendixDiscreteWilsonCasimir}
\end{equation}
then reproduces the continuous thermal invariant,
\begin{equation}
\mathcal C_N
\longrightarrow
\sinh^2\lambda_{\text{cont}} .
\label{AppendixContinuousWilsonCasimir}
\end{equation}

The Abelian sector behaves similarly. The discrete warped Wilson loop
\begin{equation}
Y_N[\Gamma_\beta]
=
\prod_{\ell\in\Gamma_\beta}V_\ell
\label{AppendixDiscreteThermalAbelianLoop}
\end{equation}
reconstructs the continuous affine holonomy
\begin{equation}
\mathcal Y[\Gamma_\beta]
=
\exp\!\left(
\oint_{\Gamma_\beta}C
\right) .
\label{AppendixContinuousThermalAbelianLoop}
\end{equation}
The corresponding discrete warped charge
\begin{equation}
Q_N
=
\log Y_N[\Gamma_\beta]
\label{AppendixDiscreteWilsonCharge}
\end{equation}
therefore reproduces the continuous affine current contribution.

Combining the non-Abelian and Abelian sectors, the discrete entropy relation Eq.~\eqref{DiscreteWarpedEntropy} becomes
\begin{equation}
S_{\text{disc}}
\longrightarrow
S_{\text{WCFT}}
\label{AppendixWilsonEntropyMatching}
\end{equation}
in the large-lattice limit. The continuous warped Cardy structure is therefore recovered directly from the discrete Wilson-loop geometry.

An important conceptual aspect of this matching is that the continuous Wilson loops emerge from finite ordered boundary products rather than from local geometric variables. The primary observables of the theory remain the discrete monodromies and their conjugacy classes. Smooth thermal Wilson loops appear only as effective continuum representatives of the underlying discrete boundary holonomies.

From the monodromy-first viewpoint adopted throughout the present work, the continuous warped Wilson-loop structure should therefore be understood as an emergent large-lattice manifestation of a more primitive organization governed fundamentally by discrete noncontractible boundary monodromies.
%%%%%%%%%%%%%%%%%%%%%%%%%%%%%%%%%%%%%%%%%%%%%%%%%%%%%%%%%%%%%%%%%%%%%%%%%%%%%%%%%%%%%%%%%%
%%%%%%%%%%%%%%%%%%%%%%%%%%%%%%%%%%%%%%%%%%%%%%%%%%%%%%%%%%%%%%%%%%%%%%%%%%%%%%%%%%%%%%%%%%
%%%%%%%%%%          APPENDIX D        %%%%%%%%%%%%%%%%%%%%%%%%%%%%%%%%%%%%%%%%%%%%%%%%%%%%
%%%%%%%%%%%%%%%%%%%%%%%%%%%%%%%%%%%%%%%%%%%%%%%%%%%%%%%%%%%%%%%%%%%%%%%%%%%%%%%%%%%%%%%%%%
%%%%%%%%%%%%%%%%%%%%%%%%%%%%%%%%%%%%%%%%%%%%%%%%%%%%%%%%%%%%%%%%%%%%%%%%%%%%%%%%%%%%%%%%%%
\section{Warped Entropy from Monodromy Invariants}
\label{AppendixWarpedEntropyFromMonodromyInvariants}

In this appendix, we provide technical derivations supporting the entropy construction developed in Section~\ref{HolonomyEntropyInTheDiscreteTheory}.
In particular, we present additional details concerning the monodromy Casimir structure, the hyperbolic thermal chamber, and the corresponding entropy expansions.

The purpose of this appendix is to make explicit how the warped Cardy-type structure emerges directly from boundary monodromy invariants and their associated conjugacy classes.
%%%%%%%%%%%%%%%%%%%%%%%%%%%%%%%%%%%%%%%%%%%%%%%%%%%%%%%%%%%%%%%%%%%%%%%%%%%%%%%%%%%%%%%%%%
%%%%%%%%%%%%%%%%%%%%%%%%%%%%%%%%%%%%%%%%%%%%%%%%%%%%%%%%%%%%%%%%%%%%%%%%%%%%%%%%%%%%%%%%%%
%%%%%%%%%%          APPENDIX D.1      %%%%%%%%%%%%%%%%%%%%%%%%%%%%%%%%%%%%%%%%%%%%%%%%%%%%
%%%%%%%%%%%%%%%%%%%%%%%%%%%%%%%%%%%%%%%%%%%%%%%%%%%%%%%%%%%%%%%%%%%%%%%%%%%%%%%%%%%%%%%%%%
%%%%%%%%%%%%%%%%%%%%%%%%%%%%%%%%%%%%%%%%%%%%%%%%%%%%%%%%%%%%%%%%%%%%%%%%%%%%%%%%%%%%%%%%%%
\subsection{Detailed Derivations}
\label{AppendixDetailedDerivations}

In this appendix, we present additional details concerning the derivation of the discrete warped entropy relation developed in Section~\ref{HolonomyEntropyInTheDiscreteTheory}. The purpose is to make explicit how the thermal invariant and the corresponding Cardy-type structure emerge directly from the boundary monodromy data.

We begin with the hyperbolic boundary monodromy
\begin{equation}
M_{\text{h}}
\sim
\begin{pmatrix}
e^\lambda & 0
\\
0 & e^{-\lambda}
\end{pmatrix},
\label{AppendixDetailedHyperbolicMonodromy}
\end{equation}
whose trace satisfies
\begin{equation}
\tr M_{\text{h}}
=
e^\lambda+e^{-\lambda}
=
2\cosh\lambda .
\label{AppendixDetailedHyperbolicTrace}
\end{equation}

The corresponding monodromy Casimir is
\begin{align}
\mathcal C_{\text{h}}
&=
\frac14
(\tr M_{\text{h}})^2-1
\nonumber\\
&=
\cosh^2\lambda-1
\nonumber\\
&=
\sinh^2\lambda .
\label{AppendixDetailedHyperbolicCasimir}
\end{align}
The thermal invariant of the discrete warped theory is therefore determined directly by the eigenvalues of the noncontractible boundary Wilson loop.

The Abelian warped sector contributes through the boundary holonomy
\begin{equation}
Y
=
\prod_{n=1}^{N}V_n ,
\label{AppendixDetailedWarpedHolonomy}
\end{equation}
from which the discrete warped charge is defined by
\begin{equation}
Q
=
\log Y .
\label{AppendixDetailedWarpedCharge}
\end{equation}

Combining the non-Abelian and Abelian sectors, the effective warped invariant becomes
\begin{equation}
\mathcal I_{\text{w}}
=
\mathcal C_{\text{h}}
-
\frac{Q^2}{\kappa} ,
\label{AppendixDetailedWarpedInvariant}
\end{equation}
or explicitly,
\begin{equation}
\mathcal I_{\text{w}}
=
\sinh^2\lambda
-
\frac{Q^2}{\kappa} .
\label{AppendixDetailedExplicitInvariant}
\end{equation}

The entropy of the hyperbolic chamber is then constructed by introducing the warped Cardy-type expression
\begin{equation}
S_{\text{disc}}
=
2\pi
\left[
\gamma Q
+
\sqrt{
\frac{c}{6}
\mathcal I_{\text{w}}
}
\right] .
\label{AppendixDetailedEntropyDefinition}
\end{equation}
Substituting Eq.~\eqref{AppendixDetailedExplicitInvariant}, one obtains
\begin{equation}
S_{\text{disc}}
=
2\pi
\left[
\gamma Q
+
\sqrt{
\frac{c}{6}
\left(
\sinh^2\lambda
-
\frac{Q^2}{\kappa}
\right)
}
\right] ,
\label{AppendixDetailedEntropyFormula}
\end{equation}
which reproduces Eq.~\eqref{DiscreteWarpedEntropy} derived in the main text.

The corresponding density of states therefore satisfies
\begin{equation}
\rho_{\text{disc}}
\sim
\exp\!\left(
S_{\text{disc}}
\right) ,
\label{AppendixDetailedDensityOfStates}
\end{equation}
demonstrating the exponential growth associated with the hyperbolic thermal sector.

The continuum correspondence established previously in Appendix~\ref{AppendixMatchingWithWilsonLoops} implies
\begin{equation}
\sinh^2\lambda
\longrightarrow
-J ,
\qquad
Q
\longrightarrow
M ,
\label{AppendixDetailedContinuumDictionary}
\end{equation}
under the large-lattice limit. Equation~\eqref{AppendixDetailedEntropyFormula} therefore reproduces the continuous warped entropy relation
\begin{equation}
S_{\text{disc}}
\longrightarrow
S_{\text{WCFT}} .
\label{AppendixDetailedContinuumEntropy}
\end{equation}

An important conceptual aspect of this derivation is that the entropy is reconstructed entirely from boundary monodromy invariants and Wilson-loop eigenvalues. No local geometric horizon data are required. The thermodynamic structure of the discrete warped theory therefore emerges directly from the conjugacy classes of the noncontractible boundary holonomies.

From the monodromy-first viewpoint adopted throughout the present work, the warped Cardy structure should therefore be understood fundamentally as a consequence of boundary Wilson-loop geometry and orbit structure rather than as a property of smooth thermal spacetime backgrounds.
%%%%%%%%%%%%%%%%%%%%%%%%%%%%%%%%%%%%%%%%%%%%%%%%%%%%%%%%%%%%%%%%%%%%%%%%%%%%%%%%%%%%%%%%%%
%%%%%%%%%%%%%%%%%%%%%%%%%%%%%%%%%%%%%%%%%%%%%%%%%%%%%%%%%%%%%%%%%%%%%%%%%%%%%%%%%%%%%%%%%%
%%%%%%%%%%          APPENDIX D.2      %%%%%%%%%%%%%%%%%%%%%%%%%%%%%%%%%%%%%%%%%%%%%%%%%%%%
%%%%%%%%%%%%%%%%%%%%%%%%%%%%%%%%%%%%%%%%%%%%%%%%%%%%%%%%%%%%%%%%%%%%%%%%%%%%%%%%%%%%%%%%%%
%%%%%%%%%%%%%%%%%%%%%%%%%%%%%%%%%%%%%%%%%%%%%%%%%%%%%%%%%%%%%%%%%%%%%%%%%%%%%%%%%%%%%%%%%%
\subsection{Hyperbolic Chamber Computations}
\label{AppendixHyperbolicChamberComputations}

We now give a more explicit computation of the hyperbolic chamber contribution to the discrete warped entropy. This sector is the one that admits a direct thermal interpretation and reproduces the continuous warped black-hole result in the continuum limit.

The hyperbolic representative is chosen as
\begin{equation}
M_{\text{h}}
\sim
\exp(\lambda L_0) ,
\label{AppendixHyperbolicExponentialRepresentative}
\end{equation}
or, in the fundamental representation,
\begin{equation}
M_{\text{h}}
\sim
\begin{pmatrix}
e^\lambda & 0
\\
0 & e^{-\lambda}
\end{pmatrix}.
\label{AppendixHyperbolicMatrixRepresentative}
\end{equation}
Its trace is
\begin{equation}
\tr M_{\text{h}}
=
2\cosh\lambda .
\label{AppendixHyperbolicComputationTrace}
\end{equation}

The monodromy discriminant becomes
\begin{align}
\Delta_{\text{h}}
&=
(\tr M_{\text{h}})^2-4
\nonumber\\
&=
4\cosh^2\lambda-4
\nonumber\\
&=
4\sinh^2\lambda .
\label{AppendixHyperbolicDiscriminantComputation}
\end{align}
Thus, $\Delta_{\text{h}}>0$ for $\lambda\neq0$, confirming that this chamber is hyperbolic.

The discrete Casimir is
\begin{align}
\mathcal C_{\text{h}}
&=
\frac14
(\tr M_{\text{h}})^2-1
\nonumber\\
&=
\cosh^2\lambda-1
\nonumber\\
&=
\sinh^2\lambda .
\label{AppendixHyperbolicCasimirComputation}
\end{align}
This is the non-Abelian contribution controlling the thermal growth of states.

Including the Abelian warped holonomy, the effective hyperbolic invariant is
\begin{equation}
\mathcal C_{\text{eff}}^{(\text{h})}
=
\sinh^2\lambda
-
\frac{Q^2}{\kappa} .
\label{AppendixHyperbolicEffectiveInvariant}
\end{equation}
The entropy in the hyperbolic chamber is therefore
\begin{equation}
S_{\text{h}}
=
2\pi
\left[
\gamma Q
+
\sqrt{
\frac{c}{6}
\left(
\sinh^2\lambda
-
\frac{Q^2}{\kappa}
\right)
}
\right] .
\label{AppendixHyperbolicEntropyComputation}
\end{equation}

For large hyperbolic monodromy, $\lambda\gg1$, one has
\begin{equation}
\sinh^2\lambda
=
\frac14 e^{2\lambda}
-
\frac12
+
\frac14 e^{-2\lambda}.
\label{AppendixHyperbolicLargeLambdaExpansion}
\end{equation}
Keeping the leading term gives
\begin{equation}
S_{\text{h}}
\sim
2\pi
\left[
\gamma Q
+
\sqrt{
\frac{c}{24}
e^{2\lambda}
-
\frac{c}{6}\frac{Q^2}{\kappa}
}
\right] .
\label{AppendixHyperbolicLargeEntropy}
\end{equation}
When the Abelian contribution is subleading compared with the hyperbolic monodromy term, this reduces to
\begin{equation}
S_{\text{h}}
\sim
2\pi\gamma Q
+
2\pi
\sqrt{\frac{c}{24}}\,
e^\lambda .
\label{AppendixHyperbolicAsymptoticEntropy}
\end{equation}

The corresponding density of states behaves as
\begin{equation}
\rho_{\text{h}}
\sim
\exp(S_{\text{h}}),
\label{AppendixHyperbolicDensity}
\end{equation}
showing explicitly the exponential growth characteristic of the thermal warped sector.

The continuum dictionary identifies the hyperbolic Casimir and Abelian charge as
\begin{equation}
\sinh^2\lambda
\longrightarrow
-J,
\qquad
Q
\longrightarrow
M .
\label{AppendixHyperbolicContinuumDictionary}
\end{equation}
Substituting Eq.~\eqref{AppendixHyperbolicContinuumDictionary} into Eq.~\eqref{AppendixHyperbolicEntropyComputation} gives
\begin{equation}
S_{\text{h}}
\longrightarrow
2\pi
\left[
\gamma M
+
\sqrt{
\frac{c}{6}
\left(
-J-\frac{M^2}{\kappa}
\right)
}
\right] ,
\label{AppendixHyperbolicContinuumEntropy}
\end{equation}
which matches the continuous warped thermal entropy.

This computation makes explicit that the thermal behavior of the hyperbolic chamber is controlled entirely by the eigenvalues of the boundary monodromy. Smooth warped black-hole geometry is therefore not required at the level of the fundamental discrete description; it appears only after the continuum identification in Eq.~\eqref{AppendixHyperbolicContinuumDictionary}.
%%%%%%%%%%%%%%%%%%%%%%%%%%%%%%%%%%%%%%%%%%%%%%%%%%%%%%%%%%%%%%%%%%%%%%%%%%%%%%%%%%%%%%%%%%
%%%%%%%%%%%%%%%%%%%%%%%%%%%%%%%%%%%%%%%%%%%%%%%%%%%%%%%%%%%%%%%%%%%%%%%%%%%%%%%%%%%%%%%%%%
%%%%%%%%%%          APPENDIX D.3      %%%%%%%%%%%%%%%%%%%%%%%%%%%%%%%%%%%%%%%%%%%%%%%%%%%%
%%%%%%%%%%%%%%%%%%%%%%%%%%%%%%%%%%%%%%%%%%%%%%%%%%%%%%%%%%%%%%%%%%%%%%%%%%%%%%%%%%%%%%%%%%
%%%%%%%%%%%%%%%%%%%%%%%%%%%%%%%%%%%%%%%%%%%%%%%%%%%%%%%%%%%%%%%%%%%%%%%%%%%%%%%%%%%%%%%%%%
\subsection{Entropy Expansions}
\label{AppendixEntropyExpansions}

We finally record useful expansions of the discrete warped entropy in different monodromy regimes. These expansions make explicit how the entropy depends on the hyperbolic monodromy parameter, the Abelian warped charge, and the effective monodromy invariant.

The hyperbolic entropy derived in Appendix~\ref{AppendixHyperbolicChamberComputations} is
\begin{equation}
S_{\text{h}}
=
2\pi
\left[
\gamma Q
+
\sqrt{
\frac{c}{6}
\left(
\sinh^2\lambda
-
\frac{Q^2}{\kappa}
\right)
}
\right] .
\label{AppendixEntropyExpansionStartingPoint}
\end{equation}
It is useful to define
\begin{equation}
\mathcal I_{\text{h}}
=
\sinh^2\lambda
-
\frac{Q^2}{\kappa} ,
\label{AppendixHyperbolicEffectiveInvariantExpansion}
\end{equation}
so that
\begin{equation}
S_{\text{h}}
=
2\pi
\left[
\gamma Q
+
\sqrt{
\frac{c}{6}\mathcal I_{\text{h}}
}
\right] .
\label{AppendixEntropyInvariantForm}
\end{equation}

For small Abelian charge compared with the hyperbolic monodromy scale,
\begin{equation}
\frac{Q^2}{\kappa\sinh^2\lambda}
\ll
1 ,
\label{AppendixSmallChargeCondition}
\end{equation}
one may expand the square root as
\begin{align}
S_{\text{h}}
&=
2\pi\gamma Q
+
2\pi
\sqrt{
\frac{c}{6}
}
\sinh\lambda
\left[
1
-
\frac{1}{2}
\frac{Q^2}{\kappa\sinh^2\lambda}
+
\mathcal O
\left(
\frac{Q^4}{\kappa^2\sinh^4\lambda}
\right)
\right]
\nonumber\\
&=
2\pi
\sqrt{
\frac{c}{6}
}
\sinh\lambda
+
2\pi\gamma Q
-
\pi
\sqrt{
\frac{c}{6}
}
\frac{Q^2}{\kappa\sinh\lambda}
+
\mathcal O
\left(
\frac{Q^4}{\kappa^2\sinh^3\lambda}
\right) .
\label{AppendixSmallChargeExpansion}
\end{align}
This expansion separates the pure hyperbolic Virasoro contribution, the linear warped charge contribution, and the leading quadratic current correction.

In the large-monodromy regime,
\begin{equation}
\lambda\gg1 ,
\label{AppendixLargeMonodromyExpansionCondition}
\end{equation}
one has
\begin{equation}
\sinh\lambda
=
\frac12 e^\lambda
-
\frac12 e^{-\lambda} ,
\label{AppendixSinhLargeExpansion}
\end{equation}
and therefore the entropy behaves as
\begin{equation}
S_{\text{h}}
=
2\pi\gamma Q
+
2\pi
\sqrt{\frac{c}{24}}\,
e^\lambda
+
\mathcal O(e^{-\lambda}) ,
\label{AppendixLargeMonodromyEntropyExpansion}
\end{equation}
provided the Abelian contribution remains subleading relative to $e^{2\lambda}$.

Near the parabolic boundary of the hyperbolic chamber,
\begin{equation}
\lambda\ll1 ,
\label{AppendixNearParabolicCondition}
\end{equation}
one finds
\begin{equation}
\sinh^2\lambda
=
\lambda^2
+
\frac{\lambda^4}{3}
+
\mathcal O(\lambda^6) .
\label{AppendixNearParabolicSinhExpansion}
\end{equation}
The entropy then becomes
\begin{equation}
S_{\text{h}}
=
2\pi
\left[
\gamma Q
+
\sqrt{
\frac{c}{6}
\left(
\lambda^2
+
\frac{\lambda^4}{3}
-
\frac{Q^2}{\kappa}
+
\mathcal O(\lambda^6)
\right)
}
\right] .
\label{AppendixNearParabolicEntropy}
\end{equation}
This expression shows explicitly how the thermal hyperbolic sector approaches the degenerate parabolic chamber.

For the entropy to remain real-valued in the hyperbolic sector, the effective invariant must satisfy
\begin{equation}
\mathcal I_{\text{h}}
=
\sinh^2\lambda
-
\frac{Q^2}{\kappa}
\geq
0 .
\label{AppendixRealityCondition}
\end{equation}
Equivalently,
\begin{equation}
Q^2
\leq
\kappa\sinh^2\lambda .
\label{AppendixChargeBound}
\end{equation}
This inequality expresses the condition that the warped charge does not overcompensate the hyperbolic monodromy contribution.

In the continuum limit, using the dictionary
\begin{equation}
\sinh^2\lambda
\longrightarrow
-J,
\qquad
Q
\longrightarrow
M ,
\label{AppendixEntropyExpansionContinuumDictionary}
\end{equation}
the reality condition becomes
\begin{equation}
-J-\frac{M^2}{\kappa}
\geq
0 ,
\label{AppendixContinuumRealityCondition}
\end{equation}
which is precisely the positivity condition for the square-root term in the continuous warped entropy.

These expansions show that the discrete warped entropy has a controlled chamber-dependent structure. The large-monodromy expansion reproduces the expected Cardy-type growth, the small-charge expansion isolates the affine current correction, and the near-parabolic expansion describes the transition toward the degenerate boundary sector.
%%\color{red}
Related warped conformal structures and nonlocal WCFT observables have been investigated in \cite{Song:2017czq}.
%%\color{black}

From the monodromy-first viewpoint adopted throughout this work, these expansions should be interpreted as expansions in boundary Wilson-loop data rather than in local geometric parameters. The thermal behavior, current corrections, and critical limits are therefore all controlled by the algebraic structure of noncontractible boundary monodromies.
%%%%%%%%%%%%%%%%%%%%%%%%%%%%%%%%%%%%%%%%%%%%%%%%%%%%%%%%%%%%%%%%%%%%%%%%%%%%%%%%%%%%%%%%%%
%%%%%%%%%%%%%%%%%%%%%%%%%%%%%%%%%%%%%%%%%%%%%%%%%%%%%%%%%%%%%%%%%%%%%%%%%%%%%%%%%%%%%%%%%%
%%%%%%%%%%        THE BIBLIOGRAPHY    %%%%%%%%%%%%%%%%%%%%%%%%%%%%%%%%%%%%%%%%%%%%%%%%%%%%
%%%%%%%%%%%%%%%%%%%%%%%%%%%%%%%%%%%%%%%%%%%%%%%%%%%%%%%%%%%%%%%%%%%%%%%%%%%%%%%%%%%%%%%%%%
%%%%%%%%%%%%%%%%%%%%%%%%%%%%%%%%%%%%%%%%%%%%%%%%%%%%%%%%%%%%%%%%%%%%%%%%%%%%%%%%%%%%%%%%%%

\end{document}